\documentclass[lettersize,journal]{IEEEtran}

\usepackage[T1]{fontenc}
\usepackage{mathtools}
\usepackage{amsfonts}
\usepackage[english]{babel}
\usepackage{booktabs}
\usepackage{tikz}
\usepackage{wrapfig}
\usepackage{xspace}
\usepackage{subcaption}
\usepackage{graphicx}
\usepackage{tabularx}
\usepackage{amsmath}
\usepackage{filecontents}
\usepackage{caption}
\usepackage{algorithm}
\usepackage{algpseudocode}
\usepackage{csquotes}
\usepackage{afterpage}
\usepackage{hyperref}
\usepackage{multicol}
\usepackage{colortbl}
\usepackage{acro}

\usepackage[inline]{enumitem}
\setlist[enumerate]{
label=\roman*)}

\usepackage{adjustbox}

\makeatletter
\let\disable\@secondoftwo
\addto\extrasenglish{%
}	
\addto\extrasenglish{%
}
\addto\extrasenglish{%
}
\addto\extrasenglish{%
}
\makeatother

\usepackage{cite}

\DeclareAcronym{defi}{
    short = DeFi,
    long = decentralized finance
}

\DeclareAcronym{tradfi}{
    short = TradFi,
    long = traditional finance
}

\DeclareAcronym{dqn}{
    short = DQN,
    long = deep Q-network
}

\DeclareAcronym{sgd}{
    short = SGD,
    long = stochastic gradient descent
}

\DeclareAcronym{ema}{
    short = EMA,
    long = exponential moving average
}

\DeclareAcronym{cdf}{
    short = CDF,
    long = cumulative distribution function
}

\DeclareAcronym{rl}{
    short = RL,
    long = reinforcement learning
}

\DeclareAcronym{ai}{
    short = AI,
    long = artificial intelligence
}

\DeclareAcronym{amm}{
    short = AMM,
    long = automated market maker
}

\DeclareAcronym{dao}{
    short = DAO,
    long = decentralized autonomous organization
}

\DeclareAcronym{mdp}{
    short = MDP,
    long = Markov Decision Process
}

\DeclareAcronym{ols}{
short = OLS,
long = ordinary least squares
}

\DeclareAcronym{kde}{
short = KDE,
long = kernel density estimation
}

\begin{document}

\title{Auto.gov: Learning-based Governance for Decentralized Finance (DeFi)}

\author{Jiahua~Xu, Yebo~Feng, Daniel~Perez, and Benjamin~Livshits

\thanks{Received December 2024; revised 12 February 2025; accepted 10 March 2025. (Corresponding author: Yebo Feng.)}

\thanks{Jiahua Xu is with Centre for Blockchain Technologies, University College London, UK, and Exponential Science. Email: jiahua.xu@ucl.ac.uk.}%

\thanks{Yebo Feng is with the School of Computer Science and Engineering, Nanyang Technological University, Singapore. Email: yebo.feng@ntu.edu.sg.}

\thanks{Daniel Perez is with the Computer Science Department, Imperial College London, UK. Email: daniel.perez@imperial.ac.uk.}%

\thanks{Benjamin Livshits is with Imperial College London, UK. Email: b.livshits@imperial.ac.uk.}%

\thanks{Digital Object Identifier \href{https://doi.org/10.1109/TSC.2025.3553700}{10.1109/TSC.2025.3553700}}

}

\maketitle

\begin{abstract}

\Ac{defi} is an integral component of the blockchain ecosystem, enabling a range of financial activities through smart-contract-based protocols. Traditional \ac{defi} governance typically involves manual parameter adjustments by protocol teams or token holder votes, and is thus prone to human bias and financial risks, undermining the system's integrity and security.
While existing efforts aim to establish more adaptive parameter adjustment schemes, there remains a need for a governance model that is both more efficient and resilient to significant market manipulations.
In this paper, we introduce \enquote{Auto.gov}, a learning-based governance framework that employs a \ac{dqn} \ac{rl} strategy to perform semi-automated, data-driven parameter adjustments.
We create a \ac{defi} environment with an encoded action-state space akin to the Aave lending protocol for simulation and testing purposes, where Auto.gov has demonstrated the capability to retain funds that would have otherwise been lost to price oracle attacks. In tests with real-world data, Auto.gov outperforms the benchmark approaches by at least 14\% and the static baseline model by tenfold, in terms of the preset performance metric---protocol profitability.
Overall, the comprehensive evaluations confirm that Auto.gov is more efficient and effective than traditional governance methods, thereby enhancing the security, profitability, and ultimately, the sustainability of \ac{defi} protocols.



\end{abstract}



\section{Introduction}
\label{sec:intro}

\IEEEPARstart{G}{overnance} is a crucial aspect of blockchain-based systems, designed to ensure their stability, integrity, and security~\cite{jensen2021decentralized}. Despite blockchains' ability to store transaction histories without relying on trusted third parties, certain decisions in blockchain systems necessitate an orchestration or {\it governance} processes. This requirement applies to both chains such as Bitcoin and Ethereum (sometimes called layer-1 chains) as well as higher-level developments, such as \acf{defi} protocols~\cite{werner2021sok}.
\Ac{defi} protocols~\cite{Gudgeon2020b} (e.g. Aave, 
Uniswap, and Curve) facilitate borrowing, lending, and exchanging crypto-assets without a financial intermediary \cite{xu2022reap}, but rather via {\it parameterized} algorithms.
For instance, an Aave-like \ac{defi} lending protocol---the focal protocol of this study---operates using various parameters, including interest rate model parameters as well as risk parameters such as {\em collateral factor} and {\em liquidation threshold}.
These asset-specific risk parameters dictate the extent of overcollateralization and liquidation incentivization of each eligible cryptocurrency, which are vital in protecting the protocol from insolvency due to loan defaults or adverse market movements.
As such, those parameters require ongoing monitoring and adjustments to ensure their suitability in the fast-changing crypto market.

\IEEEpubidadjcol

Commonly, the adjustment of protocol parameters is  determined through a governance process ~\cite{ekal2022defi,mohan2022voting,kiayias2022sok}.
A proposal on such an adjustment is first posted on the protocol's forum, where the community forming the protocol's \ac{dao} voice their opinion.
The decision of whether or not to adopt a particular proposal is generally subject to a vote by holders of the governance tokens of the corresponding protocol \cite{Xu2022FromMarket}. Voting can be done directly in the protocol's own governance forum, which typically employs single-choice voting, and/or on open platforms such as Snapshot (snapshot.org) and Tally (tally.xyz) where more sophisticated voting mechanisms such as weighted voting or quadratic voting can be supported \cite{snapshot2023}. 

\begin{figure}[t]
    \centering
    \frame{\includegraphics[width=\linewidth]{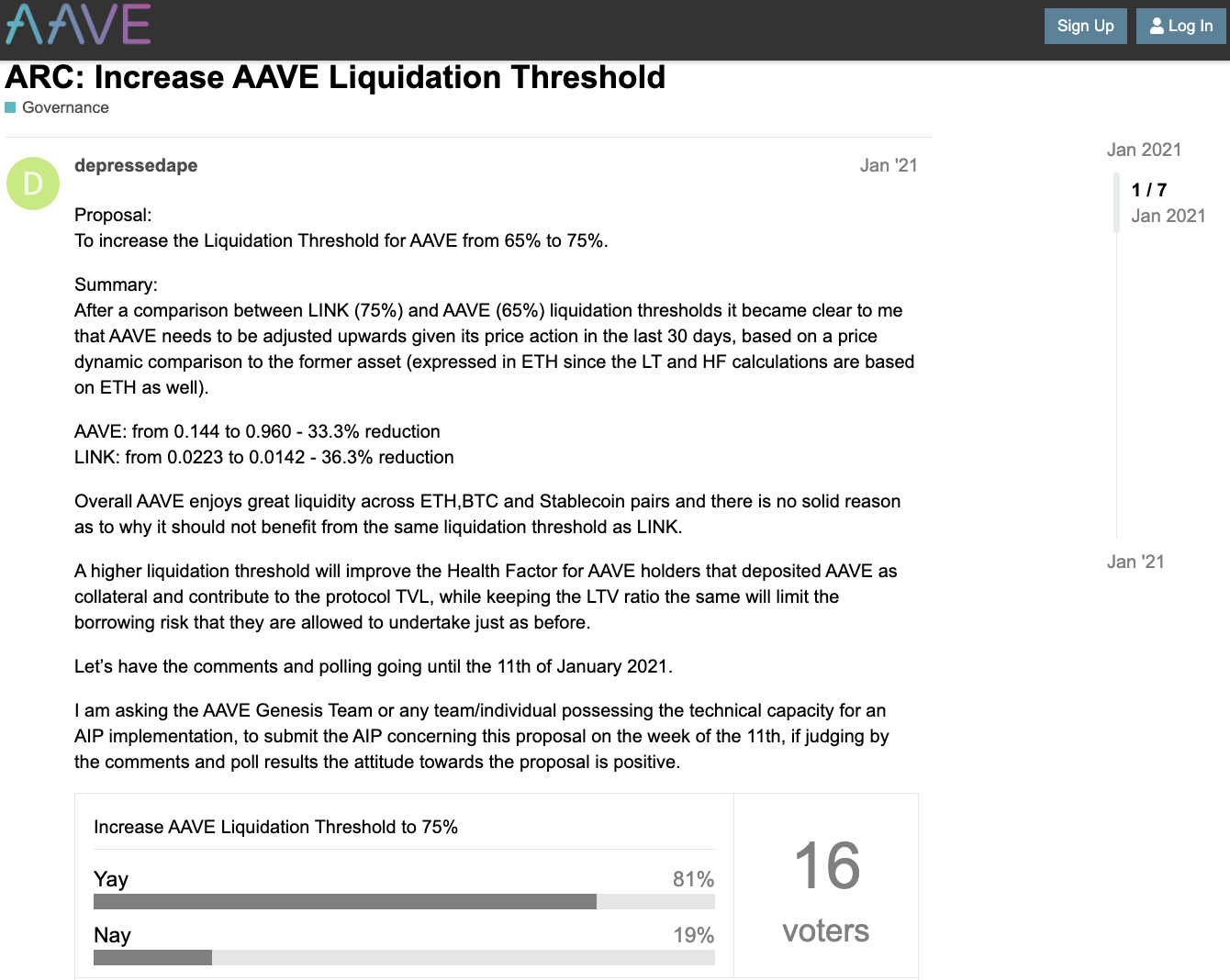}}
    \caption{A proposal in Aave governance forum to increase the {\em liquidation threshold} of the AAVE token~\cite{depressedape}.}
    \label{fig:aave-board}
\end{figure}

\paragraph{Motivation}
\label{sec:motivation}
In recent years, \ac{defi} governance forums have witnessed increased activity, particularly on risk-related matters, reflecting the community's growing recognition of the importance of \ac{defi} governance (see Appendix~\ref{app:proposal}). 
Unfortunately, the protocol's core team often holds a significant portion of governance tokens~\cite{Cousaert2022Token-BasedBlockchain} and assumes responsibility for post-voting execution, defeating the purpose of decentralized voting. In addition, the current governance process---comprising observing/ideation, proposing, discussion, voting, and eventual abortion/implementation---is labor-intensive and time-consuming. As an example, \autoref{fig:aave-board} exhibits a governance proposal on the Aave (the protocol) governance forum suggesting to increase the {\em liquidation threshold} for the AAVE (the underlying token) market.
The person justified their proposal by quoting the price performance of the AAVE token compared to a reference token LINK. The discussion of this thread lasted almost three weeks---from 3rd to 21st of January 2021, eventually resulting in the rejection of the proposal. 
This lack of agility inherent to the lengthy governance procedure is especially problematic in the rapidly evolving and volatile crypto market.
As demonstrated in Appendix~\ref{sec:aave-markets}, Aave risk parameters often remain unchanged even amid significant fluctuations in key market indicators such as price volatility, trading volume, and utilization ratio (borrow over supply). 


The rigidity of the current governance model also leaves a protocol vulnerable to malicious exploitation such as price oracle attacks \cite{Mackinga2022TWAPSaid}, where an attacker temporarily inflates the price of a collateralizing asset in order to borrow more than they should, with no intention of repayment. 
Lending protocols such as bZx Protocol and Cream Finance suffered losses of \$8 million in 2020 and \$130 million in 2021, respectively, due to price oracle exploits using \enquote{flash loans} \cite{Qian2023EmpiricalRepair}.

Overall, the current de facto centralization of governance processes and their inability to promptly respond to dynamic market changes jeopardize the integrity of \ac{defi} protocols and compromise their security. Despite recent efforts, such as \cite{Bastankhah2024ThinkingProtocol}, to establish a more adaptive \ac{defi} parameter adjustment scheme, there exists no framework for a governance model that is both more efficient and effective in general, and resilient against major market manipulations, such as price oracle attacks, in particular.

The aforementioned challenges underscore the urgent need for more agile, automated, attack-resilient, and generalizable governance mechanisms. 
In this paper, we address this need by introducing Auto.gov, a parameter adjustment framework based on \acf{dqn} \acf{rl}.
Specifically, using the lending protocol mechanism for the environment setup, we apply \ac{dqn} \ac{rl} to train a governance agent to determine the optimal adjustment policy for lending pool assets' respective {\em collateral factors}, under scenarios both with and without price oracle attacks.
Established as a proof-of-concept using an abstract \ac{defi} environment, our approach lays the groundwork for more concrete implementations.

Equipped with \Ac{rl}, Auto.gov is well-suited for this type of automation for several reasons. Firstly, the \ac{defi} environment can be modeled as a \ac{mdp}, where future states are determined solely by current states and actions, aligning well with \ac{rl} principles. Secondly, unlike many other machine learning models that rely on static datasets and require retraining to adapt, \ac{rl} excels at rapidly responding to the complex and unpredictable dynamics inherent in \ac{defi} systems, providing a governance approach that is robust, resilient, and capable of evolving alongside shifting market conditions.
Thirdly, Auto.gov takes into account the long-term consequences of decisions, a critical aspect for \ac{defi} governance where decisions can significantly impact protocol stability, user trust, and risk tolerance. Lastly, the design of Auto.gov reduces the potential for human error and mitigates the risk of \enquote{black-swan} events through the prevention of erratic system behavior as only gradual adjustments at each step are allowed.

\paragraph{Contributions} Our contributions are summarized below\footnote{The code is open-sourced at \url{https://github.com/xujiahuayz/auto-gov}.}:

\begin{itemize}
\item %
To the best of our knowledge, this paper represents the first attempt to integrate \ac{defi} governance with \ac{ai}.

\item %
We formalize and parameterize an Aave-like \ac{defi} environment that models the interaction between the protocol, user, and external market; the stylized environment can be easily modified to apply to different types of \ac{defi} protocols and to accommodate diverse user behaviors.


\item %
We quantitatively demonstrate Auto.gov's superiority in both {\em efficacy} and {\em efficiency}. Auto.gov exhibits particular effectiveness in countering oracle attacks: in a simulated training environment, it boosts the \ac{defi} protocol's final profit by over 60\% compared to the baseline governance agent; in real-world testing, it outperforms the best benchmark approach by 14\% and the static baseline approach tenfold. Besides, Auto.gov adapts swiftly to \ac{defi} environments, making robust, profitable decisions in about 4 hours of training using only a laptop, several orders of magnitude faster than the week-long voting-based governance process.
\end{itemize}


\section{Related work}
\label{sec:related}


\paragraph{\Ac{defi} modeling and management}
Our work is inspired by existing literature on modeling the behavior and risks associated with \ac{defi}, as well as effective protocol management.

Kao et al.~\cite{kao2020analysis} employed agent-based modeling to conduct a stress test on the market risks faced by participants of the Compound protocol. They recommend that the community utilize their simulation-based assessment to evaluate any new assets proposed for introduction to the protocol;
Bartoletti et al. \cite{Bartoletti2020sok} formalized the mechanisms of \ac{defi} lending protocols and describe the over- and under-collateralization risks associated with those protocols; Perez et al. \cite{Perez2020g} depicted an anecdotal liquidation event on Compound and analyze its root cause. The three aforementioned works concentrate on modeling and analyzing different aspects of \ac{defi} protocols. In contrast, our approach directly formulates governance measures aimed at improving the overall safety and profitability of the protocol.

In 2024, Bastankhah et al.~\cite{Bastankhah2024ThinkingProtocol} introduced an adaptive, data-driven protocol for \ac{defi} borrowing and lending, which is closely related to ours. This protocol features a high-frequency controller that dynamically adjusts interest rates to ensure market stability and competitiveness with external markets. However, there are three key differences between their work and ours: 
\begin{enumerate*}[label={(\roman*)}]
    \item Their primary goal is to develop a \ac{defi} protocol focusing on efficiency, whereas our focus is on revenue maximization;
    \item They design an entirely new \ac{defi} protocol, while our work involves a \ac{defi} governance agent that can be integrated into existing \ac{defi} protocols for better governance;
    \item In terms of methodology, we employ \ac{ai}, whereas their approach relies on optimization techniques.
\end{enumerate*}

\paragraph{Machine learning for \ac{defi}}
Our work is connected to the literature on machine learning for \ac{defi}. Babu et al.~\cite{Babu2024ApplicationPlatforms} proposed a decentralized interest rate swap platform using machine learning algorithms, like LSTM and SVM, to predict interest rate volatility. Palaiokrassas et al.~\cite{Palaiokrassas2024MachinePrediction} investigated the application of machine learning (e.g., logistic regression, random forest, XGBoost, CatBoost, LightGBM, and CNN) for credit risk assessment in Multichain \ac{defi}. John et al.~\cite{John2023SwarmChain} proposed a swarm-learning-based credit scoring approach for P2P lending protocols. Jiang et al.~\cite{jiang2017deep} applied a deep \ac{rl} framework to achieve a model-free portfolio optimization with crypto-assets. Hou et al.~\cite{hou2020squirrl} used deep \ac{rl} to automate attack analysis on blockchain incentive mechanisms. These works focus on predicting, optimizing, or evaluating specific elements of \ac{defi} with machine learning, offering recommendations to users as a result. However, unlike Auto.gov, none of them are directly engaged in the governance of \ac{defi}.

\paragraph{Related work in \ac{tradfi}}
Our research also intersects with the body of literature on algorithms and automation applied in the broader (traditional) finance field. Algorithmic trading in particular has been intensively studied in both finance and computer science literature~\cite{Chaboud2014RiseMarket,Weller2018DoesAcquisition,Treleaven2013AlgorithmicReview}.
Cao~\cite{Cao2023AIOpportunities} surveyed AI and data science (AIDS) techniques, as well as their challenges and opportunities in financial businesses. Huang et al.~\cite{Huang2023Machine-Learning-BasedMacro-Finance} devised a two-step machine learning algorithm---the Supervised Adaptive Group LASSO (SAGLasso)---to construct return predictors for \acs{tradfi} instruments such as government bonds. Spiegeleer et al.~\cite{DeSpiegeleer2018MachineFitting} demonstrated the usage of machine learning to achieve fast pricing, hedging, and fitting for financial derivatives. While all these studies apply machine learning to various areas of finance, our research is distinctly focused on \ac{defi} governance.

\section{Environment and Agent modeling}
\label{sec:env}

\begin{figure}[bt]
    \centering
    \includegraphics[width=\linewidth, trim={18 30 18 30}]{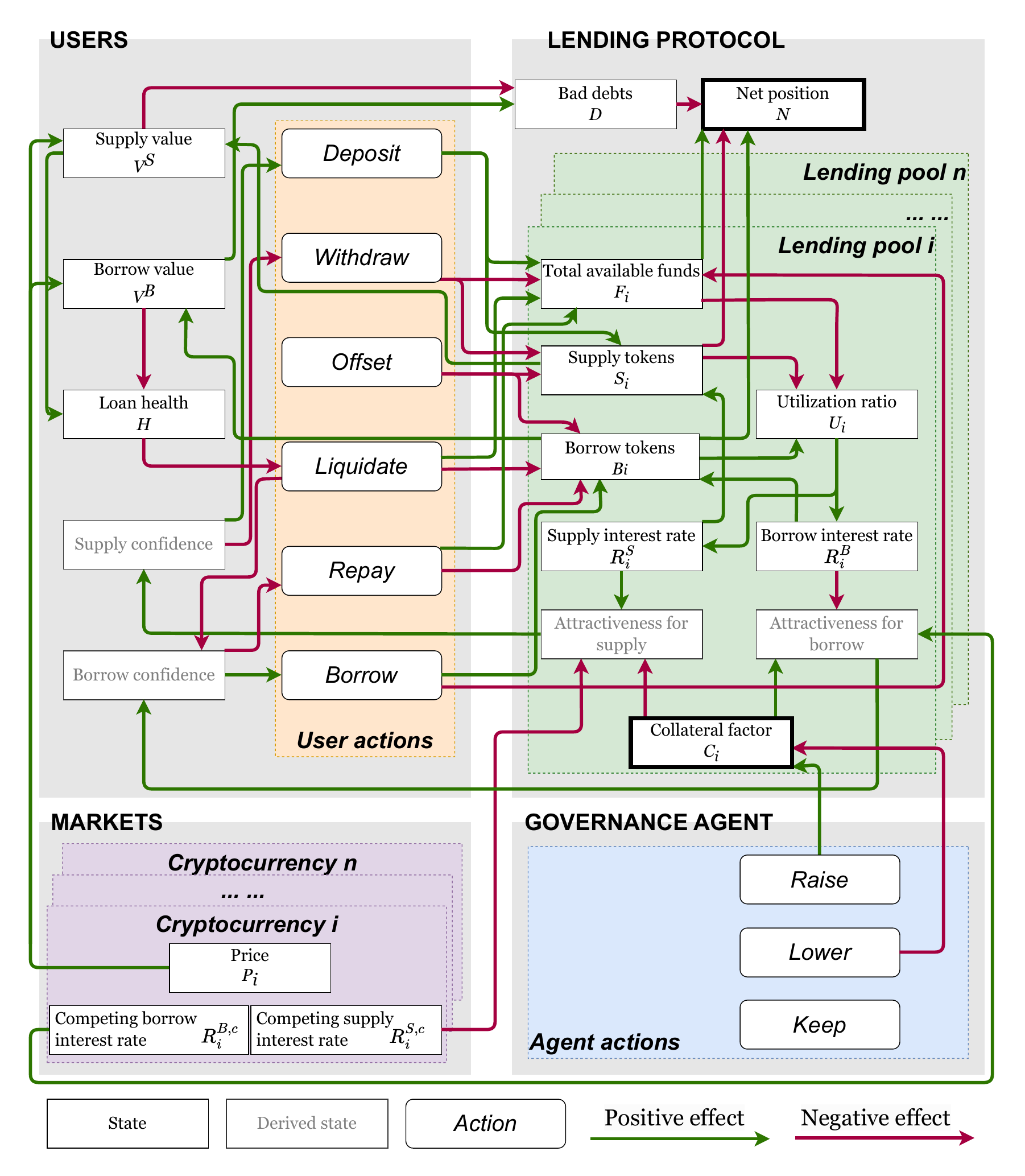}
    \caption{Action-state space encompassing users, lending pools and external markets of individual cryptocurrencies, and governance agent.}
    \label{fig:arc-protocol}
\end{figure}

For our \ac{rl} experiments, we establish a simplified \ac{defi} environment that consists of users and an Aave-like lending protocol. The \ac{rl} agent seeks to optimize the protocol's {\em net position} by adjusting each lending pool's {\em collateral factor}. \autoref{fig:arc-protocol} illustrates the complete action-state space of our \ac{rl} environment, further elaborated in~\autoref{sec:actions} and~\autoref{sec:states}.

\subsection{Actions}
\label{sec:actions}

Actions, denoted as $a$, refer to decisions made by a related party in response to the current state of the environment.

\subsubsection{Governance agent actions}
\label{sec:governance_action}
\hfill

\paragraph{Raise / lower / keep collateral factor}
\label{sec:change-coll}
When the governance agent raises, lowers, or keeps the existing collateral factor of a specific lending pool, the pool's collateral factor increases, decreases, or remains the same, respectively.
In our pursuit to develop an automated governance agent that dynamically adjusts policies using machine learning, the quality of the agent's decisions becomes crucial for the proper operation of the lending protocol. However, the inherent lack of transparency and control in machine learning models may introduce additional risks and uncertainties to the system. Therefore, to guarantee the certainty, controllability, and security of the actions made by the governance agent, we have implemented an additional layer of constraints on the agent's actions. First of all, we check the validity of agent actions (e.g., collateral factors must never exceed 1 after adjustment) 
so that only reasonable actions are applied to the protocol. Besides, our implementation caps the incremental increase or decrease at 2.5\%, meaning that even if the governance agent makes poor decisions, any potential harm to the \ac{defi} protocol would only be gradual and minimal, allowing for easy detection and correction by the community.

\subsubsection{User actions}
\label{sec:user_actions}

\hfill

\paragraph{Deposit}
When a user deposits into a specific lending pool (i.e., depositing ETH to the ETH pool, USDC to the USDC pool, and so on), the pool's {\em total available funds} increase by the deposited quantity. Simultaneously, the same amount of interest-accruing {\em supply tokens} are minted to their address, tracking the accrual of interest at the {\em supply interest rate} and reflecting the increase in the user's {\em supply value}.

\paragraph{Withdraw}
When a user withdraws from a lending pool, its {\em total available funds} decrease by the withdrawn amount, and the same quantity of interest-accruing {\em supply tokens} are burned. The quantity withdrawn cannot exceed the user's existing balance of {\em supply tokens}, and, after the withdrawal, the user's {\it loan health} $H$ (see \autoref{sec:user_states}) must not drop below 1.

\paragraph{Borrow}
When a user borrows from a specific lending pool, the pool's {\em total available funds} decrease by the borrowed quantity. Simultaneously, the same quantity of interest-accruing {\em borrow tokens} are minted, tracking the accrual of interest at the {\em borrow interest rate} and reflecting the increase in the user's {\em borrow value}. The maximum borrowable amount is such that after the borrow, the user's {\it loan health} $H$ must be at least 1.

\paragraph{Repay}
When a user repays, the pool's {\it total available funds} increase by the repaid amount, and the same quantity of interest-accruing {\it borrow tokens} are burned.

\paragraph{Liquidate}
In our stylized training environment (see Configuration~(\ref{assum:user}) in \autoref{sec:simplification}), liquidating a borrow position---on an aggregate level---updates the pool states ({\it total available funds} and interest-accruing {\it borrow tokens}) just like a repayment, but it also lowers the user's {\em borrow confidence} (see \autoref{sec:user_states}).

\paragraph{Offset}
Instead of repaying with the underlying asset, a user can also reduce the borrow position by burning the same quantity of the asset's {\it supply tokens} and {\it borrow tokens}. The action is equivalent to {\em withdrawing} and {\em repaying} combined.

\subsection{States}
\label{sec:states}

The states, collectively denoted as $\mathbb{S}$, represent the current situations or configurations of an environment that an agent interacts with. They consist of lending pool states, user states, and market states.

\subsubsection{Lending pool states}
\label{sec:lending_states}

\paragraph{\\
Total available funds $F_i$}
Total available funds of lending pool $i$ is the amount of its underlying token $i$ remaining in the pool available to be withdrawn or borrowed. From the accounting perspective, total available funds are on the asset side---equivalent to cash---of the pool's balance sheet, and hence positively contribute to the pool's {\em net position} (Equation~\ref{eq:net-position}).

\paragraph{Supply tokens $S_i$}
Supply tokens are interest-accruing tokens (such as Aave's aTokens)  that keep track of the amount of funds a lending pool owes to the user over time. From the accounting perspective, supply tokens, bearing the nature of payables, are on the liability side of the pool's balance sheet, and reduce the pool's {\em net position} (Equation~\ref{eq:net-position}) \cite{Luo2025PiercingReappraised}. The number of supply tokens also negatively influences the {\em utilization ratio} of the pool. A user's {\em supply value} within a pool equals the value of a user's allocated supply tokens within that pool times the current price of the underlying token.

\paragraph{Borrow tokens $B_i$}
Borrow tokens are interest-accruing tokens that keep track of the amount of funds the user owes to the lending pool over time. From the accounting perspective, {\em borrow tokens}, bearing the nature of receivables, are on the asset side of the pool's balance sheet, and hence enhance the pool's {\em net position} (Equation~\ref{eq:net-position}). The number of {\em borrow tokens} also positively influences the {\em utilization ratio} of the pool (see Equation~\ref{eq:utilization} below). A user's {\em borrow value} within a pool equals the value of a user's allocated borrow tokens within that pool times the current price of the underlying token.

\paragraph{Bad debts $D$}
While the lending protocol is generally secured by its overcollateralization and liquidation mechanism, under extreme market conditions, e.g. when the price of a collateral asset experiences a sudden drop, the protocol may still suffer from bad debts. Bad debts are the amount of funds that the lending protocol is unable to recover from the user; they occur when a user's loan position is undercollateralized---i.e. when a user's {\em supply value} does not suffice to cover their {\em borrow value}, leaving the user with no incentive to repay the loan and other network participants with no incentive to liquidate the position (see also \autoref{sec:default}).
From the accounting perspective, bad debts are deemed expenses, and hence negatively contribute to the pool's {\em net position} (Equation~\ref{eq:net-position}).

\paragraph{Net position $N$}
The net position of the lending protocol is calculated as the difference between the protocol's {\em total reserve} and {\em bad debts}, specifically:%
\begin{equation}
N = \Sigma_i [ P_i \cdot \underbrace{(F_i + B_i - S_i)}_{=W_i}] - D \ ,
\label{eq:net-position}
\end{equation}%
where $F_i$ denotes lending pool $i$'s total available funds (Assets: Cash), $B_i$ the borrow tokens (Assets: Receivables), $S_i$ the supply tokens (Liabilities: Payables), all denominated in token unit, and $P_i$ denotes the price of token $i$. Reserve $W_i$ reflects lending pool $i$'s net worth under the assumption that all outstanding loans will be paid back; the lending protocol is \emph{insolvent} when $\Sigma_i (P_i \cdot W_i) < D$, a state we term \enquote{bankruptcy}.

\paragraph{Utilization ratio $U_i$}
The utilization ratio of lending pool $i$ is the ratio of the number of {\em borrow tokens} to the number of {\em supply tokens}, formally%
\begin{equation}
U_i = \tfrac{B_i}{S_i}.
\label{eq:utilization}
\end{equation}%
A low utilization ratio indicates capital inefficiency, while a high utilization ratio signals a high liquidity risk of the pool. The utilization ratio is therefore used to determine the pool's supply and borrow interest rates through a pre-defined interest rate model, a monotonically increasing function of the utilization ratio. When the utilization ratio is low, both the supply and borrow interest rates are low, in order to discourage supply and encourage borrow to drive the utilization ratio higher; when the utilization ratio becomes too high, both the supply and borrow interest rates increase exponentially, in order to discourage borrow and encourage supply to drive the utilization ratio lower and to mitigate liquidity risk.

\paragraph{Collateral factor $C_i$}
The collateral factor $C_i \in [0,1)$ of a specific lending pool $i$ determines the fraction of a user's {\em supply value} in pool $i$  
counted towards the user's total borrowable value. In other words, the user's total borrowable value is given by $\sum_i (V^S_i \cdot C_i)$.
As a numerical example, consider $C_1 = 0.2$, $C_2 = 0.8$, $V^S_1 = \$100$, and $V^S_2 = \$50$; that is, the collateral factor for lending pool 1 is 0.2 and for pool 2 is 0.8, and the user has a {\em supply value} of \$100 in pool 1 and \$50 in pool 2. The user can therefore borrow up to \$60 ($ = \$100 \times 0.2 + \$50 \times 0.8 $) in total across all lending pools.

Conventionally, the collateral factor associated with each lending pool is manually adjusted by the protocol team based on the underlying token's risk profile such as price volatility and level of centralization. The riskier an asset is deemed, the lower the collateral factor is set to be. A higher collateral factor means a higher borrowing capacity, and hence a higher liquidity risk. Therefore, the collateral factor positively influences a pool's {\em attractiveness for borrow}, and negatively influences its {\em attractiveness for supply}.

\paragraph{Supply interest rate $R^S_i$}
The supply interest rate of lending pool $i$ is the interest rate that the lending pool pays to the user for supplying funds. A higher supply interest rate means a higher proliferating speed for supply tokens due to interest accrual. The supply interest rate is set such that, at each step, the supply interest to be accrued can be fully covered by the borrow interest to be accrued with some spread $\varsigma \in (0,1)$, i.e.
$ 
R^S_i = R^B_i \cdot U_i \cdot (1- \varsigma)
$.

\paragraph{Borrow interest rate $R^B_i$}
The borrow interest rate of the lending pool $i$ is the interest rate that the user pays to the lending pool for borrowing funds. A higher borrow interest rate means a higher increasing speed for borrowing tokens due to interest accrual. The borrow interest rate is set according to a pre-defined interest rate model, which in our environment, is set to be
$
    R_i^B = \frac{1}{b \cdot (1- U_i)}
$,
where $b>0$.

\paragraph{Attractiveness for supply}
The attractiveness for the supply of a lending pool is a derived state designed to model how a pool's states influence the user's willingness to supply funds to the entire lending protocol. A user's {\em supply confidence} in the lending protocol is positively affected by the aggregate attractiveness for the supply of all lending pools in the protocol (see \autoref{fig:arc-protocol}).

\paragraph{Attractiveness for borrow}
The attractiveness for borrow of a lending pool is a derived state designed to model how a pool's states influence the user's willingness to borrow funds from the entire lending protocol. A user's {\em borrow confidence} in the lending protocol is positively affected by the aggregate attractiveness for the borrow of all lending pools in the protocol.

\subsubsection{User states}
\label{sec:user_states}

\hfill

\paragraph{Supply value $V^S$}
The supply value of a user is the sum of their supply value across all lending pools:
$
    V^S = \sum_i V_i^S = \sum_i (S_i \cdot P_i)
$, where $S_i$ is the quantity of the user's token $i$ supply quantity and $P_i$ the price of token $i$.
All other things equal, a higher supply value decreases the user's loan-to-value ratio $\frac{V^B}{V^S}$, thus increasing the user's {\em loan health}.

\paragraph{Borrow value $V^B$}
The borrow value of a user is the sum of their borrow value across all lending pools:
$
    V^B = \sum_i V_i^B = \sum_i (B_i \cdot P_i)
$, where $B_i$ is the quantity of the user's token $i$ borrow quantity.
All other things equal, a higher borrow value increases the user's loan-to-value ratio $\frac{V^B}{V^S}$, and hence decreases the user's {\em loan health}.

\paragraph{Loan health $H$} 
The loan health of a user is 
the ratio between their aggregate loanable value, which equals the sum of the collateralizing asset's {\it supply value} respectively discounted with each asset's {\em collateral factor} (see numerical example under \autoref{sec:lending_states}), and their aggregate {\it borrow value}: $H 
=  \frac{\sum_i (V^S_i \cdot C_i)}{V^B}$. $H \geq 1$ signifies a healthy loan position, and $H < 1$ triggers liquidation.

\paragraph{Supply confidence}
The supply confidence of a user is a derived state designed to model their willingness to supply funds to the lending protocol. Numerically speaking in our environment, supply confidence is measured by {\em supply buffer} $\delta^S \in [0,1]$, which is the amount of funds that a user withholds as a fraction of their total funds willing to be supplied to this specific lending protocol (as opposed to competing protocols). 
High supply confidence---or low supply buffer---encourages the supply action, while low supply confidence---or high supply buffer---discourages it (see \autoref{sec:ordinary} and \autoref{sec:env_param}). 

\paragraph{Borrow confidence}
The borrow confidence of a user is a derived state designed to model their willingness to borrow funds from the lending protocol. 
Similar to supply confidence, borrow confidence is measured by {\em borrow buffer} $\delta^B \in [0,1]$.
High borrow confidence---or low borrow buffer---encourages the borrow action, while low borrow confidence---or high borrow buffer---has the opposite effect.

\subsubsection{Market states}

\hfill

\paragraph{Price $P_i$}
The price of an underlying token of a lending pool $i$ positively affects both the supply value and the borrowed value of a user within that pool. Token price movements follow a stochastic process (see \autoref{sec:price} for simulated price time series).

\paragraph{Competing supply interest rate $R^{S,c}_i$}
The competing supply interest rate of token $i$ is offered by the external market, representing the opportunity cost of supplying funds to the lending protocol. A higher competing supply interest rate reduces the {\em attractiveness for supply} of the lending pool $i$, and encourages the user to withdraw funds from the protocol (to supply to a competing protocol) (see \autoref{fig:user-action-attack} under \autoref{sec:reaction}).

\paragraph{Competing borrow interest rate $R^{B,c}_i$}
The competing borrow interest rate of token $i$ is offered by the external market. A higher competing borrow interest rate increases the {\em attractiveness for borrow} of the lending pool $i$, while a low one encourages the user to repay their loan (and to instead borrow from a competing protocol).

\subsection{Reward}
\label{sec:rewards}
We approximate the protocol's profitability using the change in the \emph{net position} $N$ after each incremental step: $\Delta N_t = N_t - N_{t-1}$. The governance agent acts in the interests of the lending protocol, whose ultimate objective is profit maximization.\footnote{%
While alternative optimization objectives (e.g. the number of users, the average duration of deposits) could be considered and the design can be somewhat arbitrary, the rationale behind our choice of \emph{net position} maximization as the objective is because the protocol ultimately belongs to its governance tokenholders. These individuals are financially incentivized by the price appreciation of governance tokens, whose fundamental value is directly tied to the protocol's profitability~\cite{Xu2022b,Xu2022e}.
}  
To evaluate the agent's performance, we compare $\Delta N_t$ in the \ac{rl} environment against the baseline environment (see \autoref{sec:baseline}). The difference quantifies the extent to which the governance agent outperforms the baseline agent. The resulting performance metric forms the reward for the governance agent, with an additional penalty applied in the case of an invalid action (see \autoref{sec:governance_action}). Specifically, the reward is given by:%
\begin{equation}
r_t = \Delta N_t^\text{RL} - \Delta N_t^\text{baseline} - \text{Penalty} \cdot {\bf 1}_{\text{invalid action}},
\label{eq:r}
\end{equation}%
where ${\bf 1}_{\text{invalid action}}$ is an indicator function that equals 1 if the action is invalid and 0 otherwise.  
This reward serves as feedback, informing the agent about the immediate consequence of its action in the given state. The discounted cumulative reward, in turn, reflects the agent's long-term optimization objective (see \autoref{sec:rl-agent}).

\subsection{Transition function}
\label{sec:trans_func}

The transition function is a key element representing the environment's response to the actions taken by the governance agent, seamlessly integrated into our modeled DeFi environment. Each action performed by the governance agent prompts a corresponding change in the environment, resulting in the generation of respective state $\mathbb{S}$.

\begin{figure}[b]
    \centering
        \includegraphics[width=\linewidth, trim={18 20 18 30}]{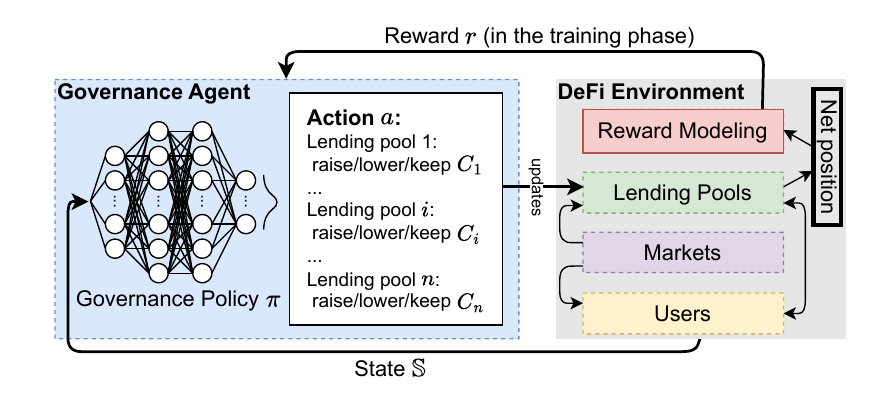}
        \caption{Architecture of Auto.gov. The detailed \ac{defi} environment is illustrated in \autoref{fig:arc-protocol}.}
        \label{fig:rl-arc}
\end{figure}

\section{Design of the governance agent}
\label{sec:method}

In this section, we present the design and training process of the governance agent, which is responsible for making appropriate decisions for \ac{defi} protocols.

\subsection{Reinforcement learning-based agent}
\label{sec:rl-agent}

We employ \acf{rl}, an artificial intelligence approach for training agents to make optimal decisions in complex and uncertain environments (i.e., \ac{defi} ecosystems) in the presence of attacks. Specifically, we use a deep \ac{rl} agent~\cite{van2016deep} to serve as the governance agent, which dynamically selects the best action based on the current state of the \ac{defi} protocols. This approach is well-suited for our purpose, as it enables the agent to learn from past experience and adapt its behavior to changing conditions in real time. By leveraging a \ac{rl} agent, we can ensure that our protocol is flexible, adaptable, resilient to attacks, and able to optimize its behavior to maximize its performance under various scenarios.

\autoref{fig:rl-arc} illustrates the architecture of the proposed approach. Our goal is to train a policy $\pi$ for the governance agent to maximize the discounted, cumulative reward $\mathbb{R}_{0}=\sum_{t=0}^\infty \gamma^{t}r_t$, where $\gamma$ denotes the discount factor. We leverage deep Q-learning~\cite{watkins1992q} to conduct the learning process for the governance agent, where a function $Q:\mathbb{S}\times a\rightarrow r$ can tell the agent what the reward would be. Then, the policy $\pi$ of the governance agent can be represented as \autoref{equ:q_value}, which always selects the action $a$ that can maximize the value of the $Q$ function. In this equation, $\mathbb{S}_t$ denotes the state $\mathbb{S}$ at time $t$.%
\begin{equation}
\label{equ:q_value}
\pi(\mathbb{S}_t) = \operatorname*{arg\,max}_{a\in\mathbf{a}} Q(\mathbb{S}_t,a)
\end{equation}%
For each training update, we enforce that every $Q$ function for the policy $\pi$ follows the Bellman equation (\autoref{equ:bellman}), where the difference between the two sides of the equality is known as the temporal difference error $\delta$ (\autoref{equ:bellman_loss}). Here, $\mathbb{S'}$ represents the state that the environment transitions to after the agent takes an action from the current state $\mathbb{S}$.
\begin{align}
    Q^{\pi}(\mathbb{S},a)&=r+\gamma Q^{\pi}(\mathbb{S'},\pi(\mathbb{S'})) \label{equ:bellman}\\
    \delta &= Q(\mathbb{S},a) - (r+\gamma \max_{a'}Q(\mathbb{S'},a')) \label{equ:bellman_loss}
\end{align}%
Additionally, considering the potentially vast state space and the expected growth with the increasing number of lending pools, we employ a \ac{dqn}~\cite{fan2020theoretical} to serve as the policy network for the governance agent, enabling it to monitor multiple pools simultaneously.
In a \ac{dqn}, a deep neural network acts as an approximator for the Q-function~\cite{Mnih2013PlayingLearning}. The \ac{dqn} combines the strengths of both deep learning and Q-learning to tackle complex reinforcement learning problems with large state spaces, providing an efficient and robust approach for learning optimal policies.

Furthermore, to expedite the process of error elimination, we utilize Adam optimization~\cite{Kingma2015Adam:Optimization} during training. Adam is an optimization algorithm that is commonly used to train deep neural networks. It is a variant of \ac{sgd} that adapts the learning rate for each weight in the network based on historical gradient information. Other optimization algorithms such as Gradient Descent and traditional \ac{sgd} were attempted and Adam outperformed them all.

\begin{algorithm}[t]
    \caption{High-level training procedure for the \ac{rl}-based governance agent.}
    \label{alg:training}
    \begin{algorithmic}[1]
    {\footnotesize
        \State Initialize \ac{dqn} $Q$
        \State Initialize target network $\hat{Q}$
        \State Initialize experience replay memory $D$
        \While{not converged}
            \State $\epsilon\leftarrow$ new epsilon with $\epsilon$-decay \Comment{Sampling}
            \State Choose an action $a$ from $\mathbb{S}$ using policy $\epsilon$-greedy($Q$)
            \State Agent takes $a$, observe reward $r$ and the next state $\mathbb{S'}$
            \State Store transition $(\mathbb{S},a,r,\mathbb{S'},is\_done)$ in $D$
            \If{enough experience in $D$} \Comment{Learning}
                \State Sample a \emph{batch} of transitions from $D$
                \For{every transition $(\mathbb{S}_i,a_i,r_i,\mathbb{S'}_i,is\_done_i)$ in \emph{batch}}
                    \If{$is\_done_i$}
                        \State $y_i \leftarrow r_i$
                    \Else
                        \State $y_i \leftarrow r_i + \gamma \max_{a'\in \mathbf{a}} \hat{Q}(\mathbb{S'}_i,a')$
                    \EndIf
                    \State $\delta \leftarrow \frac{1}{N} \sum^{N-1}_{i=0}(Q(\mathbb{S}_i,a_i)-y_i)^2$ \Comment{Error}
                    \State $m_0\leftarrow 0$, $v_0 \leftarrow 0$, $t\leftarrow 0$ \Comment{Adam optimization}
                    \While{$Q_t$ not converged}
                        \State $t \leftarrow t+1$
                        \State $g_t \leftarrow \nabla_Q\delta_{t-1}$
                        \State $m_t \leftarrow \beta_1 m_{t-1} + (1-\beta_1)g_t$
                        \State $v_t \leftarrow \beta_2 v_{t-1} + (1-\beta_2)g_t^2$
                        \State $\hat{m_t} \leftarrow \frac{m_t}{1-\beta_1^t}$ \Comment{Correcting bias}
                        \State $\hat{v_t} \leftarrow \frac{v_t}{1-\beta_2^t}$
                        \State $Q_t \leftarrow \prod_{\mathcal{F},\sqrt{\hat{v_t}}} (Q_{t-1}-\frac{\alpha\hat{m_t}}{\sqrt{\hat{v_t}}+\mu})$ \Comment{Updating}
                    \EndWhile
                \EndFor
                \State Synchronize $\hat{Q}$ with $Q$ if reaches threshold
            \EndIf
        \EndWhile
    }
    \end{algorithmic}
\end{algorithm}

Algorithm~\ref{alg:training} formalizes the training procedures including sampling, Adam optimization, and neural network updating, where $g_t$ denotes the gradient at step $t$, $m_t$ is the \ac{ema} of $g_t$, $v_t$ is the \ac{ema} of $g_t^2$, $\beta_1$ and $\beta_2$ are hyperparameters used in the moving averages of $g_t^2$ and $g_t$, $\alpha$ denotes the learning rate, and $\mu$ is a small number to prevent the denominator from becoming $0$. Additionally, $\epsilon$-greedy($Q$) is a method to balance exploration and exploitation, where $\epsilon$ refers to the probability of choosing to explore as opposed to exploiting. During the training, the input data is the states and rewards that come from the simulated market environment.

\subsection{Target network}

\begin{figure}[b]
        \centering
        \includegraphics[width=0.8\linewidth, trim={18 12 18 30}]{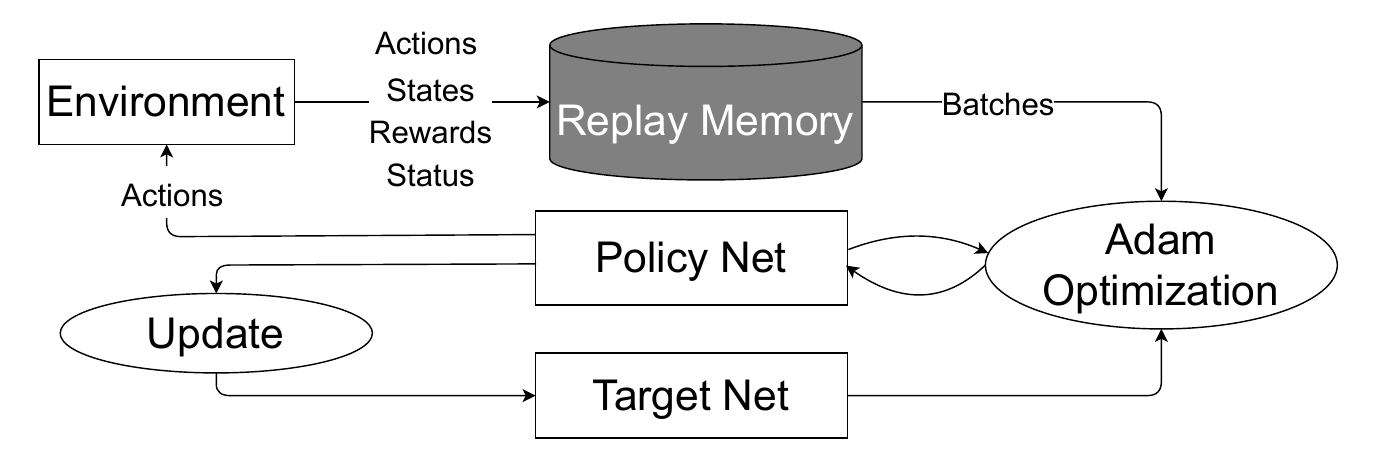}
        \caption{Operations for training the governance agent with the target network.}
        \label{fig:rl-training}
\end{figure}

To prevent overfitting during the training process and enhance the agent's resilience to potential attacks, we have implemented a target network~\cite{van2016deep}. The target network is a critical component in the training process of \acp{dqn}. It consists of a separate neural network with the same architecture as the primary or online network, but with a distinct set of weights. The primary purpose of the target network is to provide more stable and consistent target values for the Q-learning updates. In our approach, the training procedure involves periodically copying the weights from the online network to the separate target network, as illustrated in \autoref{fig:rl-training} and Algorithm~\ref{alg:training}. The target network aids the governance agent in enhancing stability and convergence, ultimately resulting in more reliable and robust learned policies.

However, simply employing the target network may bring side effects, such as conservative Q-value estimations, temporary performance drops, and slower training. To address the disadvantages associated with the target network, we propose a hybrid approach that employs the primary network and the target network during different phases of the agent training. This new method results in faster training and better agent performance. We provide more details regarding this hybrid approach in Appendix~\ref{app:target_network}.

\subsection{Prioritized experience replay}

To equip the governance agent with the ability to adapt to diverse \ac{defi} environments, expedite the policy network's training, and swiftly enhance its resistance to potential attacks, we also have incorporated prioritized experience replay into our \ac{rl} agent as an optional training feature.

Prioritized experience replay~\cite{schaul2015prioritized} enhances the standard experience replay technique, which uniformly stores and samples experiences during training. While experience replay helps break the correlation between consecutive samples and decrease update variance, prioritized experience replay assigns significance to each experience based on the discrepancy between actual and predicted Q-values. This approach increases the likelihood of sampling and learning from unexpected experiences or those with greater potential to boost the agent's performance.


Once the \ac{dqn} $Q$ is fully converged and the governance agent is properly trained, the governance agent can be used for setting parameters for the lending pool individually, even without the reward feedback from the \ac{defi} environment.
\section{Simulation setup}
\label{sec:additional}

\subsection{Environment abstraction}
\label{sec:simplification}

For ease of experimentation, we abstract the \ac{defi} lending protocol environment with the following configurations without loss of generality:
\begin{enumerate}
    \item Following the common practice in macroeconomics \cite{Geweke1985MacroeconometricAgent}, we model market user actions on an aggregate level, i.e. we have one representative agent who behaves as the entire user base collectively does.
    \label{assum:user}
    \item There is no liquidation incentive, as it is a zero-sum game among users (the liquidator wins and the borrower loses), and on an aggregate level, the inclusion or exclusion of liquidation incentive does not directly affect our optimization goal: maximizing the protocol's {\em net position} $N$ (see \autoref{eq:net-position}).
    \item We equate {\em liquidation threshold} with {\em collateral factor}.
\end{enumerate}

\subsection{Lending pool establishment}
\label{sec:lendingpool}

We experiment with how asset price volatilities might affect the \ac{rl} agent's collateral factor determination. Two assets are insufficient to model one stable asset and one volatile asset; if Asset A is volatile against B, then B must be volatile against A. Thus, we need at least three assets to demonstrate the volatility effect on the collateral factor in a self-contained modeling environment. To this end, we include in the lending protocol environment three lending pools covering all three types of crypto-assets: ETH---the numeraire or the denominating asset of the protocol, USDC---a USD-pegged stablecoin, and TKN---some arbitrary token.


\begin{figure}[tp]
     \centering
        \includegraphics[width=\linewidth, trim={18 15 18 20}]{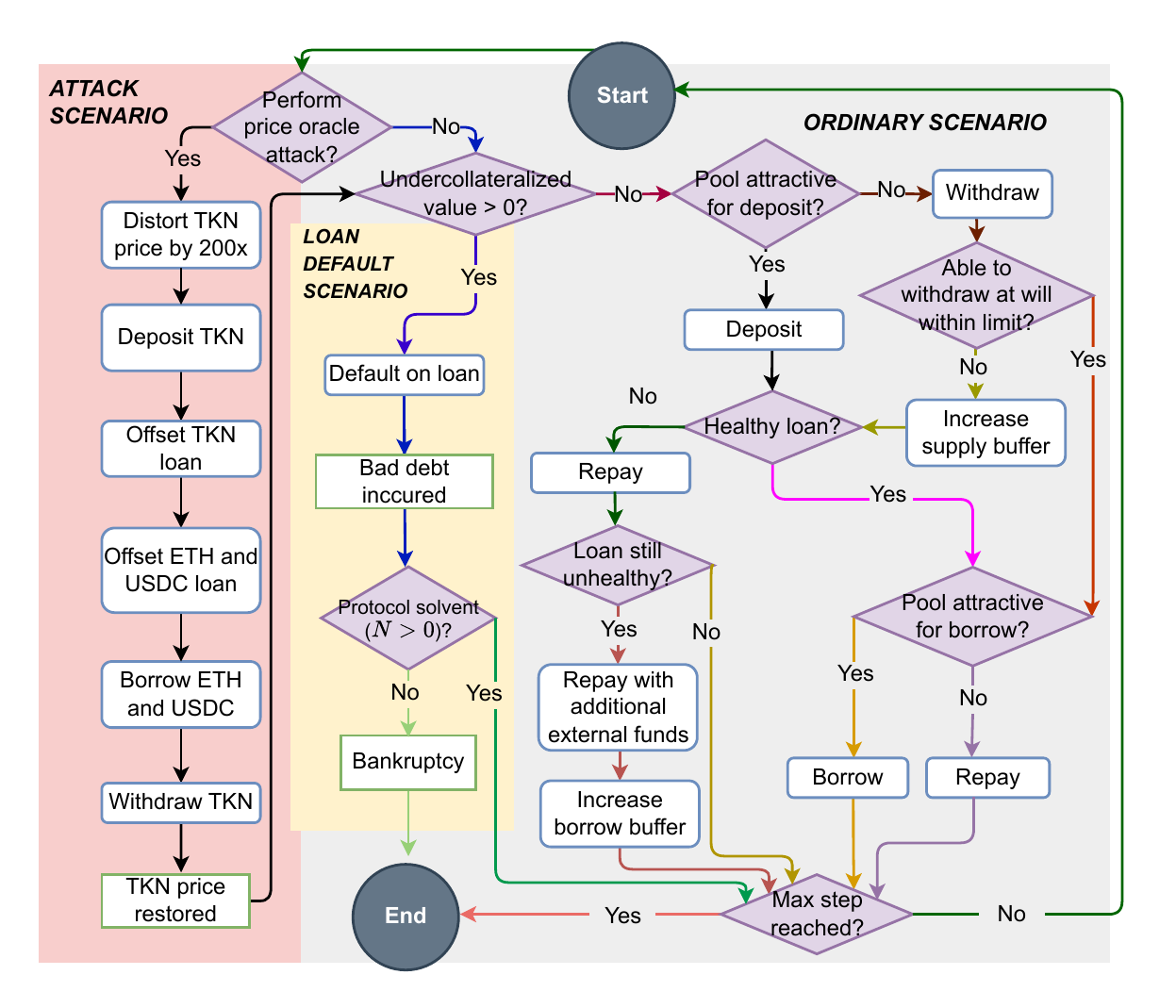}
        \caption{Series of user reactions at each step (from \enquote{Start}-node to \enquote{Max step reached?}-node) of an episode under attack (\autoref{sec:attack-scenario}), loan default (\autoref{sec:default}), and ordinary (\autoref{sec:ordinary}) scenarios, looping until episode ends (\enquote{End}-node).}
    \label{fig:user-action-attack}
\end{figure}

\subsection{User reaction simulation}
\label{sec:reaction}

\autoref{fig:user-action-attack} illustrates the series of user's reactions, pre-programmed in our model environment, according to the market condition as well as the user's own financial status.

\subsubsection{Attack scenario}
\label{sec:attack-scenario}
In each episode, the environment randomly generates 3 steps where a price oracle attack will be performed before other ordinary actions from the market user.

While Auto.gov is agnostic about {\em how} exactly the market manipulation occurs, a typical way to achieve an arbitrary price hike is through a {\bf flash-loan-funded price oracle attack} \cite{xu2021dexAmm,Qin2020AttackingTD,Arora2024SecPLF:Attacks} that involves:
\begin{enumerate}
    \item driving the price of a crypto-asset high temporarily by buying a sizable amount of that asset with funds (usually in ETH or stablecoin) borrowed through a flash loan which requires no collateralization;
    \item depositing this asset at the inflated price;
    \item borrowing out funds worth more than the true value of the deposited asset;
\item using part of the borrowed funds to repay the flashloan and profiting from the excess.
\end{enumerate}
Crypto-assets with a thinner market are generally more vulnerable to price manipulation; more widely-adopted assets such as ETH or USD-pegged stablecoins often get drained from the lending protocol under the attack described above.

At the step where an attack shall occur, the price of TKN first becomes inflated by 200 times. 
Then the market user deposits all their available TKN to the lending pool to increase their supply token balance in TKN. Next, the user offsets their TKN loan with TKN supply tokens as much as possible, such that all the remaining TKN supply tokens can be used to back other, more valuable loans. The user subsequently offsets their ETH and USDC loans with the corresponding supply tokens as much as possible, such that the remaining ETH and USDC loans will be mainly backed by TKN at an inflated price. The user then seeks to borrow out as much additional ETH and USDC as possible from the lending pools, possibly depleting the pools before the faked borrow quota is used up. Finally, the user withdraws TKN to restore their initial TKN balance as much as possible.

\subsubsection{Loan default scenario}
\label{sec:default}
A user is likely to default on their loan if their undercollateralized value---defined as their total {\em borrow value} less total {\em supply value}---is positive, leaving them with no incentive to repay the loan for collateral rescue. As intended by the attacker, loan default is most likely to occur following a price oracle attack, although it may also happen---albeit much less likely---absent malicious behavior when the market simply is too volatile.
When a loan defaults, the undercollateralized value will be treated as {\em bad debt} expenses, reducing the net position of the protocol (see Equation~\ref{eq:net-position}). Under major attacks when the attacker manages to borrow out a relatively large amount of ETH and USDC, the bad debt expenses can suffice to cause the protocol to bankrupt, i.e. having $N \leq 0$, a non-positive {\it net position} (see \autoref{sec:lending_states}), which ends the episode prematurely.

\subsubsection{Ordinary scenario}
\label{sec:ordinary}
In an ordinary scenario, the user deposits (withdraws) based on how attractive (unattractive) the lending pool is for depositing, which then depends on how much higher (lower) the lending interest rate and the collateral factor of the underlying asset of the lending pool is compared to the external market level. If the user decides that the lending pool is relatively unattractive, they will seek to withdraw liquidity. If they are not able to withdraw within the limit at will---e.g., when their deposited funds are lent out by the protocol, leaving the pool with insufficient liquidity for large withdrawal---they will withdraw as much as possible but also increase their supply buffer, meaning they will supply less liquidity in the future steps for precaution even when the lending pool is attractive for depositing.

The user subsequently checks their loan health, i.e. whether their loan value has exceeded the aggregate loanable value backed by the collateral. If healthy, the user decides whether to borrow more or repay depending on how much more or less the lending pool's borrow interest is compared to the market level. If unhealthy, the user tries to repay and restore their loan health as much as they can in the first instance. If still unhealthy, additional external funds will be injected into the user's account to repay the loan. This is to mimic the effect of liquidation, where the liquidator repays the loan on behalf of the borrower and seizes the latter's collateral. Note that as discussed in \autoref{sec:simplification}, we only model one market user to represent the aggregate user behavior's effect on the protocol. In that sense, a loan turning liquidatable can in fact directly increase the liquidity of the lending pool. On the flip side, however, the user loses some {\em borrow confidence} and consequently increases their {\em borrow buffer} every time their collateral becomes liquidatable, meaning they will borrow less in future steps for precaution even when the lending pool is attractive for borrowing. In addition, the newly injected liquidity is still subject to withdrawal in future steps should the lending protocol exhibit unfavorable conditions for depositing.

\subsection{Additional setups for the \ac{defi} lending environment}
\label{sec:add_setup}


\subsubsection{Baseline environment}
\label{sec:baseline}

The baseline environment will be initiated exactly like the main training environment, including the price trajectories of all tokens, as well as the steps at which an attack would occur. The only difference is that there is no governance agent in the baseline environment, and the collateral factors will remain at their initial value of 0.8 for all three tokens throughout each episode. 

\subsubsection{Asset price simulation}
\label{sec:price}

The initial prices---all denominated in ETH---of TKN, USDC, and ETH are all normalized to 1.  
Following the common practice in econometrics and financial stochastics \cite{Osborne1962PeriodicPrices}, we model the token price time series $P_t$ as a geometric Brownian motion with $\mu_t$ and $\sigma_t$ that can be time-variant but are piece-wise constant within every step. The price update at each step can be formally expressed as:%
\begin{equation}
P_t =  e^{(\mu_t - \frac{\sigma_t^2}{2}) t + \sigma_t \, w_t} \ ,
\label{eq:gbp}
\end{equation}
where $w_t \sim \mathbb{N}(0,1)$. 

As a numeraire, ETH has by definition $\mu_t \equiv 0, \sigma \equiv 0$, i.e. $P_t \equiv 1$.
For USDC, we set $\mu_t \equiv 10^{-4}, \sigma_t \equiv 0.05$. For contrast, we set TKN to be most volatile, with $\mu_t \equiv 10^{-5}$ and  $\sigma_t = 0.05 + \frac{(t-200)^2}{5 \times 10^5}$. Clearly, we attempt to synthesize TKN price such that it becomes gradually less volatile before day 200, and then increasingly more volatile. Note that the simulated dummy price time series are only used for model training purposes; we use real-life price data for the final test (see \autoref{sec:test_real}).

\subsubsection{Other \ac{defi} environment parameters}
\label{sec:env_param}

We set the initial collateral factor for all three assets to be 0.8. 
We assume the external competing supply and borrow interest rates to be 0.05 and 0.15 per annum, respectively, for all three tokens. We additionally set the external competing collateral factors for ETH, USDC, and TKN to be 0.7, 0.65, and 0, respectively.

The market user in the model has a starting balance of 20,000 units of ETH, USDC, and TKN each, and initiates the three lending pools by depositing 15,000 units in each at time 0. We further assume that the market user has an initial safety supply margin and borrow margin both equal to 0.5, meaning they will supply and borrow half as much as they can initially. The supply and borrow buffers increase when the user experiences withdrawal restrictions and liquidations as described in \autoref{sec:reaction}. The buffers also increase when the collateral factor is adjusted due to the resultant unstable image of the protocol perceived by the user. The buffer values decrease if the user experiences smooth supplies and borrows without restrictions or liquidations for over 20 steps consecutively.

\subsubsection{Random events and stochastic processes}
\label{sec:random_events}

For each episode, new price trajectories (see \autoref{eq:gbp}) and attack time points (see \autoref{sec:attack-scenario}) are generated and equally applied to both the \ac{rl} and baseline environments. 
By incorporating random events and stochastic processes in each episode, this approach prevents the \ac{rl} agent from overfitting to specific price trajectories or attack timings, thereby enhancing the robustness and generalizability of the trained model.

\begin{table}[t]
\centering
\caption{Key hyperparameters of the \ac{rl} agent and its \ac{dqn}.}
\label{tab:para}
\resizebox{\linewidth}{!}{
\begin{tabular}{@{}lr|lr@{}}
\toprule
Total number of layers    & 5   & Input neurons               & 10$\times$pool\_num \\
DQN hidden layers         & 3   & Number of training episodes & 7,301       \\
Neurons in hidden layer 1 & 256 & Target net switch on point  & 0.3        \\
Neurons in hidden layer 2 & 256 & Learning rate ($\alpha$)    & 0.001      \\
Neurons in hidden layer 3 & 256 & Gamma ($\gamma$)            & 0.5        \\
Epsilon ($\epsilon$) start  & 1  & Batch size (with attack)    & 32\\
Epsilon ($\epsilon$) end    & 2e-3 & Batch size (without attack) & 256\\
Epsilon ($\epsilon$) decay  & 5e-7 & & \\
\bottomrule
\end{tabular}
}
\end{table}

\section{Results}
\label{sec:results}

In this section, we outline the implementation specifics of our approach and present the evaluation results.
We record training times to gauge the model's efficiency under various configurations. The effectiveness of the model is directly quantified by the training score, defined as the cumulative rewards (see Equation \ref{eq:r} for reward function) at the end of each episode. Since profit maximization is the predefined objective (see \autoref{sec:rewards}), we also evaluate \emph{net position} $N$ to assess the model's efficacy, which closely aligns with the training score.\footnote{The training score differs from the {\em net position} $N$ in that the former reflects the \ac{rl}'s excess {\em net position} relative to the baseline model's {\em net position} after deducting any applicable penalties.%
}
To demonstrate the model's resilience to adversarial conditions, we report training score and \emph{net position} $N$ in scenarios both with and without attacks.%

A noteworthy mention is that \autoref{sec:add_setup} provides details on how we initialize the \ac{defi} environment.

\subsection{Hyperparameter tuning}

Our approach is implemented using PyTorch
for the governance agent and OpenAI Gym
for the \ac{defi} environment. The training evaluation processes are performed on a Mac laptop with an M2 Pro CPU and 24 GB of RAM.


%
Fine-tuning hyperparameters is an essential part of building the governance agent. \autoref{tab:para} provides a detailed list of the key hyperparameters used in the governance agent and its \ac{dqn}. Due to space constraints, we skip the lengthy process of fine-tuning hyperparameters and place the justifications for key hyperparameter selections in Appendix~\ref{app:parameter}

\begin{figure}[t]
    \centering
    \begin{subfigure}{0.45\linewidth}
        \centering
    \includegraphics[width=\linewidth]{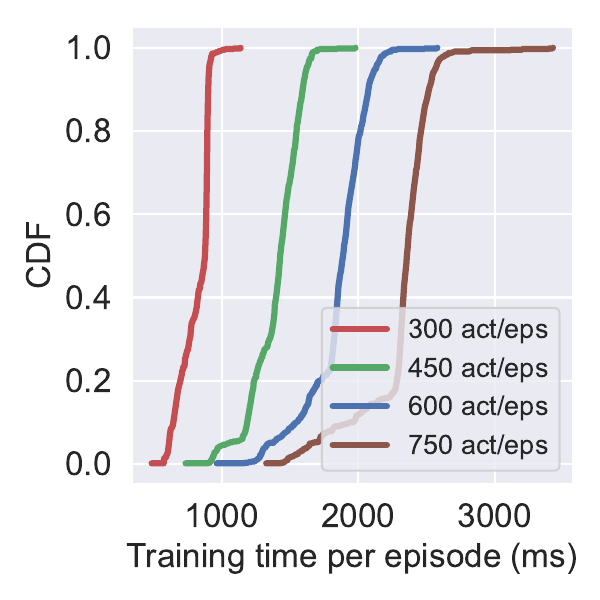}
    \caption{Without target networks.}
    \label{fig:delay_without_target}
    \end{subfigure}
    \begin{subfigure}{0.45\linewidth}
        \centering
    \includegraphics[width=\linewidth]{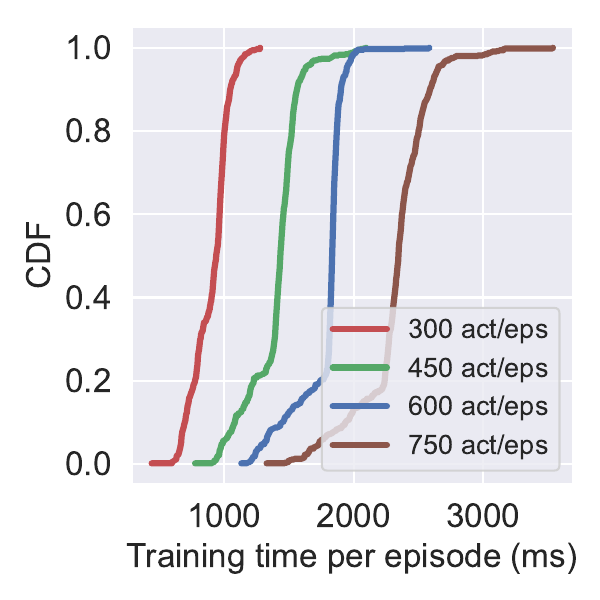}
    \caption{With target networks.}
    \label{fig:delay_with_target}
    \end{subfigure}
    \caption{Cumulative distribution functions (CDFs) of training time per episode with different numbers of actions per episode.}
    \label{fig:delay}
\end{figure}

\subsection{Training of the governance agent}

\begin{figure}[t]
    \centering
    \begin{subfigure}{\linewidth}
        \centering
        \includegraphics[width=\linewidth, trim={50 52 50 50},clip]{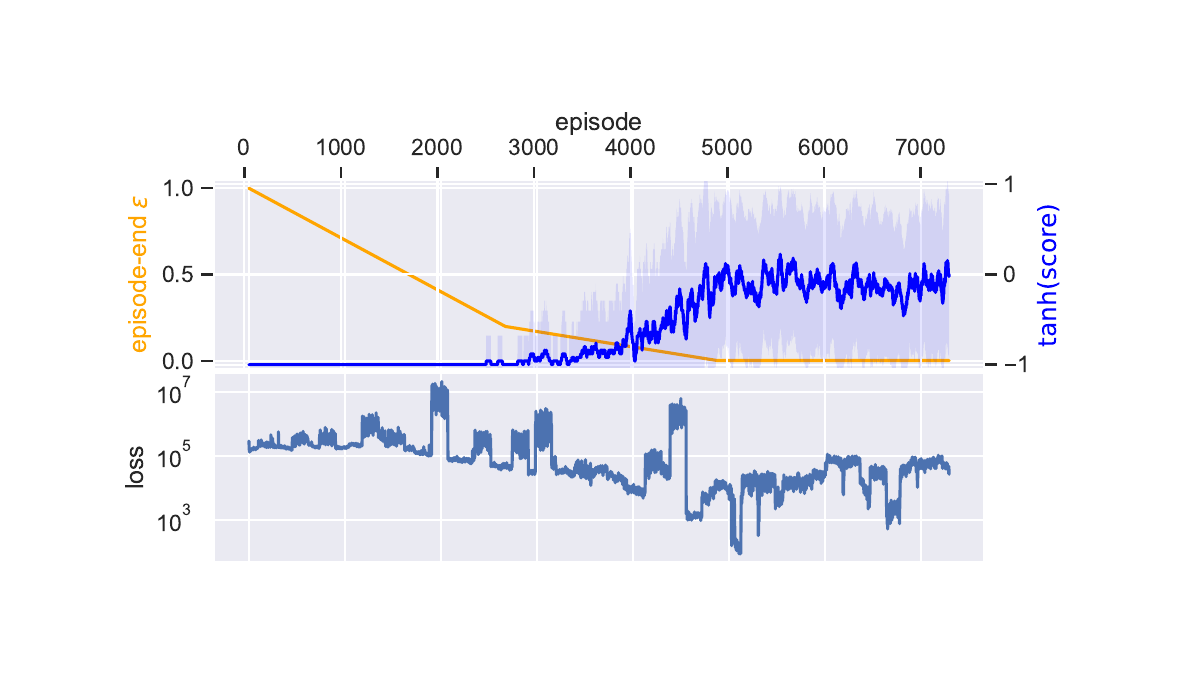}
        \caption{Without attacks.}
        \label{fig:training_no_attacks}
    \end{subfigure}
    \hfill
    \begin{subfigure}{\linewidth}
        \centering
        \includegraphics[width=\linewidth, trim={50 52 50 50},clip]{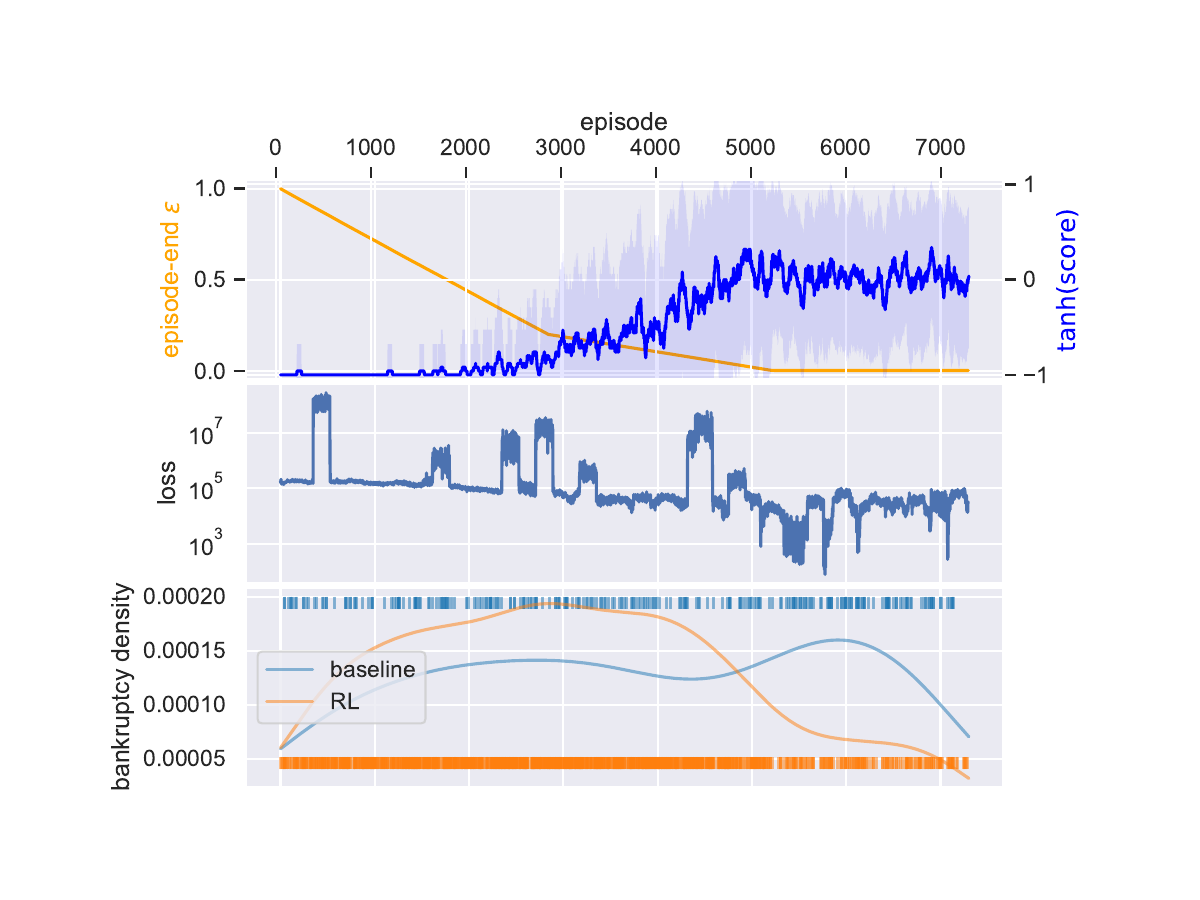}
        \caption{With attacks.}
        \label{fig:training_attacks}
    \end{subfigure}
    \caption{Changes in episode-end epsilon $\epsilon$, score, and loss with increasing training episodes. Bankruptcy occurs exclusively (though not always) in the presence of attacks; in the last plot of (b), the occurrence of bankruptcy at each episode is marked by a short vertical line \enquote{|}, and its density approximated by \ac{kde} curves.}
    \label{fig:training}
\end{figure}


Our initial evaluation focuses on the training of the governance agent, particularly examining the speed and effectiveness of its training process.

\begin{figure*}[t]
\centering
\begin{subfigure}{0.245\linewidth}
    \centering
    \includegraphics[width=\linewidth,trim={0.8cm 0.8cm 0.8cm 0.8cm},clip]{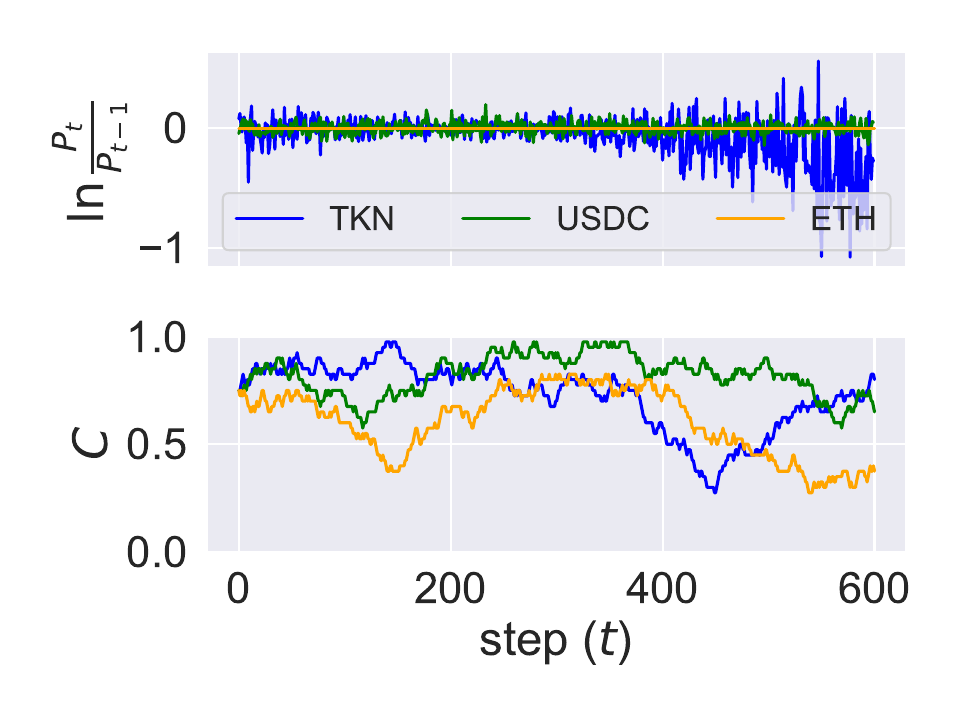}
    \caption{\footnotesize Early episode, no attacks.}
    \label{fig:colfact187}
\end{subfigure}
\begin{subfigure}{0.245\linewidth}
    \centering
    \includegraphics[width=\linewidth,trim={0.8cm 0.8cm 0.8cm 0.8cm},clip]{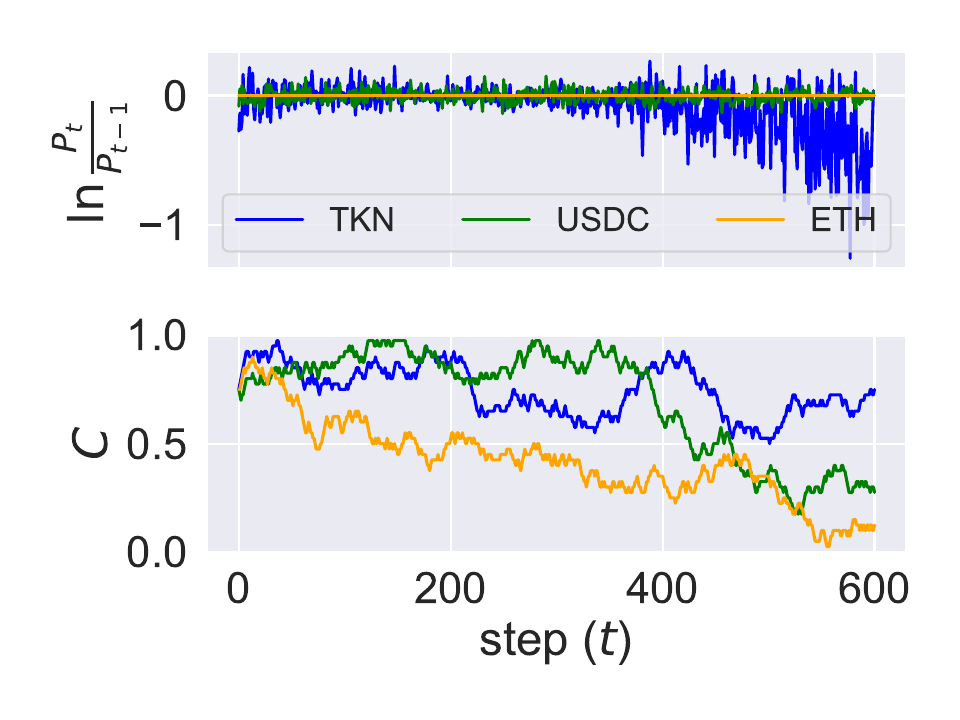}
    \caption{\footnotesize Early episode, with attacks.}
    \label{fig:colfact256}
\end{subfigure}
\begin{subfigure}{0.245\linewidth}
    \centering
    \includegraphics[width=\linewidth,trim={0.8cm 0.8cm 0.8cm 0.8cm},clip]{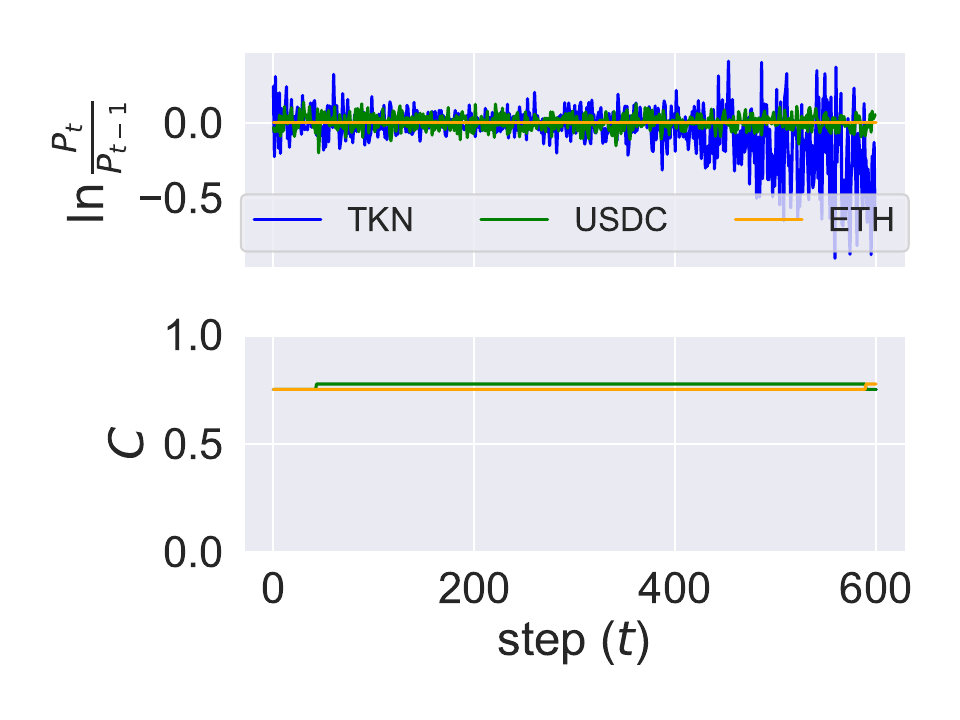}
    \caption{\footnotesize Well-trained episode, no attacks.}
    \label{fig:colfact703}
\end{subfigure}
\begin{subfigure}{0.245\linewidth}
    \centering
    \includegraphics[width=\linewidth,trim={0.8cm 0.8cm 0.8cm 0.8cm},clip]{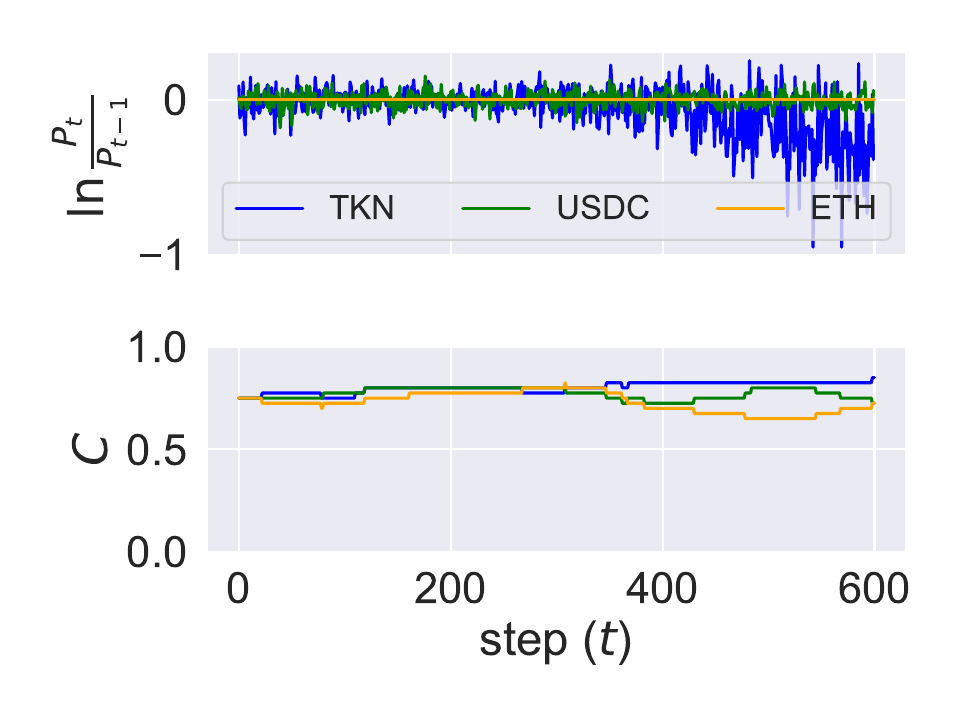}
    \caption{\footnotesize Well-trained episode, with attacks.}
    \label{fig:colfact681}
\end{subfigure}
\caption{ETH-denominated price change (expressed in log return $\ln\tfrac{P_{t}}{P_{t-1}}$) and collateral factor ($C$) adjustment of all tokens in the \ac{rl} environment throughout selected episodes.}
\label{fig:colfact}
\end{figure*}

First of all, the training speed is crucial due to the need for retraining in changing \ac{defi} environments or protocol updates.
\autoref{fig:delay} depicts the \acp{cdf} of training time per episode for varying numbers of actions. Training speed is crucial for the practicality of our approach, as the governance agent may need updating or retraining in response to different \ac{defi} environments or protocol changes. The results demonstrate that for all scenarios (i.e., 300, 450, 600, and 750 actions per episode), the training time remains below 3000ms per episode. With 300 actions per episode, the \ac{dqn} can finish decision-making and training for each episode within just 1000ms. Our experiments indicate that a higher number of actions per training step can slightly enhance training quality but can also considerably prolong the training duration. Based on our findings, we determined that training with 600 actions per episode is adequate for the rapid and efficient training of the governance agent. Consequently, we utilize 600 actions per training episode in all subsequent evaluations, consistently yielding satisfactory results. Moreover, by comparing Figures~\ref{fig:delay_without_target} and~\ref{fig:delay_with_target}, we observe that using target networks does not noticeably increase training time, as their time costs are very similar. 
The remaining evaluation results are derived from training that incorporates both the primary network and the target network, as detailed in Appendix~\ref{app:target_network}. The target network is activated only when $\epsilon$ is less than 0.3 (indicated in~\autoref{tab:para}).
Moreover, as outlined in \autoref{tab:para}, the agent requires 7,301 episodes for full training, resulting in a total training time of approximately 4 hours.

After landing on the appropriate hyperparameters, we train the governance agent for two scenarios: one without attacks and one with price oracle attacks occurring at random steps within each training episode (see \autoref{sec:reaction}). \autoref{fig:training} illustrates the changes in epsilon, losses, and scores as the number of training episodes increases in both scenarios.
Epsilon ($\epsilon$) in \ac{rl} represents the probability of the agent selecting a random action to explore; otherwise, it chooses the best-known action to exploit. 
As shown in \autoref{tab:para} and \autoref{fig:training}, we initialize $\epsilon$ at 1 to emphasize exploration during the early training phase. Over the first 5,000 episodes, $\epsilon$ decays gradually, shifting the agent's focus towards exploitation as it learns optimal actions. Afterwards, $\epsilon$ stabilizes at $5 \times 10^{-7}$, prescribing the agent to concentrate on exploitation.

To enhance interpretability and mitigate fluctuations caused by randomness, we employ the $\tanh$ function to normalize scores to the range $(-1,1)$~\cite{Phan-Minh2023DriveIRL:Learning}.
Our experiment results in \autoref{fig:training} indicate that the governance agent converges effectively after approximately 4,800 episodes, with $\tanh(\text{score})$ initially increasing before plateauing; correspondingly, while losses briefly increase midway through training, they stabilize and decrease as training progresses.
In the absence of adversarial attacks, the lending protocol can remain solvent (\autoref{fig:training_no_attacks}). However, when subjected to price oracle attacks, bankruptcy likely occurs (\autoref{fig:training_attacks}). In the baseline environment, the appearance of bankruptcies is distributed relatively evenly throughout the episodes. In contrast, in the \ac{rl} environment, bankruptcies arise more frequently during the initial episodes but the frequency decreases substantially after approximately 5,000 episodes, demonstrating the effectiveness of the training process. In addition, we observe that even in the presence of attacks, the governance agent's convergence speed is not slower than in their absence. This is attributed to the target network, which helps prevent overfitting to the environment and delivers a more robust policy.

Additionally, when combined with the delay information in \autoref{fig:delay}, the agent can rapidly adapt to a specific \ac{defi} environment and make robust, profitable decisions within several hours, using only the computing resources of a personal laptop. Overall, our findings suggest that our approach can effectively train a governance agent capable of making profitable and robust decisions in various \ac{defi} environments in a timely and efficient manner.

\begin{figure*}[!tpb]
\centering
\begin{subfigure}{0.245\linewidth}
    \centering
\includegraphics[width=\linewidth,trim={0.8cm 0.8cm 0.8cm 0.8cm},clip]{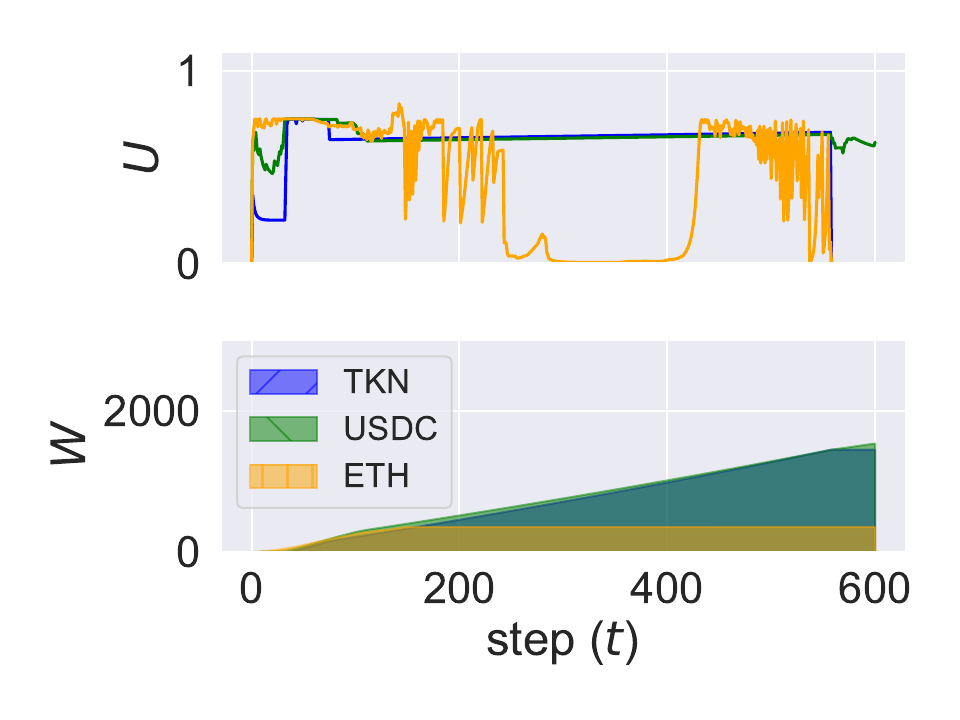}
\caption{\footnotesize Early episode, no attacks.}
\label{fig:state187}
\end{subfigure}
\begin{subfigure}{0.245\linewidth}
    \centering
\includegraphics[width=\linewidth,trim={0.8cm 0.8cm 0.8cm 0.8cm},clip]{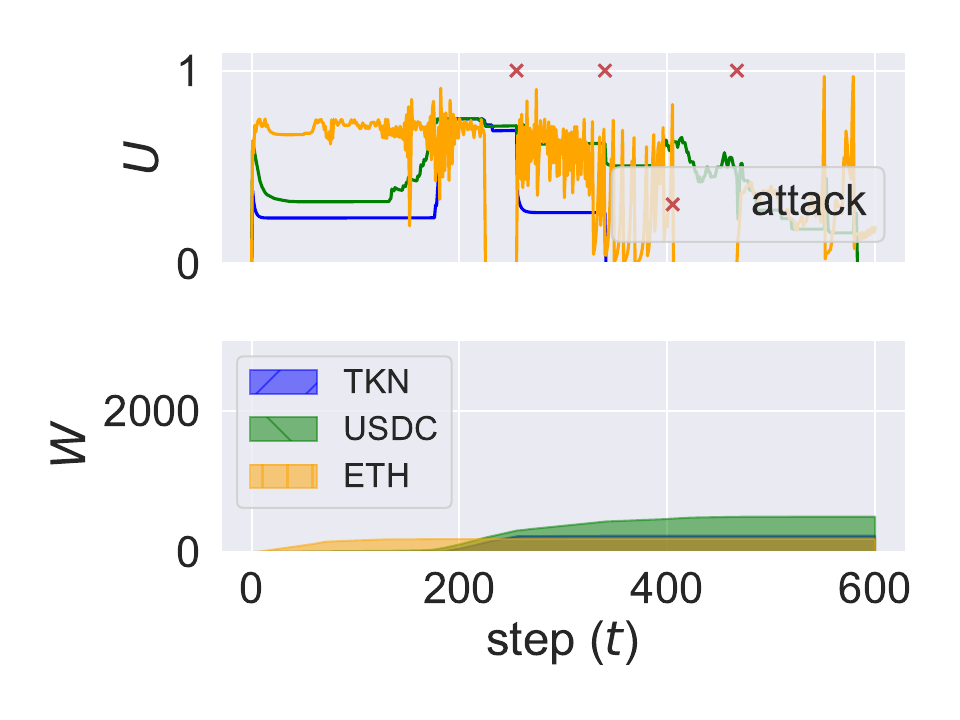}
\caption{\footnotesize Early episode, with attacks.}
\label{fig:state256}
\end{subfigure}
\begin{subfigure}{0.245\linewidth}
    \centering
\includegraphics[width=\linewidth,trim={0.8cm 0.8cm 0.8cm 0.8cm},clip]{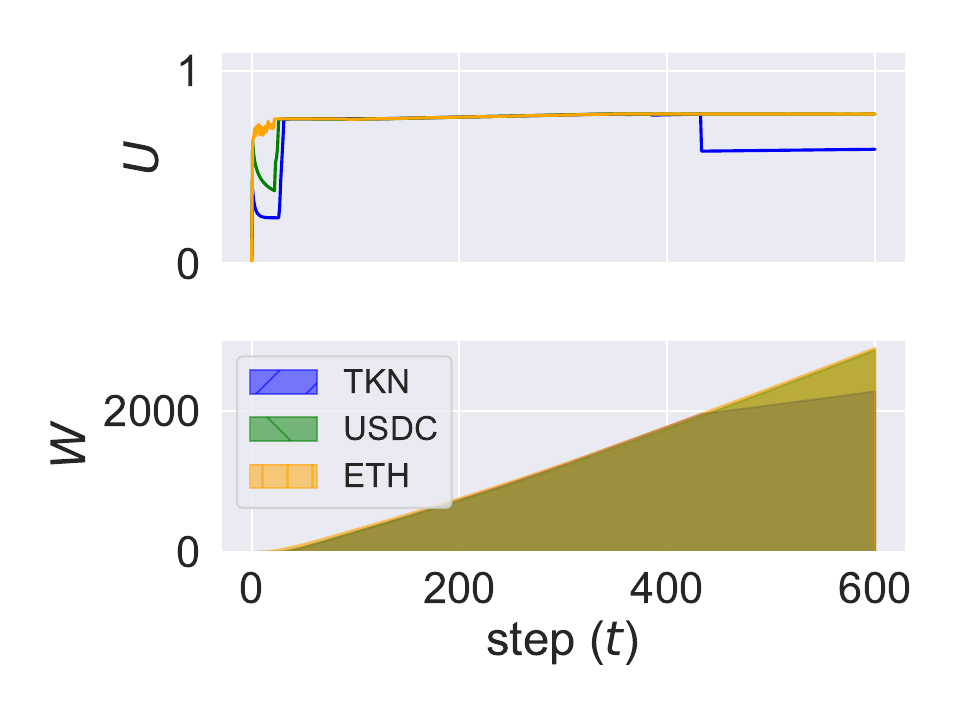}
\caption{\footnotesize Well-trained episode, no attacks.}
\label{fig:state703}
\end{subfigure}
\begin{subfigure}{0.245\linewidth}
    \centering
\includegraphics[width=\linewidth,trim={0.8cm 0.8cm 0.8cm 0.8cm},clip]{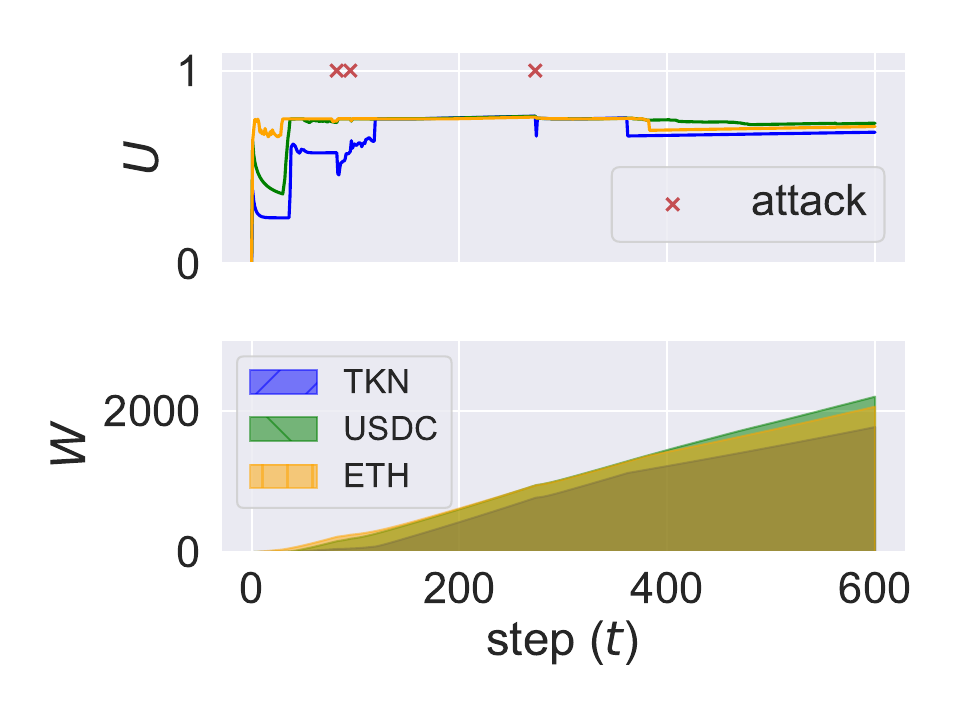}
\caption{\footnotesize Well-trained episode, with attacks.}
\label{fig:state681}
\end{subfigure}
\caption{Utilization ratio ($U$) and reserve ($W$) of the \ac{rl} environment's lending pools throughout each of the selected episodes. The red crosses (\textcolor{red}{$\times$}) in (b) and (d) indicate the steps ($t$) at which attacks occur.}
\label{fig:state}
\end{figure*}

\begin{figure*}[bt]
\centering
\begin{subfigure}{0.245\linewidth}
    \centering
\includegraphics[width=\linewidth,trim={0.8cm 0.8cm 0.8cm 0.8cm},clip]{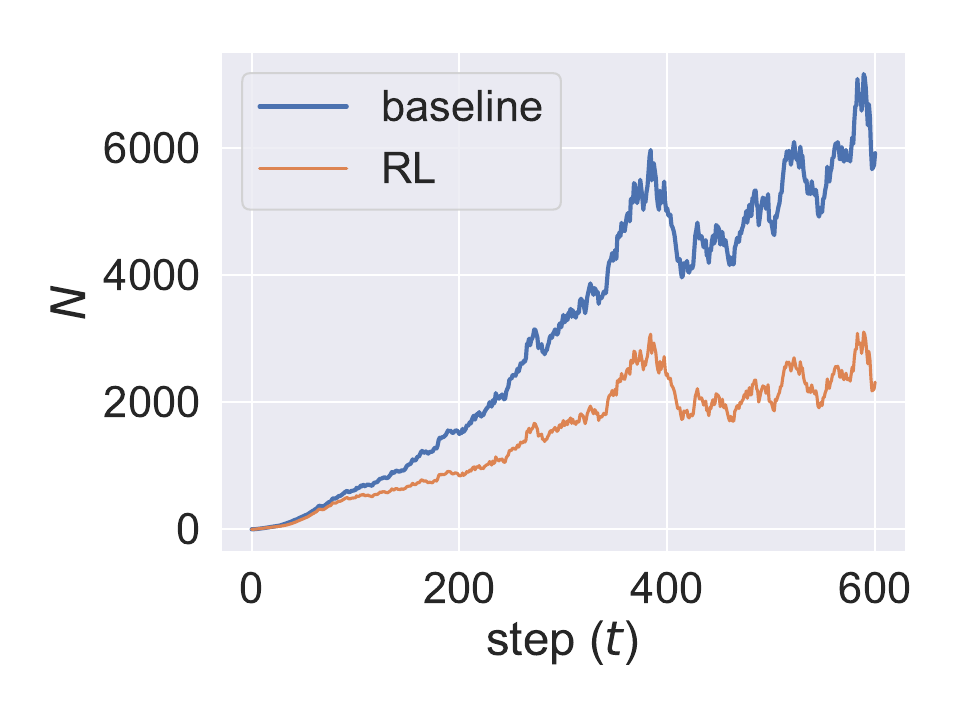}
\caption{\footnotesize Early episode, no attacks.}
\label{fig:netpos187}
\end{subfigure}
\begin{subfigure}{0.245\linewidth}
    \centering
\includegraphics[width=\linewidth,trim={0.8cm 0.8cm 0.8cm 0.8cm},clip]{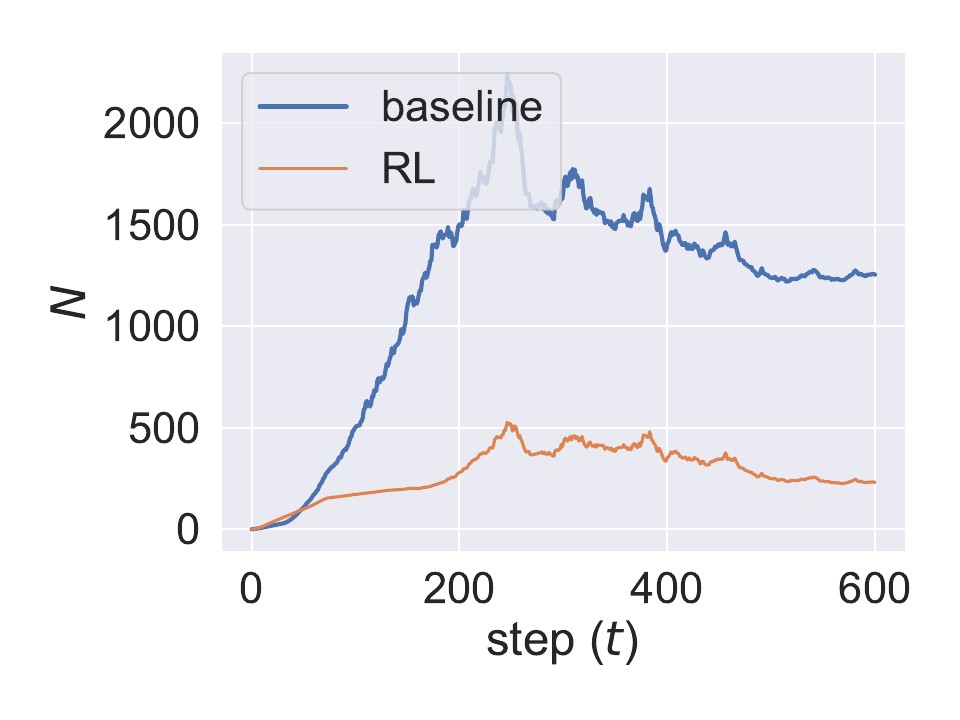}
\caption{\footnotesize Early episode, with attacks.}
\label{fig:netpos256}
\end{subfigure}
\begin{subfigure}{0.245\linewidth}
    \centering
\includegraphics[width=\linewidth,trim={0.8cm 0.8cm 0.8cm 0.8cm},clip]{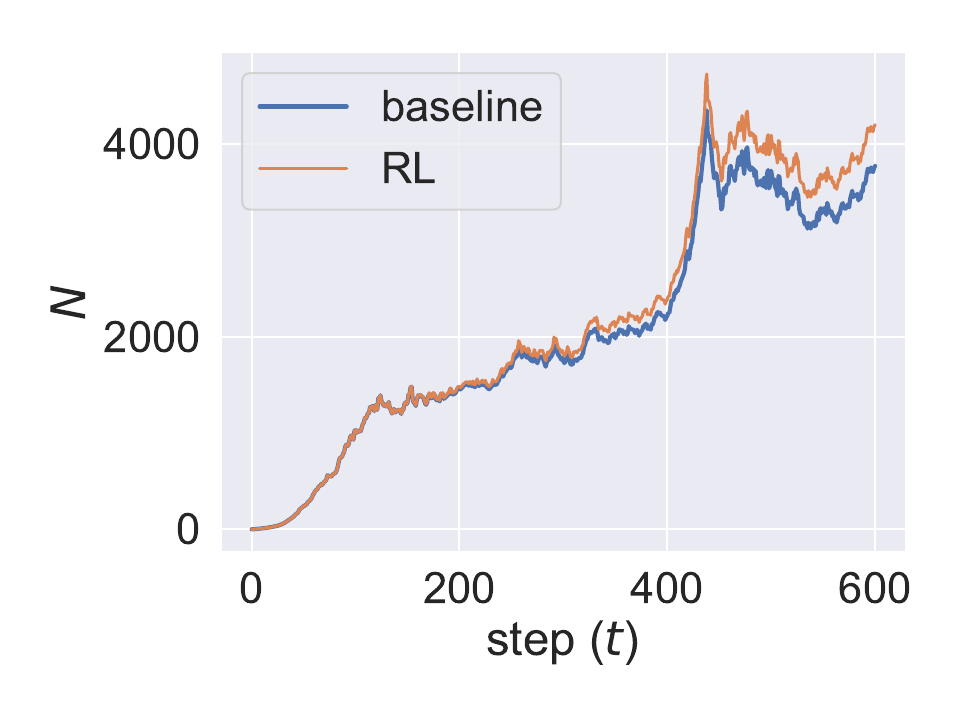}
\caption{\footnotesize Well-trained episode, no attacks.}
\label{fig:netpos703}
\end{subfigure}
\begin{subfigure}{0.245\linewidth}
    \centering
\includegraphics[width=\linewidth,trim={0.8cm 0.8cm 0.8cm 0.8cm},clip]{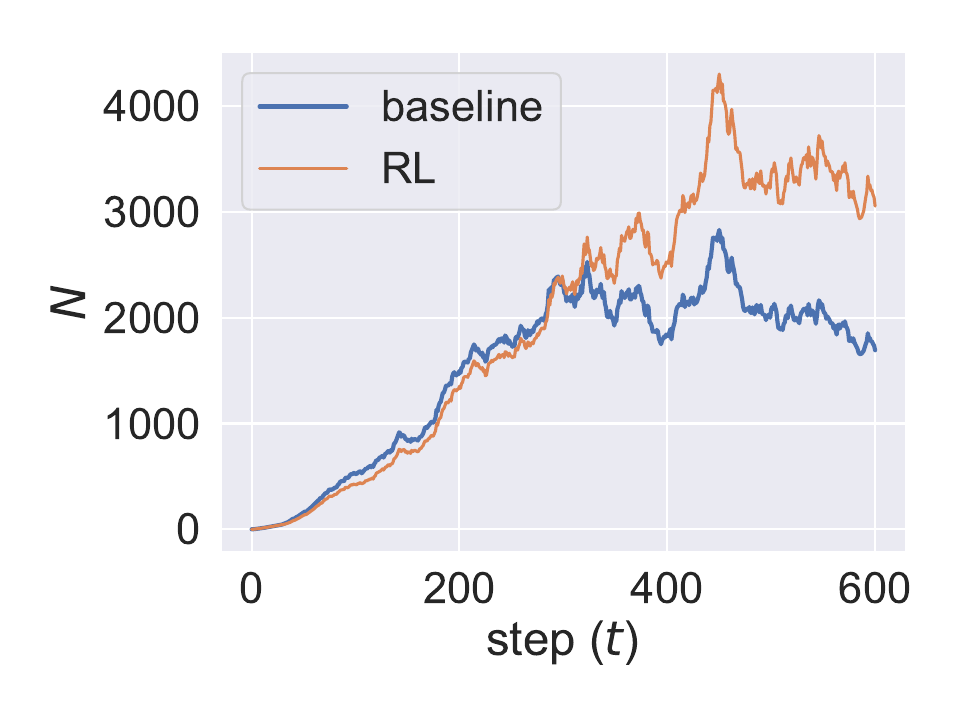}
\caption{\footnotesize Well-trained episode, with attacks.}
\label{fig:netpos681}
\end{subfigure}
\caption{Protocol's net position ($N$)
of the \ac{rl} environment
compared with the baseline environment (see \autoref{sec:baseline}).}
\label{fig:netpos}
\end{figure*}

\subsection{Governance results}
\label{sec:train_results}

In this subsection, we analyze the governance decisions implemented by the agent to determine the effectiveness of the governance agent in managing the lending protocol.

For each scenario with or without attacks, we select one early episode and one well-trained episode to demonstrate the governance results.
The early episode is the first episode whose final score equals the medium score among all the episodes before the target net is switched on. The well-trained episode is the first episode that has the top 25 percentile final score among all the episodes with a positive score after the target net is switched on.

\autoref{fig:colfact} shows for selected episodes the \ac{rl} governance agent's collateral factor determination for each token vis-à-vis the tokens' price movements\footnote{
Borrowing a common practice from \ac{tradfi} (see e.g. \cite{Bollerslev1987AReturn}), we use the logarithmic price return, expressed as $\ln \tfrac{P_t}{P_{t-1}}$, to represent price changes. Using returns makes price movements across assets comparable, unaffected by absolute price levels; using logged values symmetrically reflects the extent of price increases and decreases (in contrast, unlogged values can produce extreme positive returns that are theoretically infinite, while extreme negative returns are capped at $-100$\%).
}. 
We have noticed that in early episodes either with or without attacks, the agent exhibits a tendency to alter the collateral factor in a random and frequent manner. Note that during this phase, the agent has not yet undergone training. After the agent has been trained, the change of collateral factor becomes infrequent. This is the correct learning outcome reacting to the design of the market user behavior that a change of collateral factor increases their supply buffer, which further decreases the funds available for borrowing and ultimately reduces the protocol's revenue.

In addition, the agent tends to adopt a more aggressive strategy, i.e. maintaining collateral factors at a consistently high level throughout the episode, in the scenario without attacks than in the scenario with attacks. This behavior is intuitive, as opting for a relatively dynamic collateral factor can enhance the lending protocol's resilience against price attacks, although it may result in reduced profitability. The agent also appears to understand that, even when the possibility of attacks is present, as long as they are infrequent, the collateral factor can still be set at a moderately high level to attract deposits and allow borrows, ultimately leading to higher profits.
Being able to balance between keeping the protocol safe and maximizing profits is the key learning outcome of the agent.

\autoref{fig:state} illustrates the utilization ratio and reserve quantity of the three token assets in the lending protocol for selected episodes. Under price oracle attacks whose occurrences are marked with \enquote{\textcolor{red}{$\times$}} in the two subfigures, the utilization ratio of TKN typically experiences a dip as the attacker always first offsets TKN loans (see \autoref{sec:reaction}).
Under no attacks, especially in well-trained episodes when collateral factors do not undergo frequent and random alteration, the utilization ratios of all three tokens almost always remain at an optimal level of between 0.6 and 0.8. 
This is likely driven by the correctly encoded market user reactions that tend to be {\em borrow} when the utilization ratio is low (hence low {\em borrow interest rate}) and {\em supply} when the utilization ratio is high (hence high {\em supply interest rate}) (see \autoref{sec:lending_states}), thus keeping the utilization ratio at a steady, equilibrated level.
Comparing \autoref{fig:state703} and~\ref{fig:state681} with \autoref{fig:state187} and~\ref{fig:colfact256}, it is evident that the governance agent has learned to effectively increase the protocol reserve after being trained in scenarios both with and without attacks.

\begin{figure*}[t]
\centering
\begin{subfigure}{0.32\linewidth}
    \centering
\includegraphics[width=0.9\linewidth,trim={0.8cm 0.8cm 0.8cm 0.8cm},clip]{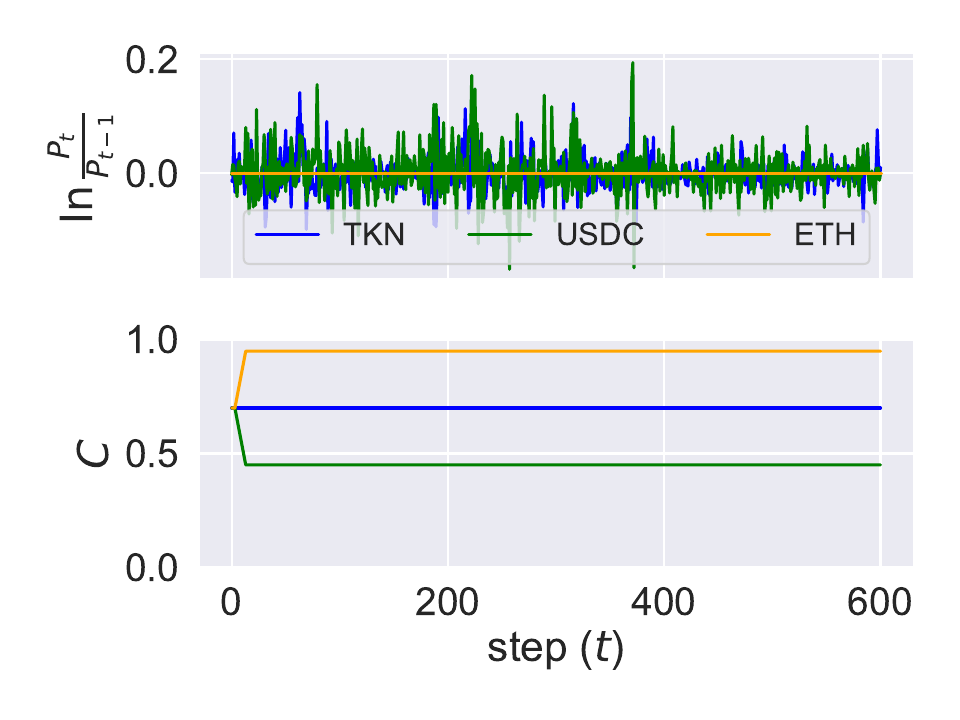}
\caption{Price change in log return  ($\ln\tfrac{P_{t}}{P_{t-1}}$) and collateral factor ($C$) generated by the \ac{rl} governance agent.}
\label{fig:test_colfact}
\end{subfigure}
\hfill
\begin{subfigure}{0.32\linewidth}
    \centering
\includegraphics[width=0.9\linewidth,trim={0.8cm 0.8cm 0.8cm 0.8cm},clip]{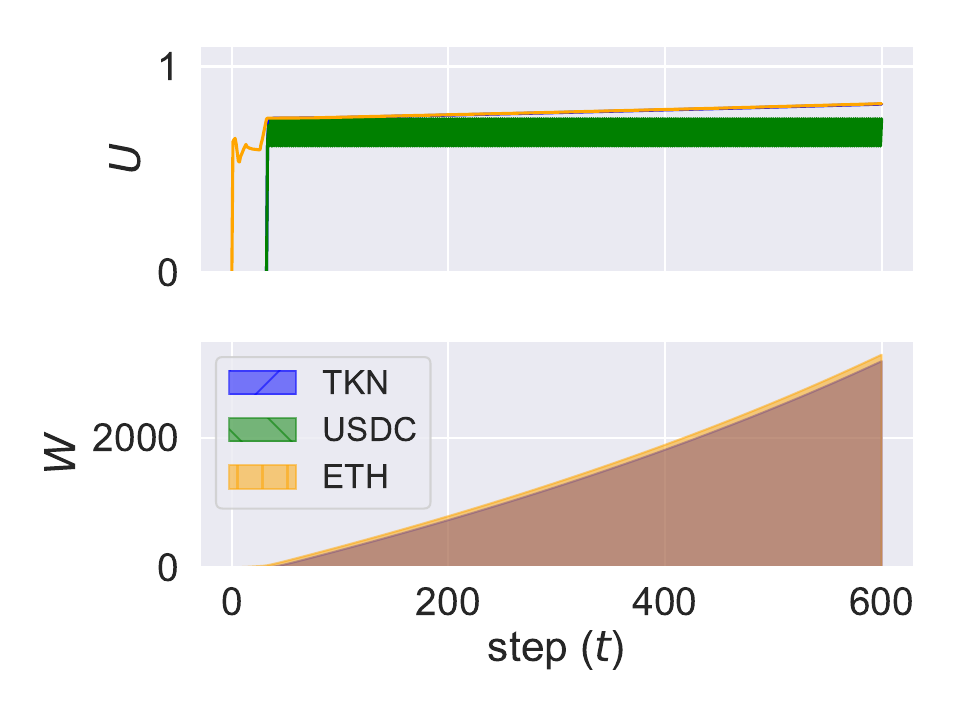}
\caption{Lending pool states, utilization ratio ($U$) and reserve ($W$), over time in the \ac{rl} environment.}
\label{fig:test_state}
\end{subfigure}
\hfill
\begin{subfigure}{0.32\linewidth}
    \centering
\includegraphics[width=0.9\linewidth,trim={0.8cm 0.8cm 0.8cm 0.8cm},clip]{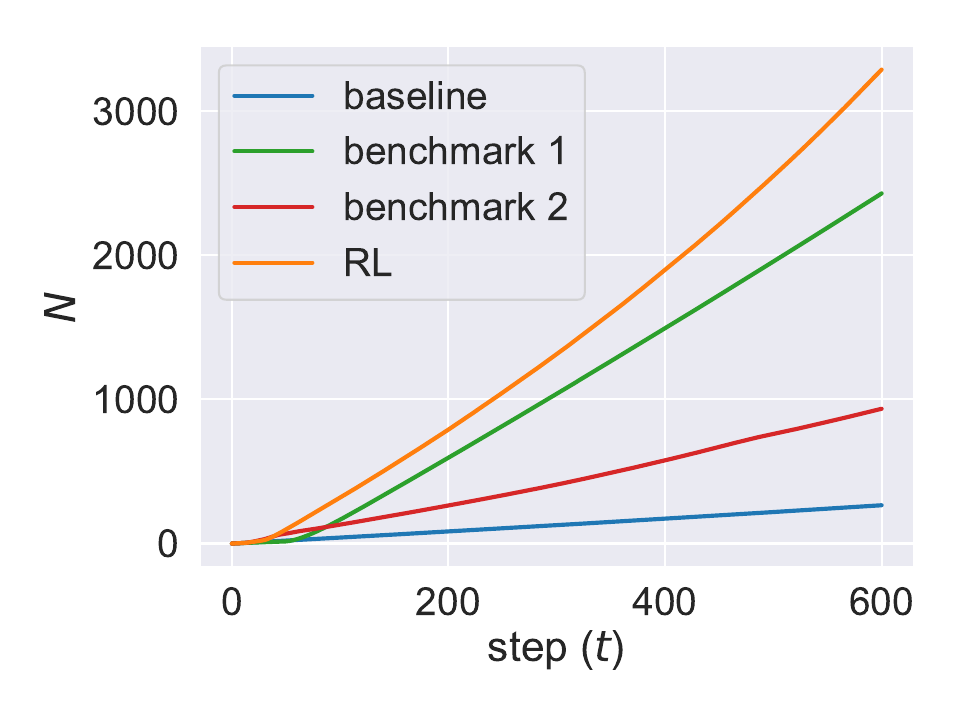}
\caption{Protocol's net position ($N$) in the \ac{rl}, benchmarks 1 \& 2, and baseline environments.}
\label{fig:test_netpos}
\end{subfigure}
\caption{Test with trained \ac{rl} model using real historical price data.}
\label{fig:realworld_test}
\end{figure*}

The positive outcomes of training and learning are particularly evident in \autoref{fig:netpos}, which illustrate the protocol's net position. 
In early episodes, the undertrained \ac{rl} agent underperforms the baseline either without (\autoref{fig:netpos187}) or with (\autoref{fig:netpos256}) the existence of attacks.
The performance rank flips in later episodes when the \ac{rl} agent becomes well-trained.
In scenarios without attacks (\autoref{fig:netpos703}), the \ac{rl} agent consistently produces a higher net position (as calculated with \autoref{eq:net-position}). In the presence of attacks (\autoref{fig:netpos681}), the \ac{rl} agent slightly lags behind the baseline in terms of net position during the initial steps. However, as the number of steps increases, the agent remarkably outperforms the baseline, showcasing the robust effectiveness of our model. 

We further observe that the superiority of \ac{rl} becomes more pronounced in the presence of attacks: while the \ac{rl} agent outperforms the baseline agent by approximately 10\% absent attacks (\autoref{fig:netpos703}), this advantage surges to over 60\% when attacks are present (\autoref{fig:netpos681}), demonstrating the particular advantage of Auto.gov in the attack-prone \ac{defi} ecosystem.
Another point worth noting is that the agent's reward value is approximately equal to the difference between the orange line and the blue line in \autoref{fig:netpos}. It is evident that well-trained agents achieve positive cumulative rewards over time, whereas under-trained agents end up with negative cumulative rewards.

\subsection{Tests with real-world data}
\label{sec:test_real}

We further evaluate our well-trained governance agent by testing it against other approaches using real-world data. 

Regarding the real-world test data, we fetch from CryptoCompare.com the time series of daily ETH-denominated prices of USDC and LINK, where LINK---the protocol token of Chainlink---plays the role equivalent to TKN's in our training stage. We do not plant any attack event, as we believe that all price manipulation activities---if any---have already been manifested in the historical price data retrieved.

We test the policy generated by the well-trained agent in comparison with not only the static baseline strategy (see \autoref{sec:baseline}); additionally, we compare Auto.gov with two benchmark approaches: a statistical approach and a Q-learning-based approach. Since there are no existing approaches for automated \ac{defi} governance, we have designed and implemented these two benchmark approaches as follows.

\paragraph{Benchmark 1: statistical approach} This approach utilizes a dynamic governance strategy derived from the analysis of empirical data.
The environment of Benchmark 1 is initiated like Auto.gov and the baseline environments, and the strategy used by Benchmark 1 to adjust the collateral factor in response to market volatility following the two steps below:
\begin{enumerate}
    \item Every 7 days, the governance agent in the benchmark 1 environment first calculates the theoretical collateral factor:%
\begin{equation*}
    C_{i,t} = \max (C_{i,0} - b \cdot \sigma_{i,t},0),
\end{equation*}%
where $C_{i,0} = 75\%$, $b = 0.08$ and $\sigma_{i,t}$ denotes the standard deviation of daily log returns from $t-6$ to $t$. The value of $C_{i,0}$ corresponds to the historical collateral factor value of ETH (see \autoref{fig:consensus-tokens}) and the value of coefficient $b$ corresponds to the coefficient of the 7-day volatility in a simple \ac{ols} regression with the collateral factor being the dependent variable using historical data (the same data used to generate \autoref{fig:aave-corr}).
\item The theoretical collateral factor $C_{i,t}$ is compared with the current collateral factor to see whether an incremental adjustment of 0.25 (see \autoref{sec:change-coll}) is needed: if the theoretical value exceeds or falls below the current one by over 0.25, then a downward or upward adjustment of 0.25 is executed, respectively; otherwise, the current collateral factor remains for another 7-day cycle.
\end{enumerate}

\paragraph{Benchmark 2: Q-learning-based approach} We implement an alternative automated \ac{defi} governance approach based on \ac{rl}. The environment and agent modeling methods are identical to those used in Auto.gov. However, Benchmark 2 relies on the original tabular Q-learning algorithm, which uses a Q-table~\cite{watkins1992q} to store state-action pairs. Since the state is represented by continuous values, which could make the Q-table infinitely large, we discretize the state values to ensure they fit within the Q-table.

\autoref{fig:realworld_test} illustrates the test result. The collateral factor adjustment policy determined by the trained \ac{rl} agent (Auto.gov) is depicted in \autoref{fig:test_colfact}. The adjustment is only concentrated within the first few steps and the collateral factors quickly stabilize for the rest of the testing episode. The infrequent adjustment aligns with what we observe in the well-trained episodes from the training stage (\autoref{fig:colfact703} and \ref{fig:colfact681}). \autoref{fig:test_state} shows that all three lending pools in the \ac{rl} environment consistently build up reserve over time, and maintain a healthy utilization ratio of between 0.6 and 0.8 quickly after the start of the testing episode. Finally in \autoref{fig:test_netpos}, we demonstrate our \ac{rl} model's superior capability in meeting its pre-set objective of profit maximization, i.e. to increase $N$ as much as possible (see \autoref{sec:rewards}): the \ac{rl} strategy outperforms not only the static baseline strategy but also the two benchmark approaches. Although Benchmark 2 is also based on \ac{rl}, the Q-table struggles to handle large state spaces effectively, leading to significantly worse performance compared to Auto.gov and Benchmark 1. This limitation arises from the Q-table's inability to generalize across continuous state values and its exponential growth with increasing state dimensions. These shortcomings highlight the necessity of using \ac{dqn} to power Auto.gov, as it can efficiently manage large and continuous state spaces through function approximation.

\section{Discussion}
\label{sec:discussion}

This section discusses the limitations of Auto.gov and proposes avenues for future improvements.

\subsection{Limitations}
\label{sec:limitations}

\paragraph{Environment variations}
For the ease of result interpretation and due to facility constraints, we used a stylized \ac{defi} environment (see \autoref{sec:simplification}) to achieve demonstration purposes. Depending on the actual complexity of the specific \ac{defi} protocol and desired automation level, the training time may increase, enhanced computational resources may be required, and the agent efficacy may vary.

\paragraph{Operator risks}
Our approach may be subject to risks stemming from the training operator. Currently, we are agnostic about the training operator in our framework.
Depending on the exact design of the operator (e.g., \ac{dao} and elected party), they may be able to employ biased training strategies to instruct the governance agent in a manner that could generate profits for the operators themselves.

\paragraph{Adversarial machine learning attacks} Our approach might be exposed to threats stemming from adversarial machine learning tactics~\cite{Huang2011AdversarialLearning}. Firstly, adversaries might deduce the agent model by probing or observation~\cite{Barreno2006CanSecure}, subsequently forecasting the upcoming collateral factor for conducting malicious activities.
Conversely, by identifying weaknesses of the model, adversaries may destabilize the agent with certain input modifications by altering the environment. Addressing this concern entails more rigorous training, regular validation, real-time monitoring, and timely model updates~\cite{McDaniel2016MachineSettings}.

\subsection{Adjustments and expansions}
\label{sec:adjustment}

For improvement and to address the limitations listed above, our framework may be further adjusted and expanded as follows: 

\paragraph{More training dimensions}
Instead of only training the governance agent to find the optimal collateral factor adjustment policy, we can also allow it to adjust other parameters such as other risk parameters and the interest rate model parameters.

\paragraph{More training scenarios}
We can train the governance agent under various scenarios such as different market conditions (e.g. time-varying competing interest rates) and different user behaviors (e.g. various levels of borrow confidence decrease after a collateral factor drop).

\paragraph{More sophisticated machine learning techniques}
We can apply more sophisticated machine learning models such as multi-agent reinforcement learning by allowing users to also be reinforcement learning agents with their own objectives (e.g. maximizing their terminal wealth).

\paragraph{Model adaptability and transferability}
\label{sec:extension}
With some tweaks and modifications, our model can be easily extended to other lending protocols such as Compound and dForce.
Upon proper amendment, the model can also be applied to other types of \ac{defi} protocols such as \acp{amm}. For example, the governance agent can dynamically adjust \ac{amm} pool parameters---such as Curve's amplification coefficient~\cite{nagaking}---based on market conditions.

\section{Conclusion}
\label{sec:conclusion}
In this paper, we attempt to build an optimal, resilient \ac{defi} governance solution by first abstracting a simplified yet comprehensive \ac{defi} environment to model the governance agent's reward function, and then applying \ac{dqn} \ac{rl} to train the agent to find the optimal policy for protocol parameter adjustment. We use an Aave-like lending protocol, with the likelihood of price oracle attacks as an example, but our environment can be easily adjusted to apply to other protocol categories and account for different types of attacks.

Our results clearly show that the trained agent successfully adapts to various scenarios, including those with and without attacks, demonstrating its capability to understand the consequences of setting different values and the necessity of assigning appropriate values based on the specific situation. The trained Auto.gov agent also demonstrates its superiority by outperforming the static baseline approach and two benchmark approaches (one statistical and one tabular Q-learning-based) in environments with both simulated and real-world data.

Our experiment demonstrates the potential to replace the existing lengthy governance procedure, which is fully manual and entails human bias, with an \ac{rl}-based approach that emphasizes security, profitability, and operating efficiency. We anticipate that more \ac{defi} protocols will adopt (a variation of) our model for their governance process, leading to a more secure and efficient ecosystem.

\section*{Acknowledgments}
We thank Alice Ng, Yitian Wang, Su Farn Fong, Kamil Tylinski and Ali Irzam Kathia for reviewing the manuscript.

This research / project is partially supported by the National Research Foundation, Singapore, and the Cyber Security Agency under its National Cybersecurity R\&D Programme (NCRP25-P04-TAICeN). Any opinions, findings and conclusions or recommendations expressed in this material are those of the author(s) and do not reflect the views of National Research Foundation, Singapore and Cyber Security Agency of Singapore.

This material is based upon work partially supported by Ripple under the University Blockchain Research Initiative (UBRI). Any opinions, findings, and conclusions or
recommendations expressed in this material are those of the authors and do not necessarily reflect the views of Ripple.

\bibliographystyle{IEEEtran}
\bibliography{references}
\newpage

\vspace{-10mm}
\begin{IEEEbiography}[{\adjustbox{raise=7.5mm}{\includegraphics[width=1in, height=1in, trim ={0 100 0 10}, clip]{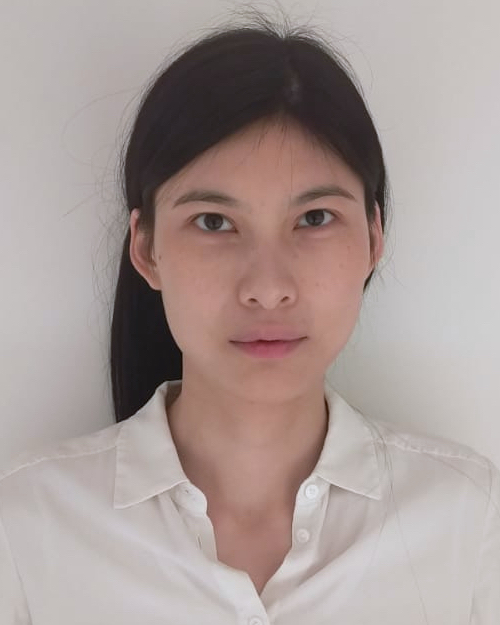}}}]
{Jiahua Xu}
    is an Associate Professor in Financial Computing and Programme Director of the MSc Emerging Digital Technologies at UCL. Her research focuses on blockchain economics and decentralized finance, with publications in Usenix Security, ACM IMC, ACM ASIACCS, FC, IEEE ICDCS and IEEE COMST. 
\end{IEEEbiography}
\vspace{-20mm}
\begin{IEEEbiography}
[{\adjustbox{raise=7.5mm}{\includegraphics[width=1in,height=1in,trim={0 250 0 100}, clip]{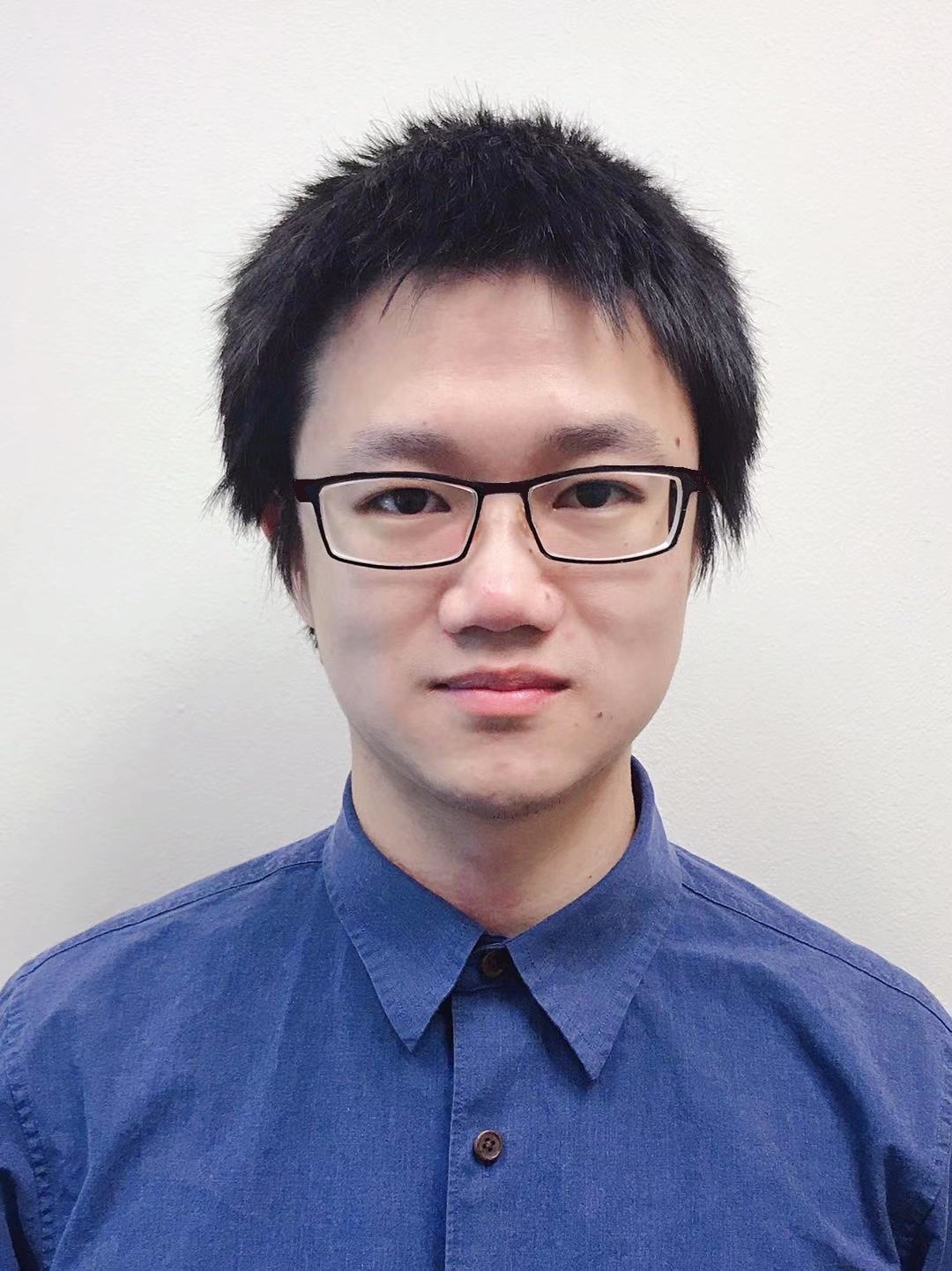}}}]{Yebo Feng}
    is a research fellow at Nanyang Technological University (NTU).
    He received his Ph.D. degree in Computer Science from the University of Oregon (UO).
    His research interests include network security, blockchain security, AI security, and anomaly detection.
\end{IEEEbiography}
\vspace{-20mm}
\begin{IEEEbiography}
[{\adjustbox{raise=7.5mm}{\includegraphics[width=1in,height=1in, clip]{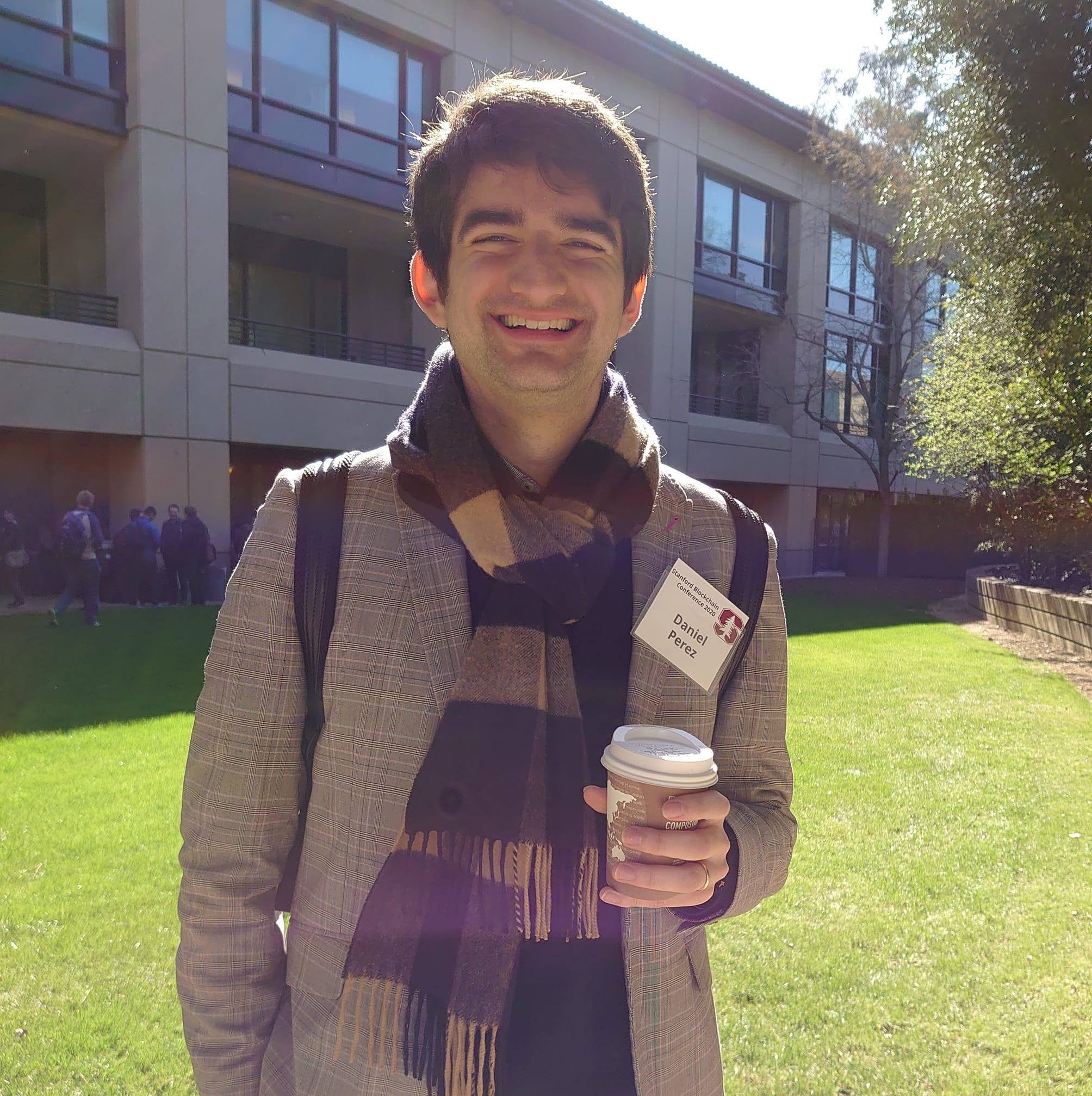}}}]{Daniel Perez}
    is a software engineer and researcher in cybersecurity and decentralized finance. He is the co-founder Gyroscope, an all-weather stablecoin protocol. He received his PhD at Imperial College London sponsored by the Ethereum Foundation.
    He has published in Usenix Security, ACM IMC, NDSS, FC and MSR. 
\end{IEEEbiography}
\vspace{-20mm}
\begin{IEEEbiography}
[{\adjustbox{raise=7.5mm}{\includegraphics[width=1in,height=1in, trim={50 0 0 0},  clip]{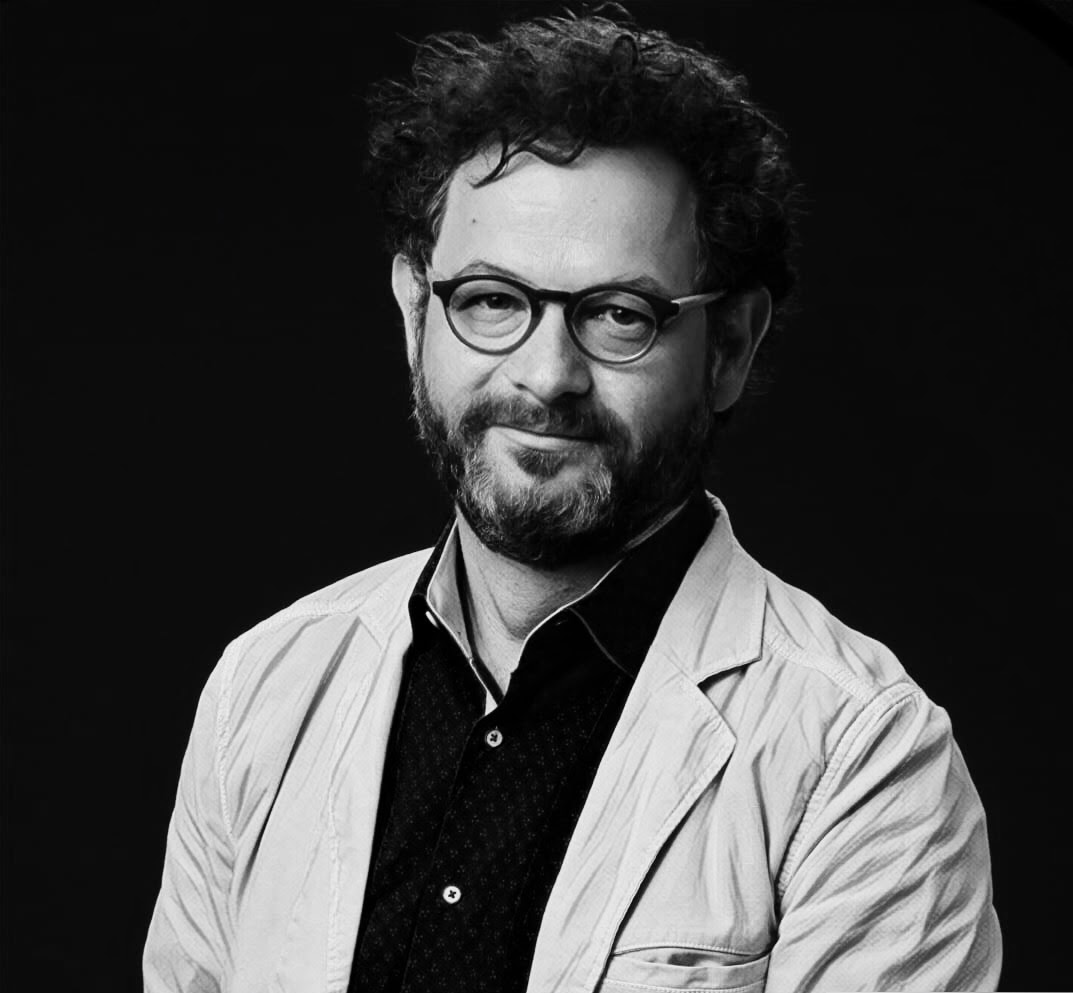}}}]{Benjamin Livshits}
    is a Reader at Imperial College London and an affiliate professor at the University of Washington in Seatte, USA. 
    Previously, he was a research scientist at Microsoft Research in Seattle for about ten years.
    He received a bachelor's degree in Computer Science and Math from Cornell University in 1999, and his M.S. and Ph.D. in Computer Science from Stanford University in 2002 and 2006.
\end{IEEEbiography}

\clearpage
\newpage
\appendix
\subsection{Trend of discussion in \ac{defi} governance forum}
\label{app:proposal}

\begin{figure}[b]
    \centering
    \includegraphics[width =\linewidth]{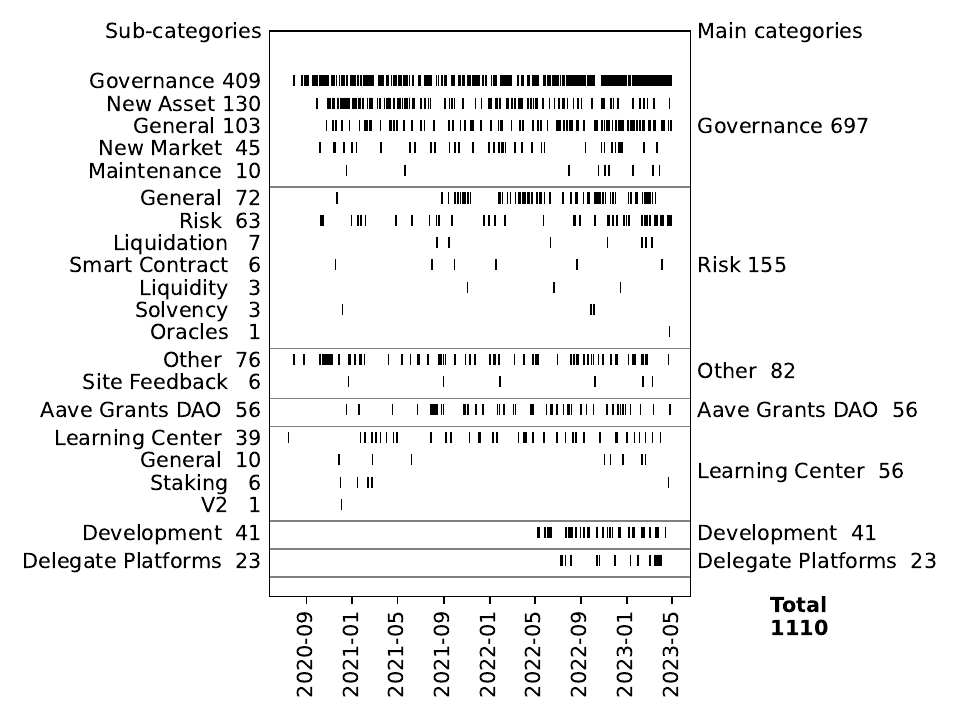}
    \caption{Statistics of Aave governance forum posts.}
    \label{fig:aavegovposts}
\end{figure}

\autoref{fig:aavegovposts} illustrates the occurrences of forum posts over time across various categories, totaling 1,110 posts from July 2020 until April 2023. The discussion on smart-contract-level adjustment---e.g., the value of risk parameters, acceptance or deprecation of an asset as collateral---is typically under \enquote{Governance} and \enquote{Risk}, two of the most active main categories. Evidently from \autoref{fig:aavegovposts}, the governance forum, especially the above-mentioned two categories, has been increasingly used during the past year, indicating the growing significance in \ac{defi} governance perceived by the community.

\subsection{Training with both the primary network and the target network}
\label{app:target_network}

In this appendix, we introduce our hybrid approach to leveraging both the primary network and the target network in agent training.

In \ac{rl}, applying the target network is an effective approach to preventing overfitting during training and enhancing the agent's resilience to potential attacks. However, simply employing the target network may bring the following side effects:
\begin{itemize}
\item \textbf{Stability vs. Responsiveness}: The target network increases stability but reduces responsiveness, causing temporary mismatches between the online network's estimates and target values.
\item \textbf{Delayed learning}: The infrequent updating of the target network may lead to slower training and temporary performance drops as the online network readjusts.
\item \textbf{Overestimation bias}: Using a separate target network can help mitigate overestimation bias but may result in conservative Q-value estimations, causing temporary performance drops.
\end{itemize}

\begin{algorithm}[tb]
            \caption{Training with both primary networks and target networks.}
            \label{alg:target_net}
            \begin{algorithmic}[1]
                \State Input $total\_episode\_num$
                \State Input $decay\_func1()$, $decay\_func2()$ \Comment{Input two different decay functions}
                \State Input $target\_switch\_on\_point$
                \State $Ctr \leftarrow 0$ \Comment{Initialize the counter}
                \State $\epsilon \leftarrow 1$
                \While{$Ctr \leq total\_episode\_num$}
                    \State $Ctr \leftarrow Ctr+1$
                    \If{$\epsilon < target\_switch\_on\_point$}
                        \State Switch on the target network
                        \State $\epsilon \leftarrow decay\_func2(\epsilon)$
                        \State $train()$ \Comment{Train with target net}
                    \Else
                        \State Switch off the target network
                        \State $\epsilon \leftarrow decay\_func1(\epsilon)$
                        \State $train()$ \Comment{Train without target net}
                    \EndIf
                \EndWhile
            \end{algorithmic}
\end{algorithm}

To address the disadvantages associated with the target network, we propose a novel approach for its implementation. Initially, we train the agent using only the primary network, resulting in accelerated early learning. After a certain amount of training, we introduce the target network and allow $\epsilon$ to decrease at a more gradual pace. This slower reduction in $\epsilon$ means that the agent explores the environment more slowly. This strategy can aid the agent in discovering more optimal solutions. By incorporating the target network later in the training process, we aim to balance the trade-offs among stability, responsiveness, and learning speed.
Algorithm~\ref{alg:target_net} illustrates this training procedure, where we employ different $\epsilon$ decay functions for the primary and target networks, with the latter being slower.
The evaluation results presented in \autoref{sec:results} of the main body are derived from the training using Algorithm~\ref{alg:target_net}.

\subsection{RL key hyperparameters}
\label{app:parameter}

Our research included extensive experiments to determine the optimal hyperparameters for our methodology. In this appendix, we present the rationale behind key hyperparameter selections.

For the structure of a neural network, particularly the number of hidden layers and neurons per layer, it is generally observed that higher numbers can enhance the network's ability to learn and interpret complex patterns. However, this increase in complexity is not without its drawbacks. Key challenges include the potential for overfitting, where the model excessively learns from the training data to the detriment of its performance on new data; heightened computational demands during both the training and inference phases; and the phenomenon of diminishing returns, where additional complexity fails to significantly improve model performance. Our experiments involved testing various combinations of hyperparameters, such as networks with 1, 2, 3, 4, or 5 hidden layers and varying numbers of neurons per layer (i.e., 64, 128, 256, 512). Through this rigorous testing process, we have identified an optimal configuration: a neural network architecture comprising 3 hidden layers, each with 256 neurons. This structure strikes a balanced compromise, delivering robust performance while maintaining reasonable computational efficiency.

Another key hyperparameter that has a substantial impact on neural network training is the batch size, which is the number of training examples used in each iteration, and influences several aspects of learning dynamics, including learning speed, generalization, and memory constraints. Through extensive testing of various batch sizes (16, 32, 64, 128, 256, 512, and 1024), we noted distinct optimal sizes for different scenarios in \ac{defi} environments. For scenarios involving attacks, a batch size of 32 proved the most effective, striking a balance between rapid convergence and model generalization. In contrast, for scenarios without attacks, a larger batch size of 256 was optimal, utilizing computational resources more efficiently while still maintaining effective memory management. These chosen batch sizes complement our learning rate well, fostering stable and efficient training dynamics. They represent a tailored approach, recognizing the unique demands of different operational environments in \ac{defi} systems.

Last but not least, a hyperparameter worth mentioning in the context of reinforcement learning is gamma ($\gamma$), the discount factor. This parameter is crucial for balancing the importance of immediate versus future rewards. Lower gamma values, such as 0.1, lead the model to prioritize immediate rewards, while higher values like 0.9 shift the focus toward long-term benefits. In our model, we set $\gamma = 0.5$, indicating that we value immediate gains and long-term outcomes equally. Such a balanced choice is particularly strategic in environments where it is essential to weigh short-term actions against their potential long-term impact.
Selecting a mid-range gamma value thus allows our model to be responsive to current situations while still considering the repercussions of its actions in the future. 

\subsection{Crypto-asset markets on Aave}
\label{sec:aave-markets}

\autoref{fig:atokens} presents the market condition and lending pool state history of tokens in the Aave protocol from January 2020 to April 2023.
For each token, we plot the time series of the daily log return of its ETH-denominated price, as well as the trading volume against ETH. We additionally delineate the historical values of each lending pool's three risk parameters: liquidity incentive, liquidation threshold, and collateral factor. Finally, we show the history of each pool's total borrows and total liquidity---both in the unit of the underlying token, as well as its actual and optimal utilization ratios.

In general, assets with higher price volatility and lower liquidity are more likely to cause protocol insolvency, all other factors being equal. Consequently, risk parameters are set on an asset-by-asset basis and may be adjusted, albeit with low historical frequency, in the event of a significant change in the asset's risk profile. The collateral factor and liquidation threshold should theoretically be set lower for assets with higher price volatility, while the liquidation incentive should be set higher for assets with lower market liquidity.
Empirically, we observe a similar pattern. \autoref{fig:aave-corr} displays the correlation matrix between various risk parameters and asset risk metrics of assets listed on Aave, calculated based on asset-day-level observations from January 2021 to April 2023. As expected, the collateral factor and liquidation threshold are negatively correlated with the asset's volatility---measured by the 7-day standard deviation of daily logarithmic returns, and the liquidation incentive is negatively correlated with the asset's market liquidity---measured by the 7-day average daily trading volume.

\begin{table}[!t]
  \centering
  \setlength{\tabcolsep}{0pt}
  \caption{Spearman's Rank Correlation Coefficient \cite{spearman1904} between Risk Parameters and Asset Risk Metrics of a
  Assets of Aave, Based on Asset-day-observations from January 2021 to April 2023.
  \label{fig:aave-corr}}%
  \tiny
\begin{tabularx}{\linewidth}{r@{\hspace{1mm}}rX@{\hspace{0.5mm}}rX@{\hspace{0.5mm}}rX@{\hspace{0.5mm}}rX@{\hspace{0.5mm}}rX}
    \toprule
& \multicolumn{2}{c}{liquidation} & \multicolumn{2}{c}{liquidation} & \multicolumn{2}{c}{collateral} & \multicolumn{2}{c}{7-day} & \multicolumn{2}{c}{7-day average} \\
& \multicolumn{2}{c}{incentive} & \multicolumn{2}{c}{threshold} & \multicolumn{2}{c}{factor} & \multicolumn{2}{c}{volatility} & \multicolumn{2}{c}{volume} \\
liquidation incentive & \cellcolor[rgb]{ .855,  .588,  .58}$1.000$ & \cellcolor[rgb]{ .855,  .588,  .58}$^{***}$ & \cellcolor[rgb]{ .984,  .992,  .992}$-0.036$ & \cellcolor[rgb]{ .984,  .992,  .992}$^{}$ & \cellcolor[rgb]{ .965,  .894,  .894}$0.258$ & \cellcolor[rgb]{ .965,  .894,  .894}$^{*}$ & \cellcolor[rgb]{ .98,  .945,  .941}$0.140$ & \cellcolor[rgb]{ .98,  .945,  .941}$^{}$ & \cellcolor[rgb]{ .898,  .953,  .965}$-0.231$ & \cellcolor[rgb]{ .898,  .953,  .965}$^{*}$ \\
liquidation threshold & \cellcolor[rgb]{ .984,  .992,  .992}$-0.036$ & \cellcolor[rgb]{ .984,  .992,  .992}$^{}$ & \cellcolor[rgb]{ .855,  .588,  .58}$1.000$ & \cellcolor[rgb]{ .855,  .588,  .58}$^{***}$ & \cellcolor[rgb]{ .906,  .725,  .722}$0.670$ & \cellcolor[rgb]{ .906,  .725,  .722}$^{***}$ & \cellcolor[rgb]{ .949,  .976,  .98}$-0.114$ & \cellcolor[rgb]{ .949,  .976,  .98}$^{}$ & \cellcolor[rgb]{ .965,  .89,  .89}$0.269$ & \cellcolor[rgb]{ .965,  .89,  .89}$^{**}$ \\
collateral factor & \cellcolor[rgb]{ .965,  .894,  .894}$0.258$ & \cellcolor[rgb]{ .965,  .894,  .894}$^{*}$ & \cellcolor[rgb]{ .906,  .725,  .722}$0.670$ & \cellcolor[rgb]{ .906,  .725,  .722}$^{***}$ & \cellcolor[rgb]{ .855,  .588,  .58}$1.000$ & \cellcolor[rgb]{ .855,  .588,  .58}$^{***}$ & \cellcolor[rgb]{ .863,  .933,  .953}$-0.320$ & \cellcolor[rgb]{ .863,  .933,  .953}$^{**}$ & \cellcolor[rgb]{ .988,  .992,  .996}$-0.021$ & \cellcolor[rgb]{ .988,  .992,  .996}$^{}$ \\
7-day volatility & \cellcolor[rgb]{ .98,  .945,  .941}$0.140$ & \cellcolor[rgb]{ .98,  .945,  .941}$^{}$ & \cellcolor[rgb]{ .949,  .976,  .98}$-0.114$ & \cellcolor[rgb]{ .949,  .976,  .98}$^{}$ & \cellcolor[rgb]{ .863,  .933,  .953}$-0.320$ & \cellcolor[rgb]{ .863,  .933,  .953}$^{**}$ & \cellcolor[rgb]{ .855,  .588,  .58}$1.000$ & \cellcolor[rgb]{ .855,  .588,  .58}$^{***}$ & \cellcolor[rgb]{ 1,  .996,  .992}$0.019$ & \cellcolor[rgb]{ 1,  .996,  .992}$^{}$ \\
7-day average volume & \cellcolor[rgb]{ .898,  .953,  .965}$-0.231$ & \cellcolor[rgb]{ .898,  .953,  .965}$^{*}$ & \cellcolor[rgb]{ .965,  .89,  .89}$0.269$ & \cellcolor[rgb]{ .965,  .89,  .89}$^{**}$ & \cellcolor[rgb]{ .988,  .992,  .996}$-0.021$ & \cellcolor[rgb]{ .988,  .992,  .996}$^{}$ & \cellcolor[rgb]{ 1,  .996,  .992}$0.019$ & \cellcolor[rgb]{ 1,  .996,  .992}$^{}$ & \cellcolor[rgb]{ .855,  .588,  .58}$1.000$ & \cellcolor[rgb]{ .855,  .588,  .58}$^{***}$ \\
    \bottomrule
\end{tabularx}
*, **, and *** denote the 1\%, 5\%, and 10\% significance levels, respectively.
\end{table}%

\begin{figure*}
\centering

\begin{subfigure}{\textwidth}
    \centering
    \includegraphics[width=.39\textwidth]{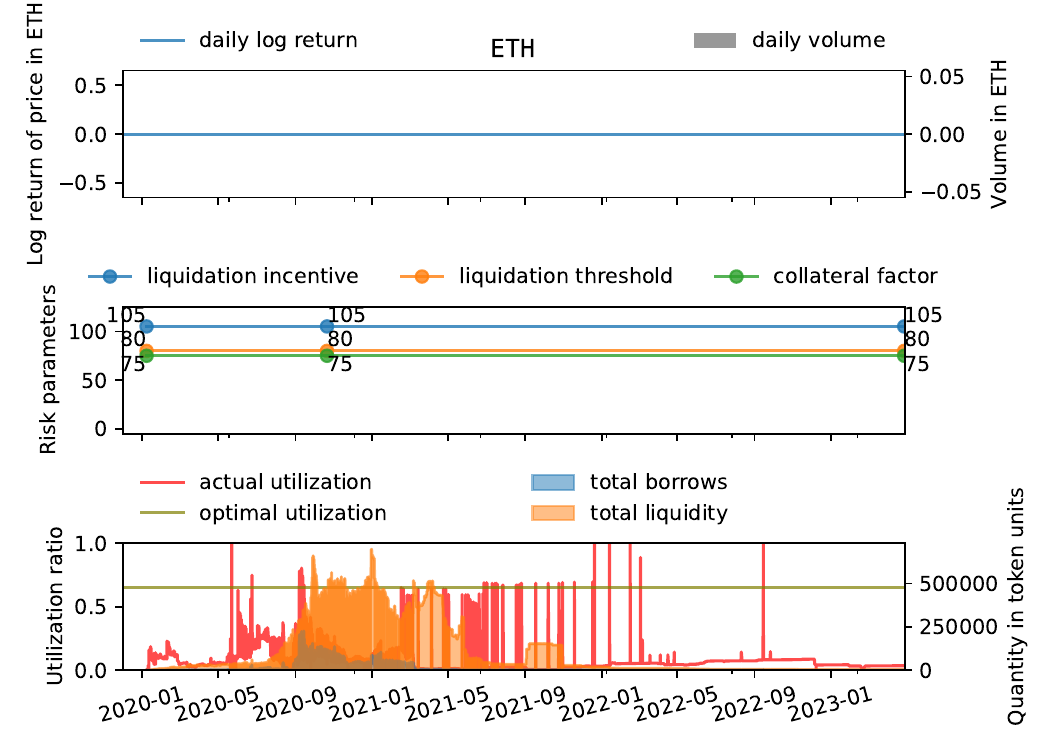}
    \qquad
    \includegraphics[width=.39\textwidth]{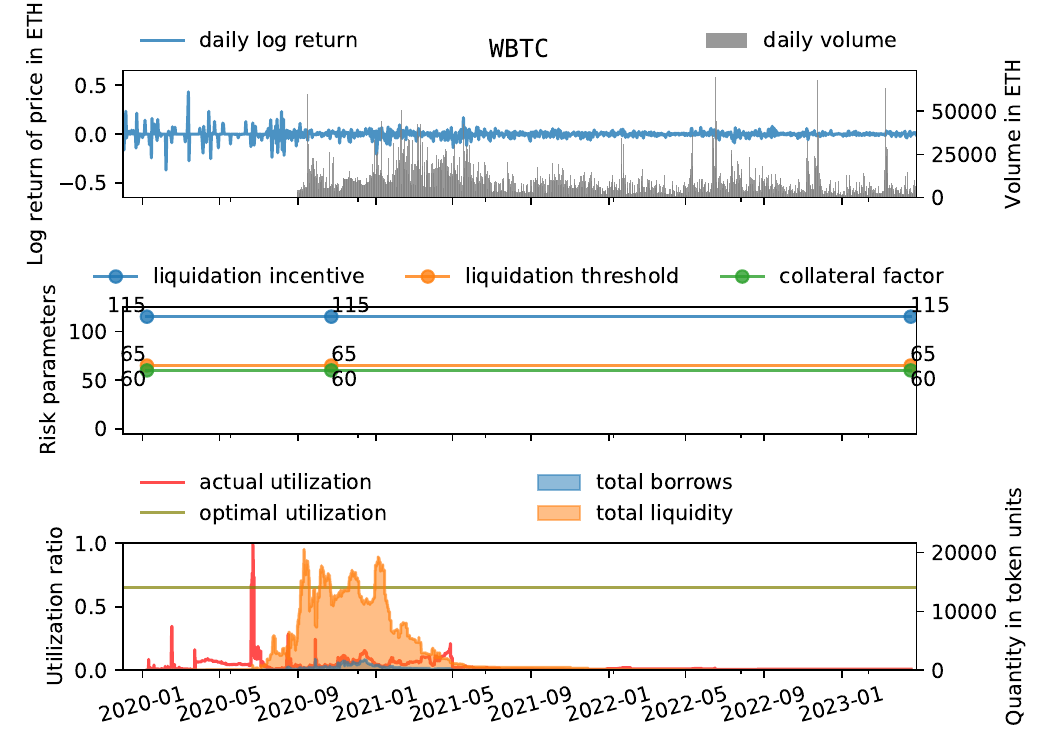}
    \caption{Distributed ledger consensus-layer tokens}
    \label{fig:consensus-tokens}
\end{subfigure}

\begin{subfigure}{\textwidth}
\centering
\vspace{10pt}
\includegraphics[width=.39\textwidth]{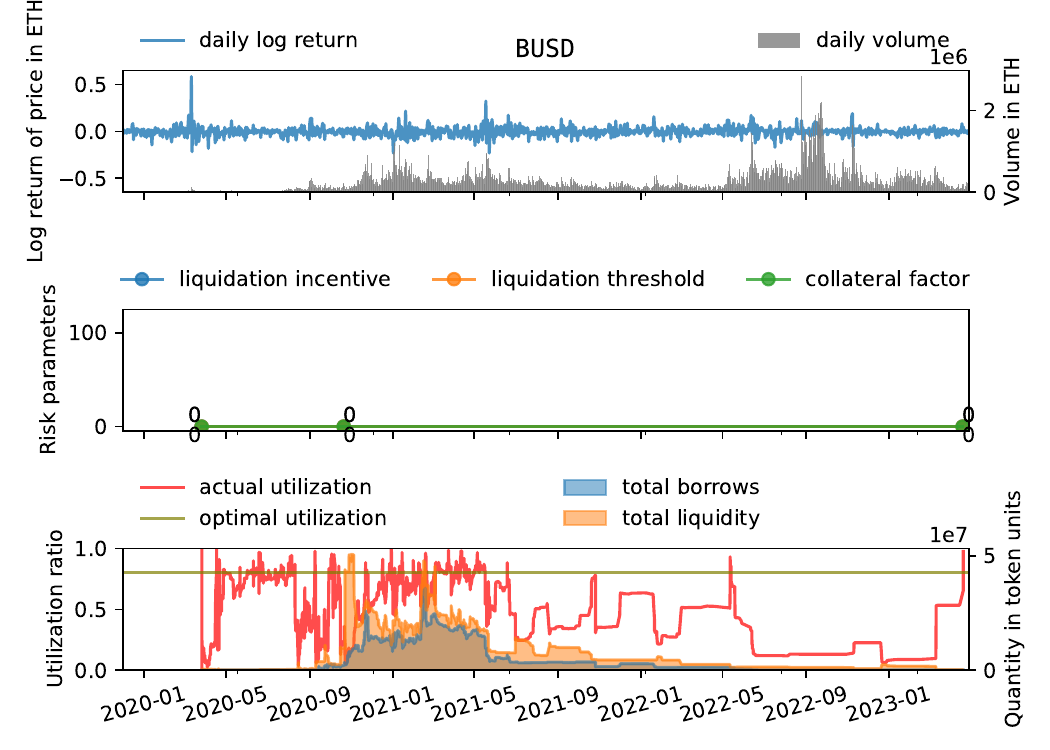}
\qquad
\includegraphics[width=.39\textwidth]{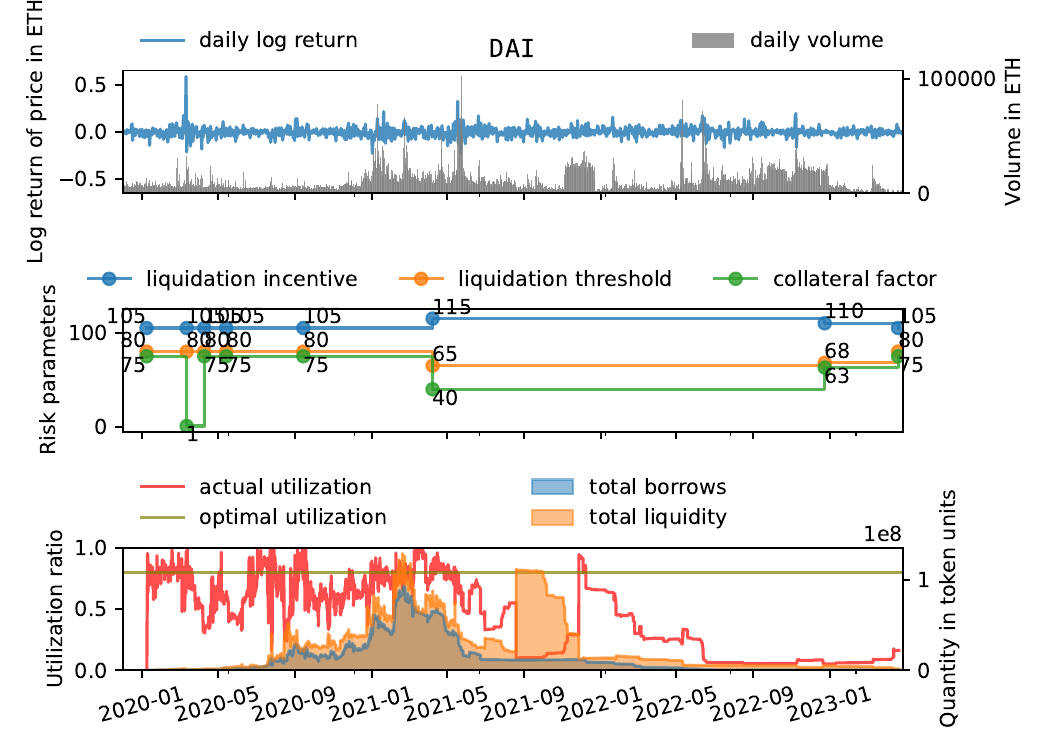}

\vspace{10pt}
\includegraphics[width=.39\textwidth]{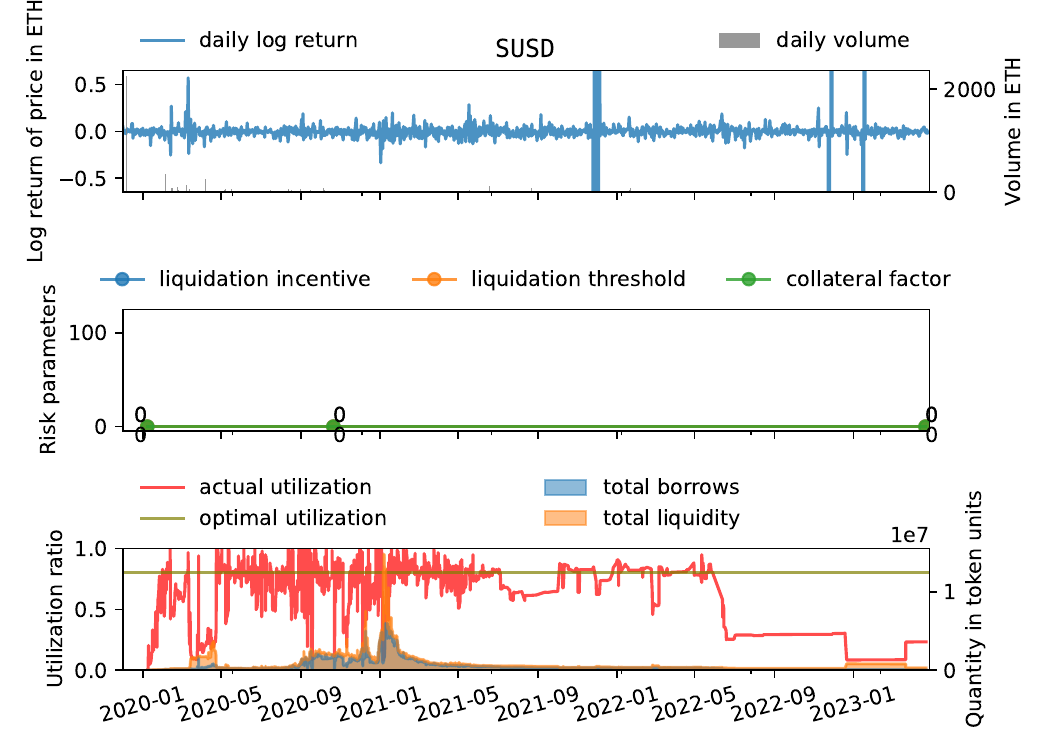}
\qquad
\includegraphics[width=.39\textwidth]{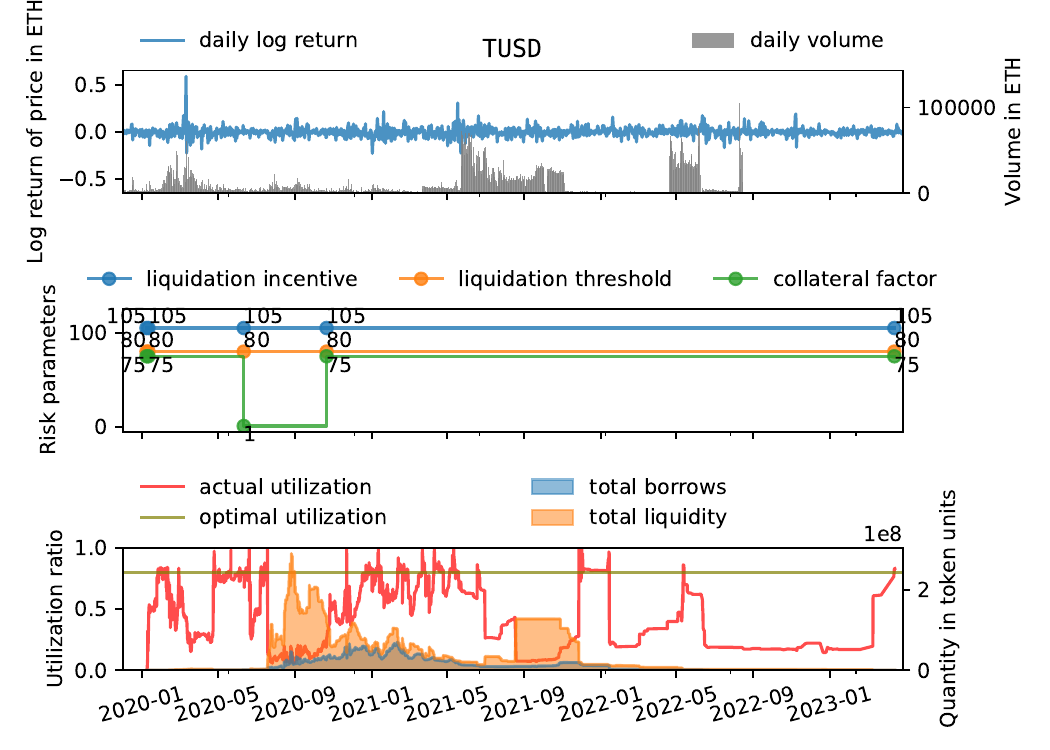}

\vspace{10pt}
\includegraphics[width=.39\textwidth]{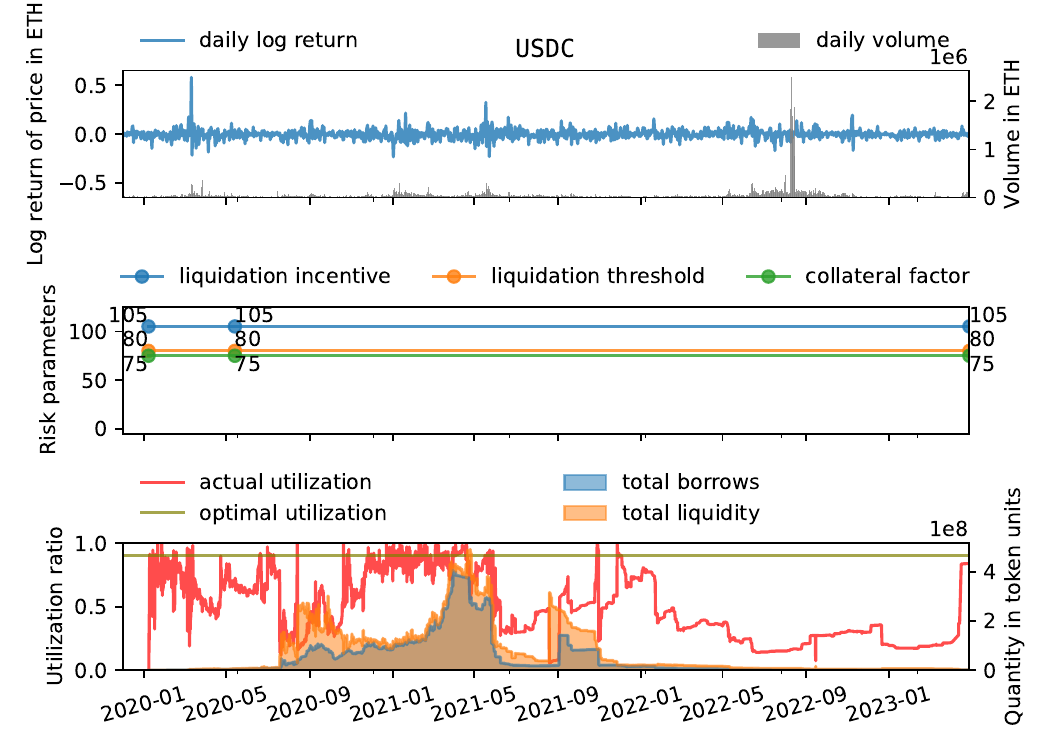}
\qquad
\includegraphics[width=.39\textwidth]{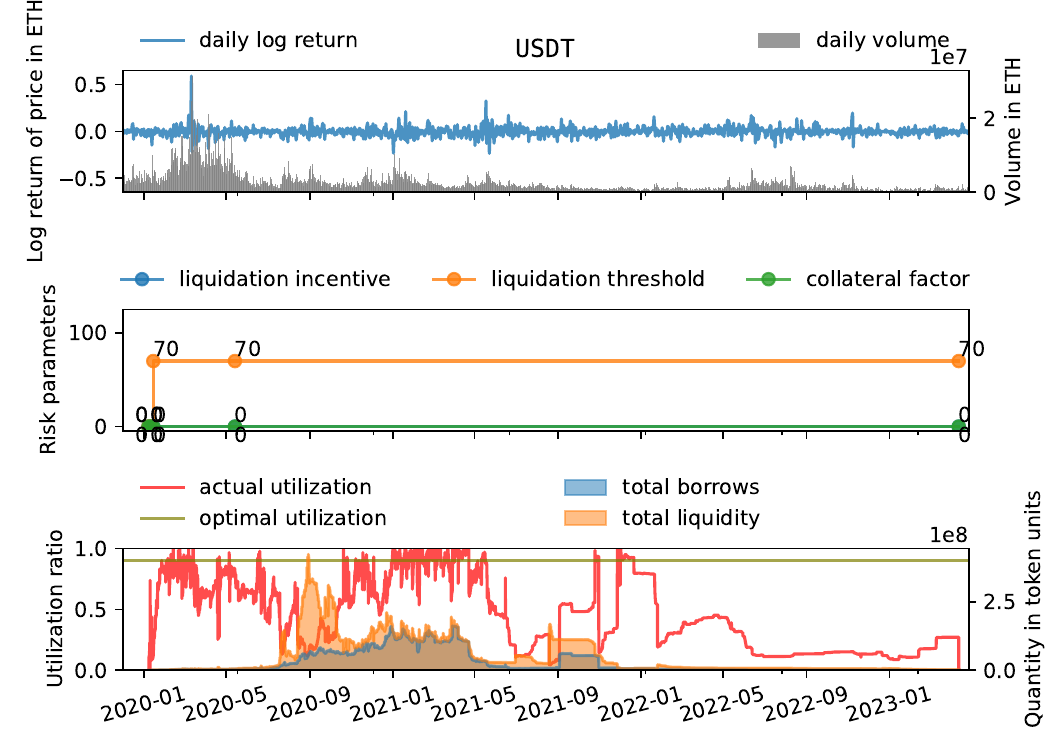}
\caption{USD-pegged stablecoins}
\end{subfigure}
\end{figure*}

\begin{figure*}
\centering
\ContinuedFloat
\begin{subfigure}{\textwidth}
\centering
\includegraphics[width=.39\textwidth]{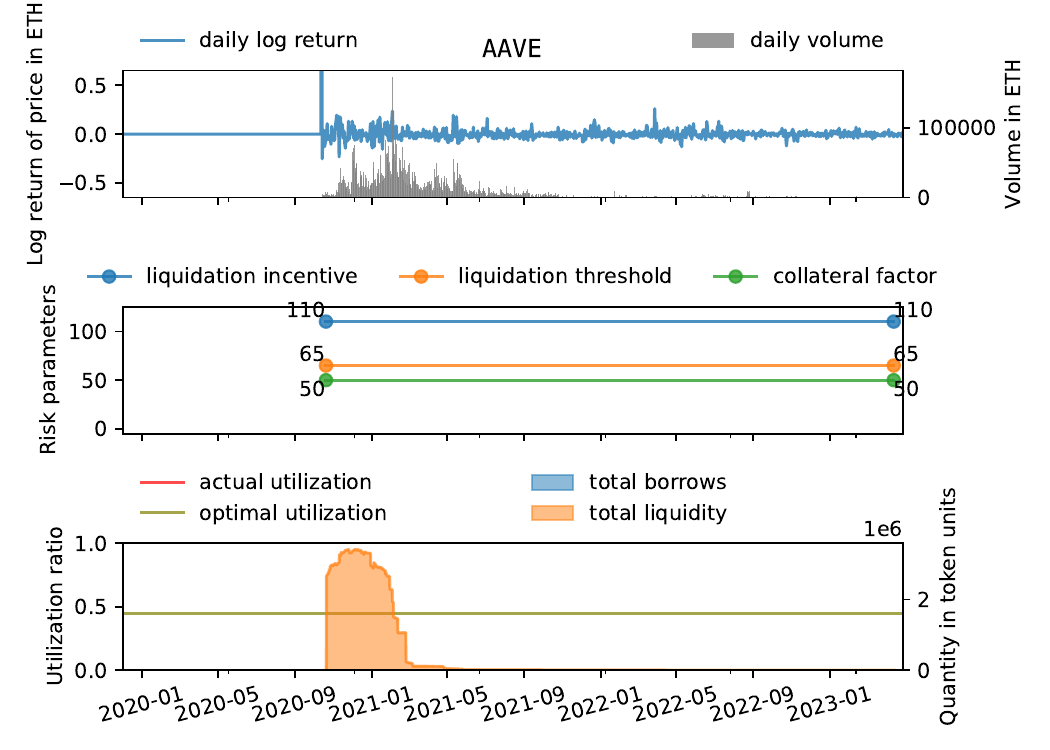}
\qquad
\includegraphics[width=.39\textwidth]{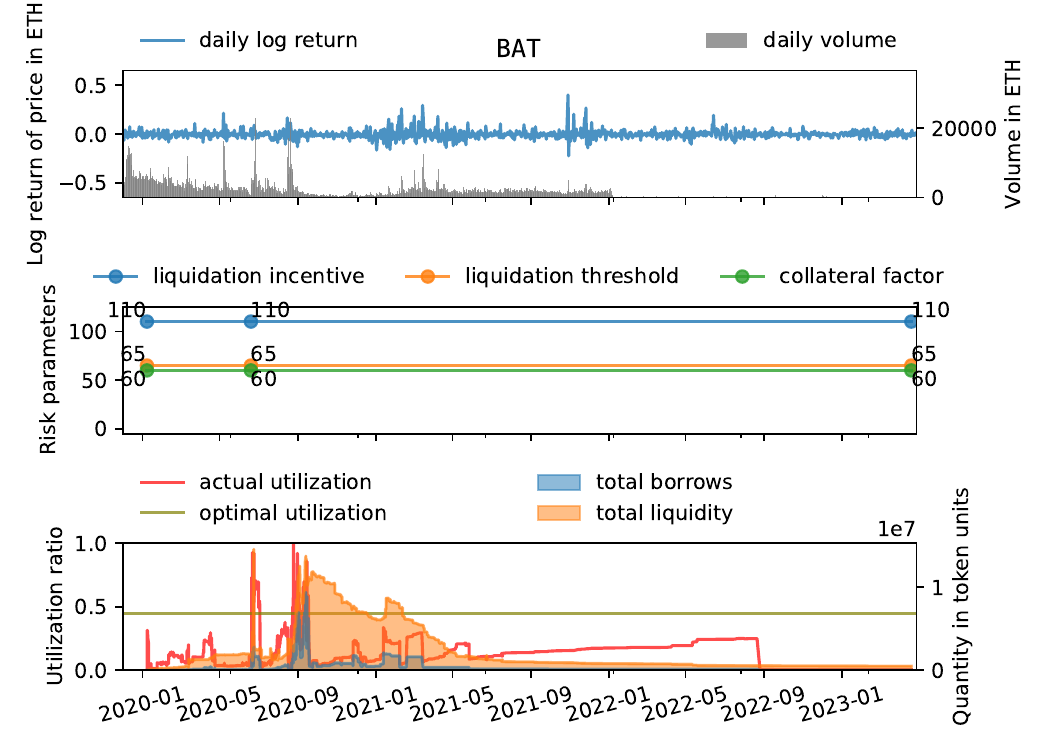}

\vspace{10pt}
\includegraphics[width=.39\textwidth]{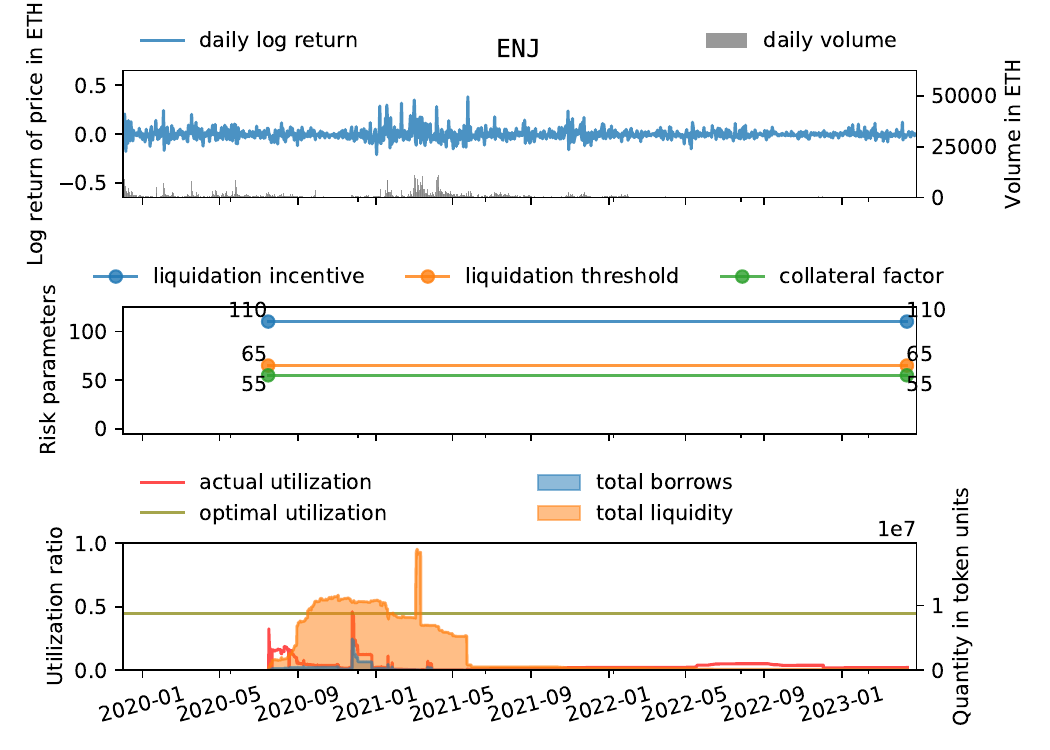}
\qquad
\includegraphics[width=.39\textwidth]{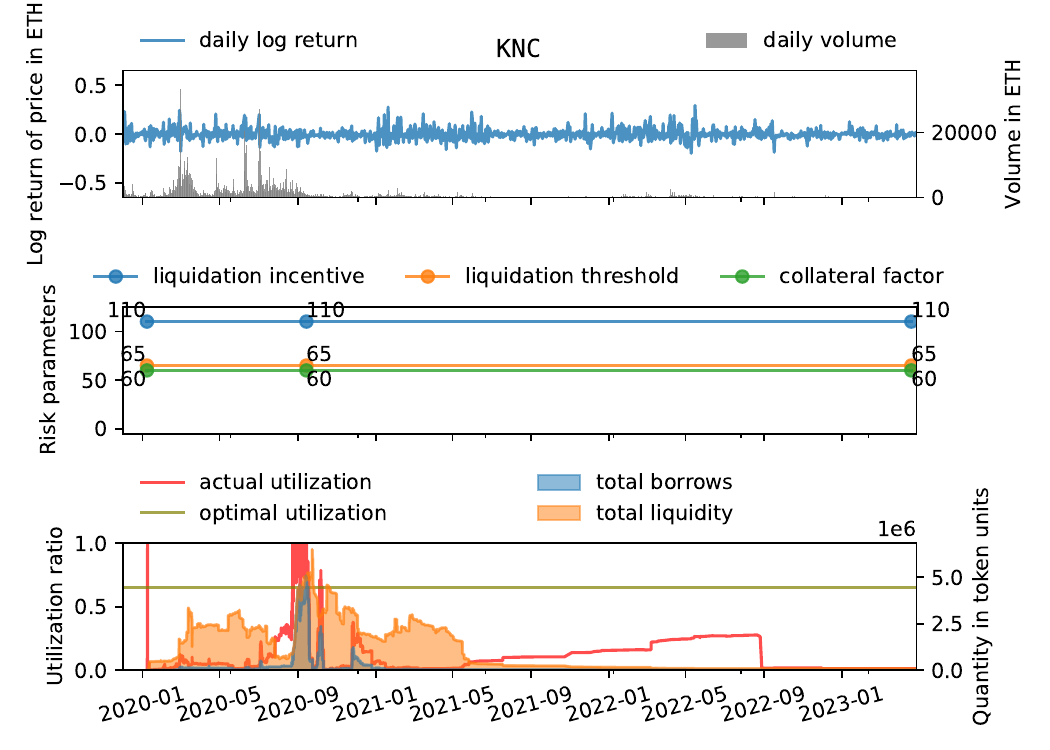}

\vspace{10pt}
\includegraphics[width=.39\textwidth]{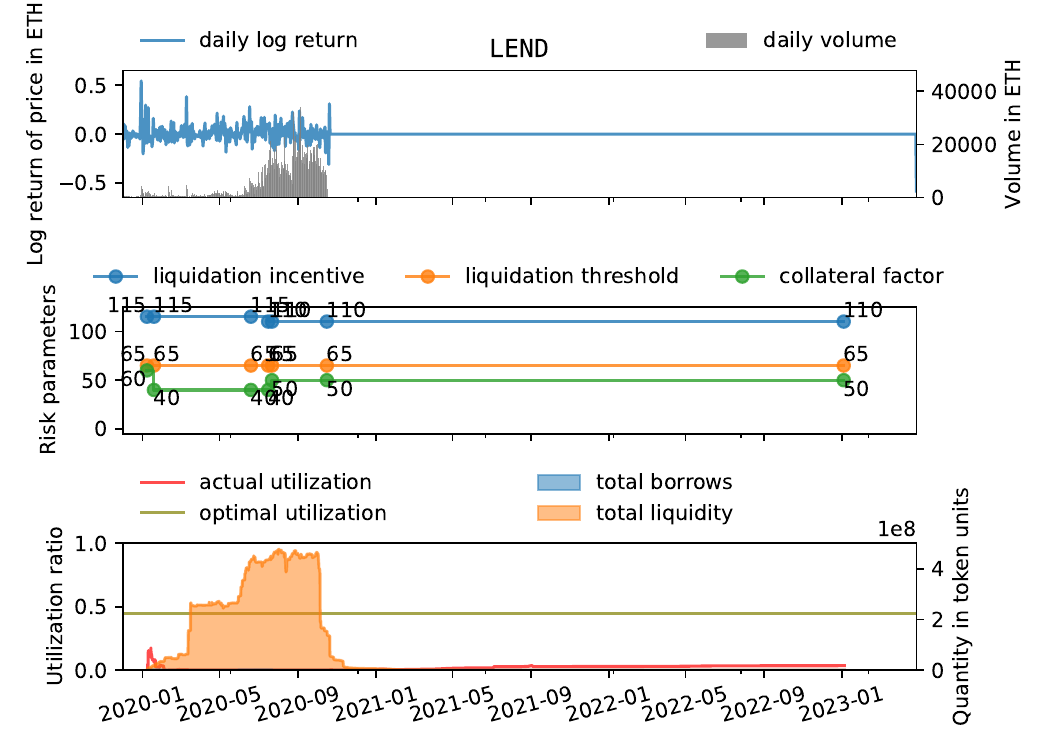}
\qquad
\includegraphics[width=.39\textwidth]{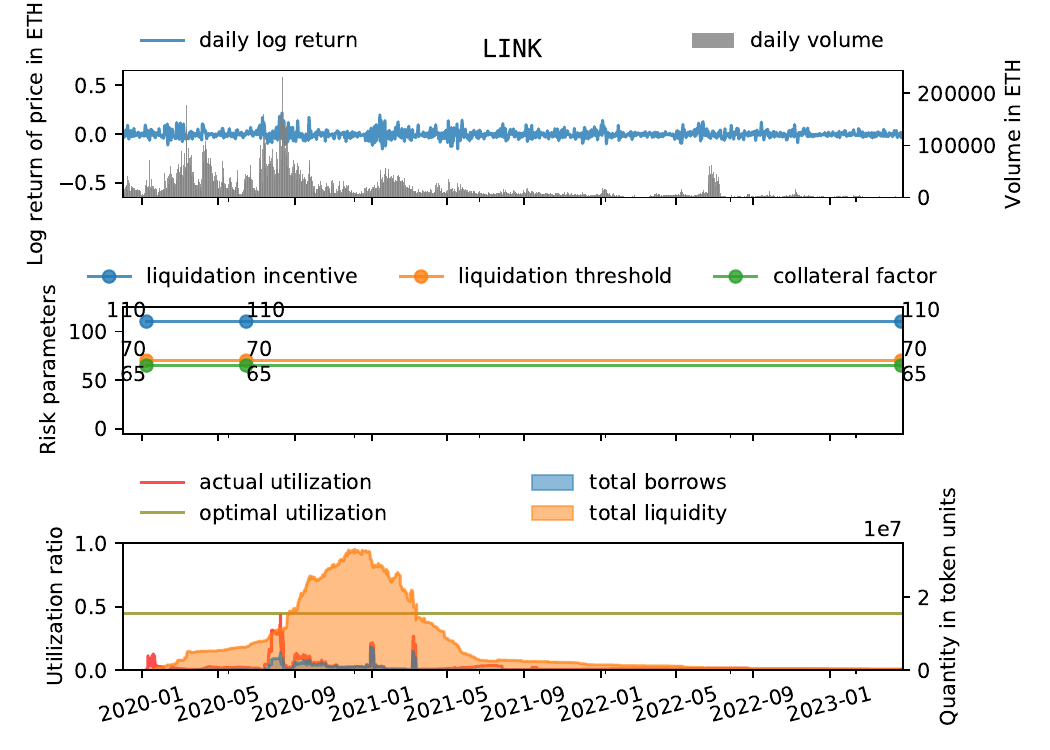}

\vspace{10pt}
\includegraphics[width=.39\textwidth]{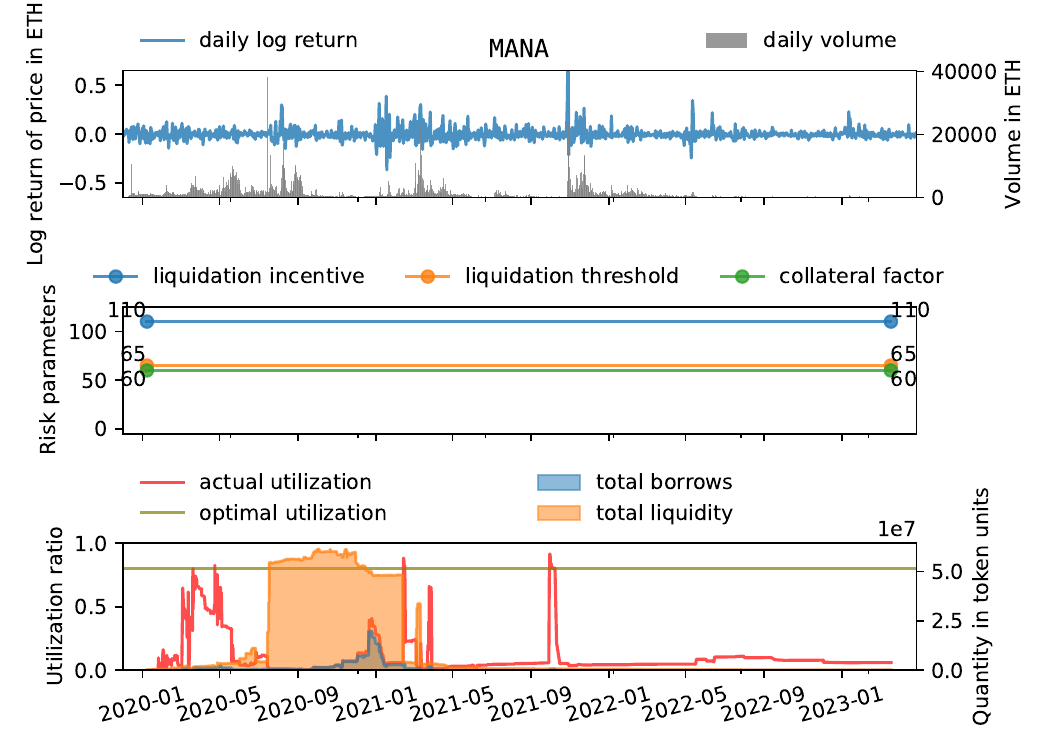}
\qquad
\includegraphics[width=.39\textwidth]{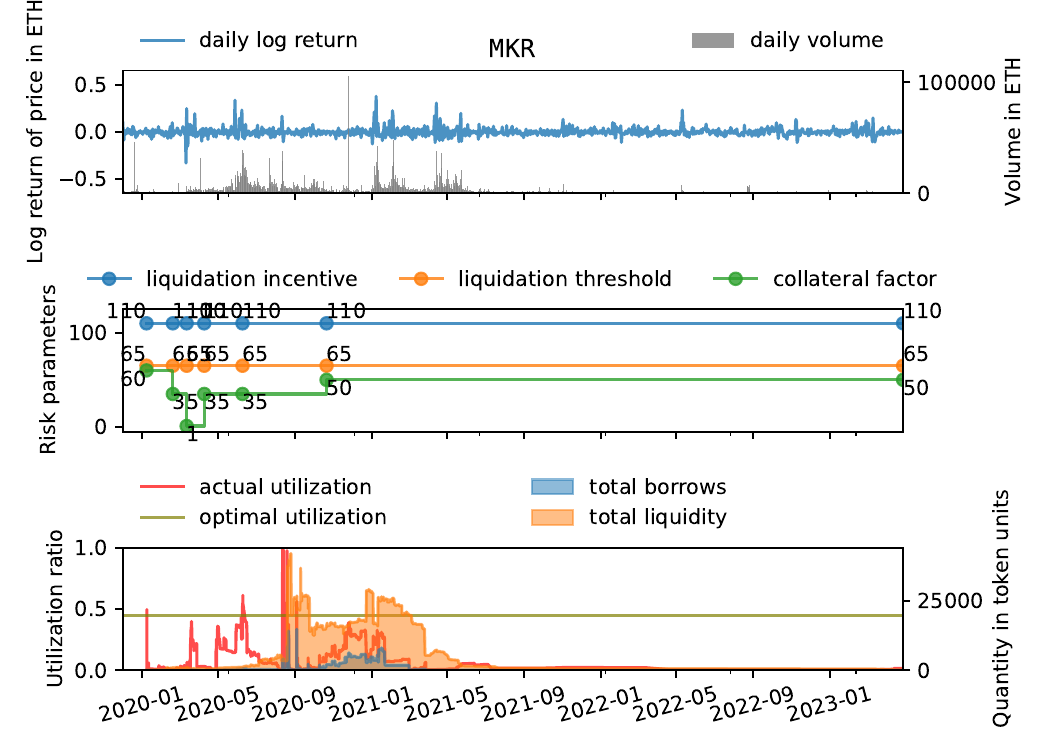}
\end{subfigure}
\end{figure*}

\begin{figure*}
\centering
\ContinuedFloat
\begin{subfigure}{\textwidth}
\centering
\includegraphics[width=.39\textwidth]{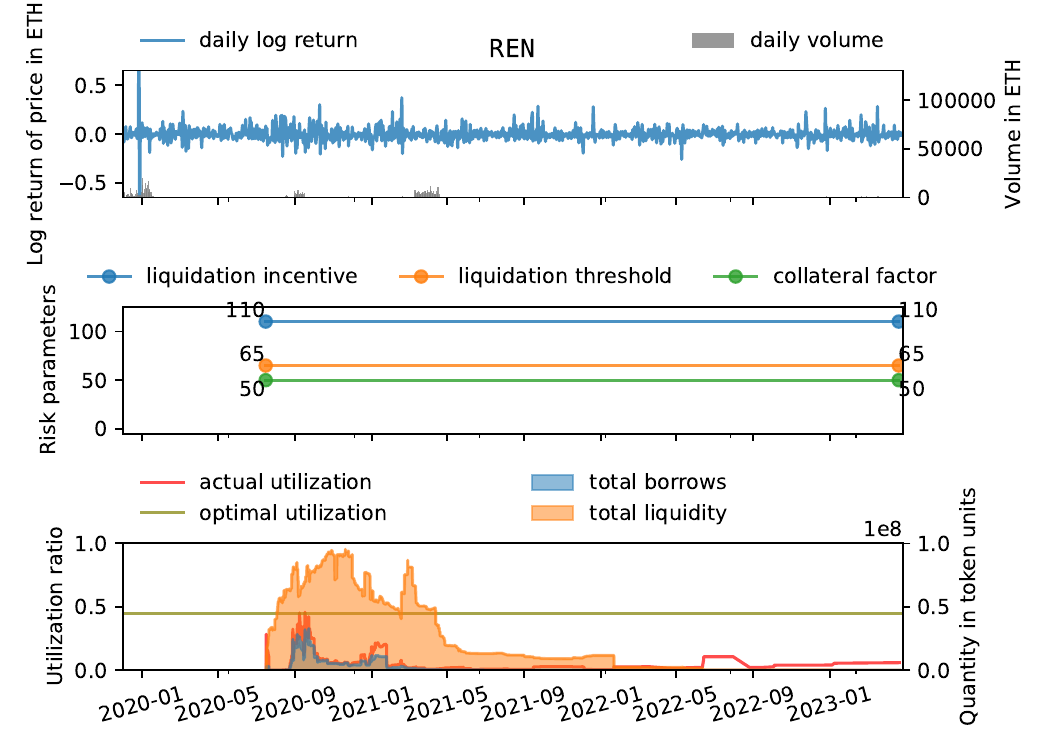}
\qquad
\includegraphics[width=.39\textwidth]{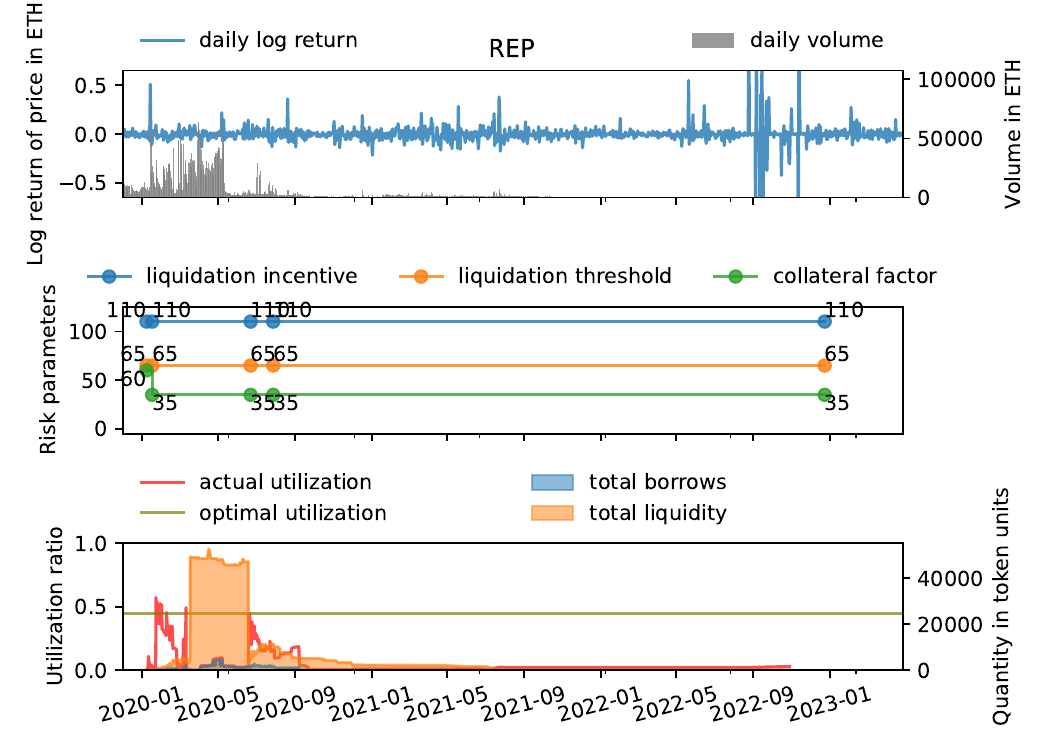}

\vspace{10pt}
\includegraphics[width=.39\textwidth]{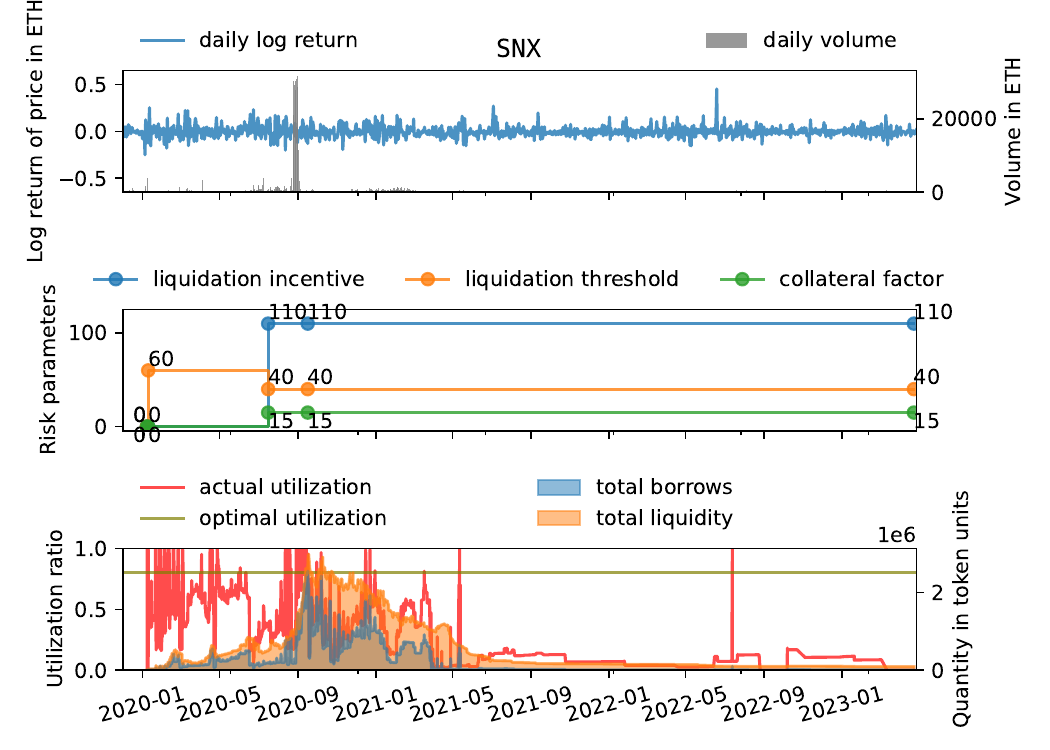}
\qquad
\includegraphics[width=.39\textwidth]{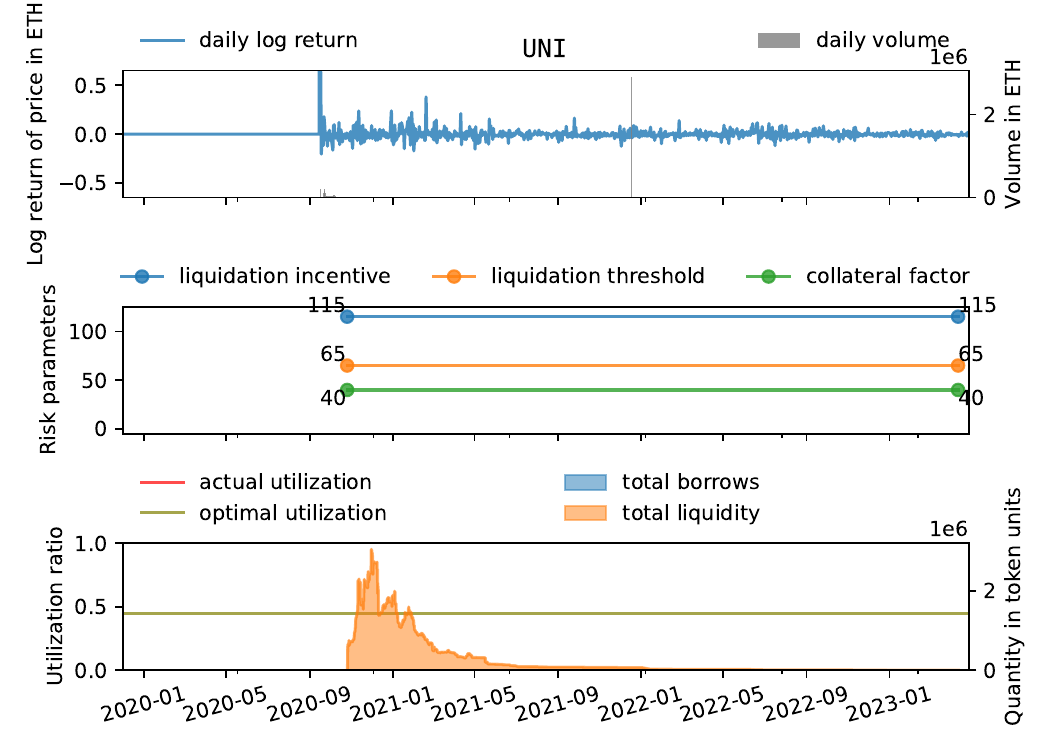}

\vspace{10pt}
\includegraphics[width=.39\textwidth]{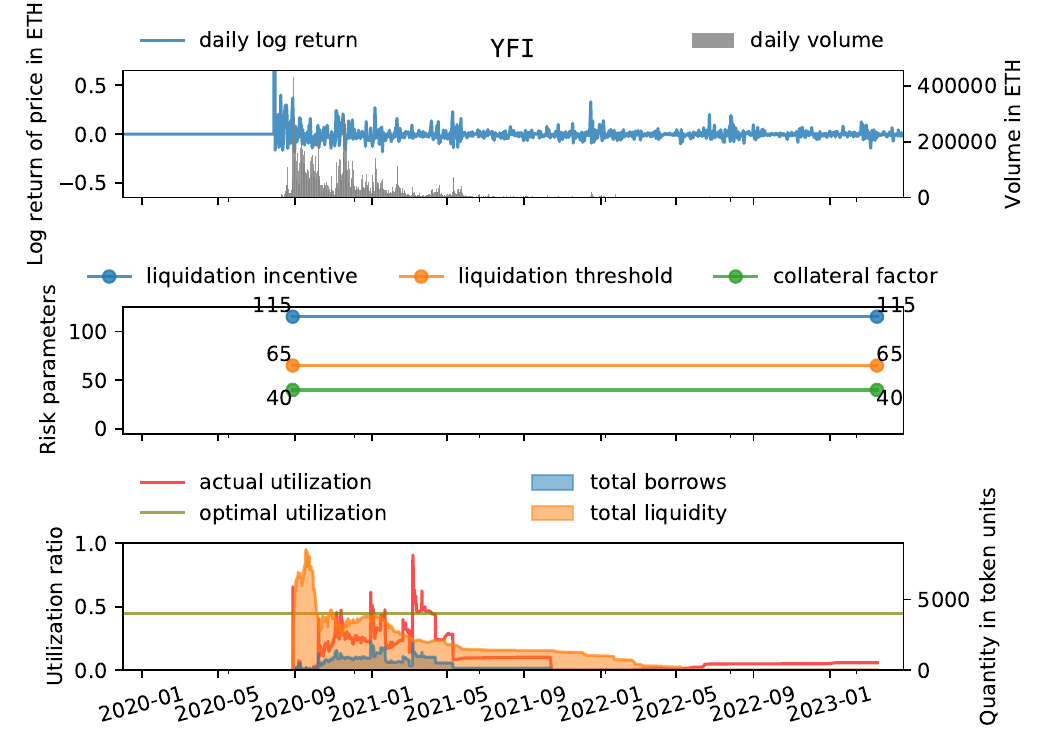}
\qquad
\includegraphics[width=.39\textwidth]{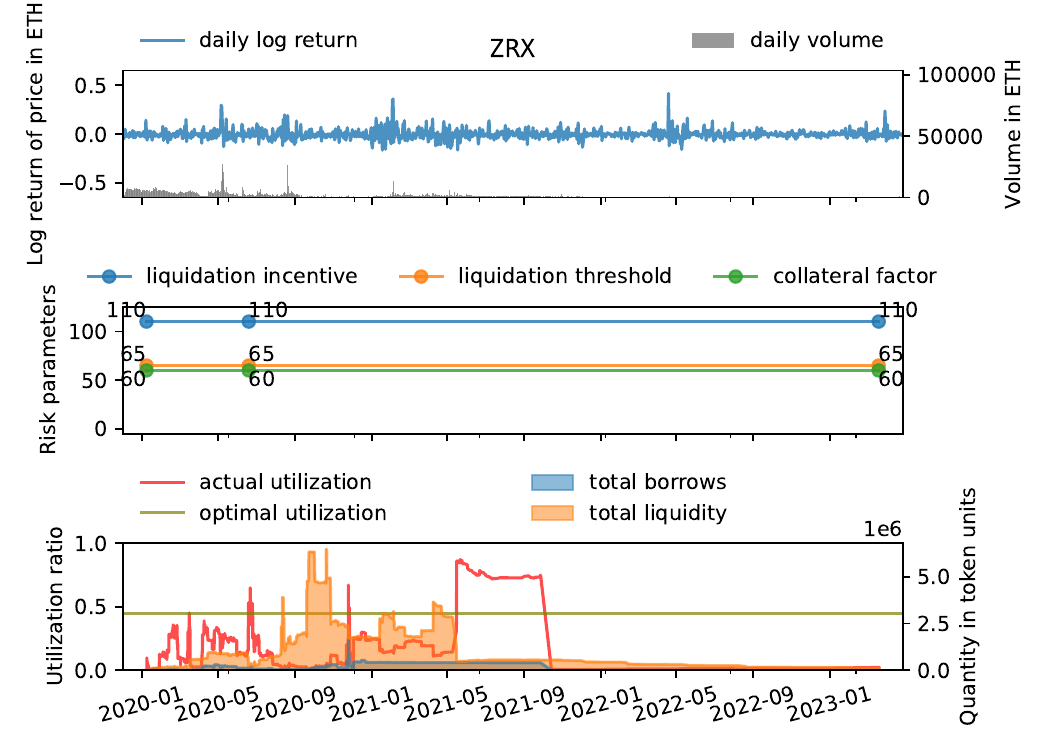}
\caption{Application-layer protocol governance tokens}
\end{subfigure}

\caption{Time series of various crypto-assets' market conditions (return, volume) juxtaposed with their corresponding risk parameter (liquidation incentive, liquidation threshold, collateral factor) values, as well as other state variable values (utilization ratio, total borrow, total liquidity) of lending pools on Aave.}
\label{fig:atokens}
\end{figure*}

\end{document}